\documentclass[fleqn,usenatbib,useAMS]{mnras}
\usepackage{eso-pic}

\AddToShipoutPictureBG*{%
  \AtPageUpperLeft{%
    \hspace{0.75\paperwidth}%
    \raisebox{-3.5\baselineskip}{%
      \makebox[0pt][l]{\textnormal{DES-2019-0466}}
}}}%

\AddToShipoutPictureBG*{%
  \AtPageUpperLeft{%
    \hspace{0.75\paperwidth}%
    \raisebox{-4.5\baselineskip}{%
      \makebox[0pt][l]{\textnormal{FERMILAB-PUB-20-663-E-SCD-V}}
}}}%

\usepackage[T1]{fontenc}
\usepackage{graphicx}
\DeclareGraphicsExtensions{.pdf,.png}
\usepackage{xcolor}
\usepackage{amsmath}
\usepackage{amssymb}
\usepackage{comment}
\usepackage[normalem]{ulem}
\usepackage{footnote}
\usepackage{widetext}
\usepackage{dsfont}
\usepackage{enumitem}
\usepackage{bbold}
\usepackage[switch,mathlines]{lineno}

\title[DES Y3: Covariance validation]{Dark Energy Survey Year 3 Results: Covariance Modelling and its Impact on Parameter Estimation and Quality of Fit}

\newcommand{\etc}{etc$.$}

\newcommand{\eg}{e.g$.$}

\newcommand{\ie}{i.e$.$}
\newcommand{\cf}{cf$.$}
\newcommand{\resp}{resp$.$}
\newcommand{\dd}{\ensuremath{\mathrm{d}}}

\newcommand{\figref}[1]{Figure \ref{fig:#1}}
\newcommand{\figrefnospace}[1]{Figure \ref{fig:#1}}
\newcommand{\Figref}[1]{Figure \ref{fig:#1}}

\newcommand{\eqnref}[1]{Equation \ref{eq:#1}}
\newcommand{\eqnrefnospace}[1]{Equation \ref{eq:#1}}
\newcommand{\Eqnref}[1]{Equation \ref{eq:#1}}

\newcommand{\appref}[1]{appendix \ref{app:#1}}
\newcommand{\apprefnospace}[1]{appendix \ref{app:#1}}

\newcommand{\secref}[1]{Section \ref{sec:#1}}
\newcommand{\secrefnospace}[1]{Section \ref{sec:#1}}
\newcommand{\Secref}[1]{Section \ref{sec:#1}}

\newcommand{\tabref}[1]{Table \ref{tab:#1} }
\newcommand{\Tabref}[1]{Table \ref{tab:#1} }
\newcommand{\tabrefnospace}[1]{Table \ref{tab:#1}}

\usepackage[mathlines]{lineno}

\usepackage{etoolbox} 

\newcommand*\linenomathpatch[1]{%
  \expandafter\pretocmd\csname #1\endcsname {\linenomath}{}{}%
  \expandafter\pretocmd\csname #1*\endcsname{\linenomath}{}{}%
  \expandafter\apptocmd\csname end#1\endcsname {\endlinenomath}{}{}%
  \expandafter\apptocmd\csname end#1*\endcsname{\endlinenomath}{}{}%
}
\newcommand*\linenomathpatchAMS[1]{%
  \expandafter\pretocmd\csname #1\endcsname {\linenomathAMS}{}{}%
  \expandafter\pretocmd\csname #1*\endcsname{\linenomathAMS}{}{}%
  \expandafter\apptocmd\csname end#1\endcsname {\endlinenomath}{}{}%
  \expandafter\apptocmd\csname end#1*\endcsname{\endlinenomath}{}{}%
}

\expandafter\ifx\linenomath\linenomathWithnumbers
  \let\linenomathAMS\linenomathWithnumbers
  \patchcmd\linenomathAMS{\advance\postdisplaypenalty\linenopenalty}{}{}{}
\else
  \let\linenomathAMS\linenomathNonumbers
\fi

\linenomathpatch{equation} 
\linenomathpatchAMS{gather}
\linenomathpatchAMS{multline}
\linenomathpatchAMS{align}
\linenomathpatchAMS{alignat}
\linenomathpatchAMS{flalign}

\author[DES Collaboration]{
\parbox{\textwidth}{
\Large
O.~Friedrich,$^{1,2}$
F.~Andrade-Oliveira,$^{3,4}$
H.~Camacho,$^{3,4}$
O.~Alves,$^{5,3}$
R.~Rosenfeld,$^{6,4}$
J.~Sanchez,$^{7}$
X.~Fang,$^{8}$
T.~F.~Eifler,$^{8,9}$
E.~Krause,$^{8}$
C.~Chang,$^{10,11}$
% ...
\begin{center} (DES Collaboration) \end{center}
}
\vspace{-1cm}
\\
}

\defcitealias{Planck2020}{Planck Collaboration 2020}

\begin{document}
\maketitle
\date{\today}
\begin{abstract}
We describe and test the fiducial covariance matrix model for the combined 2-point function analysis of the Dark Energy Survey Year 3 (DES-Y3) dataset. Using a variety of new ansatzes for covariance modelling and testing we validate the assumptions and approximations of this model. These include the assumption of Gaussian likelihood, the trispectrum contribution to the covariance, the impact of evaluating the model at a wrong set of parameters, the impact of masking and survey geometry, deviations from Poissonian shot-noise, galaxy weighting schemes and other, sub-dominant effects. We find that our covariance model is robust and that its approximations have little impact on goodness-of-fit and parameter estimation. The largest impact on best-fit figure-of-merit arises from the so-called $f_{\mathrm{sky}}$ approximation for dealing with finite survey area, which on average increases the $\chi^2$ between maximum posterior model and measurement by $3.7\%$ ($\Delta \chi^2 \approx 18.9$). Standard methods to go beyond this approximation fail for DES-Y3, but we derive an approximate scheme to deal with these features. For parameter estimation, our ignorance of the exact parameters at which to evaluate our covariance model causes the dominant effect. We find that it increases the scatter of maximum posterior values for $\Omega_m$ and $\sigma_8$ by about $3\%$ and for the dark energy equation of state parameter by about $5\%$.
\end{abstract}

\makeatletter
\def \blfootnote{\xdef\@thefnmark{}\@footnotetext}
\makeatother

\blfootnote{$^{\dagger}$ E-mail: of259@ast.cam.ac.uk}

${}$

\section{Introduction}
\label{sec:intro}

Our understanding of the Universe has become much more accurate in the past decades due to a massive amount of observational data collected through different probes, such as the cosmic microwave background \citepalias[CMB; see \eg][]{Planck2020}, Big Bang Nucleosynthesis \citep[BBN; see \eg][]{Fields2020}, type IA supernovae \citep[see \eg][]{Riess2017, Smith2020}, number counts of clusters of galaxies \citep[see \eg][]{Mantz2014, Costanzi2019, DES_Clusters_2020}, the correlation of galaxy positions, and that of their measured shape \citep[see \eg][]{DES2018, Heymans2020}. From the study of that data a standard cosmological model has emerged characterized by a small number of parameters \citep[see \eg\ ][]{Frieman:2008sn,2012ARA&A..50....1P,Blandford:2020omc}. Current spectroscopic and photometric surveys of galaxies such as the Extended Baryon Oscillation Spectroscopic Survey (eBOSS\footnote{\tt www.sdss.org/surveys/eboss}) and earlier phases of the Sloan Digital Sky Survey (SDSS), the Hyper Suprime-Cam Subaru Strategic Program (HSC-SSP\footnote{\tt hsc.mtk.nao.ac.jp/ssp}), the Kilo-Degree Survey (KiDS\footnote{\tt kids.strw.leidenuniv.nl}) and the Dark Energy Survey (DES\footnote{\tt www.darkenergysurvey.org}) have become instrumental in testing this standard model at a new front: the growth of density perturbations in the late-time Universe. And future surveys, such as the Dark Energy Spectroscopic Instrument (DESI\footnote{\tt www.desi.lbl.gov}), the Vera Rubin Observatory Legacy Survey of Space and Time (LSST\footnote{\tt www.lsst.org}), Euclid\footnote{\tt www.euclid-ec.org} and the Nancy Grace Roman Space Telescope \footnote{\tt nasa.gov/content/goddard/nancy-grace-roman-space-telescope} will push this test to a precision exceeding that provided by other cosmological probes.

An important part of this program is the Dark Energy Survey, a
state-of-the-art galaxy survey that completed its six-year observational campaign in January 2019 \citep{Diehl:2019dmi} collecting data on position, color and shape for more than 300 million galaxies. This makes DES the most sensitive and comprehensive photometric galaxy survey ever performed. The main cosmological analyses of the first year (Y1) of DES data have been concluded \citep{DES2018,Abbott:2018wzc} and analyses of the first three years of data (Y3) are under way. The study of the large-scale structure (LSS) of the Universe based on the DES-Y3 data set has the potential to become the most stringent test of our understanding of cosmological physics to date.

To achieve this goal the DES team is comparing different theoretical models characterized by a range of cosmological parameters to the measured statistics of the LSS in order to determine the model and range of parameters that are in best agreement with the data. The statistics of the LSS considered in the main DES-Y3 analysis are 2-point correlation functions of the galaxy density field (galaxy clustering), the weak gravitational lensing field (cosmic shear) and the cross-correlation functions between these fields (galaxy-galaxy lensing) in real space and measured in different redshift bins. These three types of 2-point correlation functions are combined into one data vector - the so-called 3x2pt data vector.

A key ingredient in analyzing these statistics is a model for the likelihood of a cosmological model given the measured correlation functions. Under the assumption of Gaussian statistical uncertainties (which is to be validated) this likelihood is completely characterized by the covariance matrix that describes how correlated the uncertainties of different data points in the 3x2pt data vector are. Validating the quality of the covariance model for the DES-Y3 2-point analyses is the main focus of this paper.

There are several methods to estimate covariance matrices that can roughly be divided into four main categories: covariance estimation from the data itself \citep[\eg through jackknife or sub-sampling methods, \cf][]{Norberg2009, Friedrich2016}, covariance estimation from a suite of simulations \citep[\eg][]{Hartlap:2006kj, Dodelson2013, Taylor2013, Percival2014, Taylor2014, Sellentin2017, Joachimi2017, Avila:2017nyy, Shirasaki2019}, theoretical covariance modelling \citep[\eg][]{Schneider2002, Eifler2009, Krause2017} or hybrid methods combining both simulations and theoretical covariance models \citep[\eg][]{Pope2008, Friedrich_Eifler, Hall2019}.

For the DES-Y3 3x2pt analysis we adopt a theoretical covariance model as our fiducial covariance matrix. This fiducial covariance model is based on a halo model and includes a dominant Gaussian component, a non-Gaussian component (trispectrum and super-sample covariance), redshift space distortions, curved sky formalism, finite angular bin width, non-Limber computation for the clustering part, Gaussian shape noise, Poissonian shot noise and $f_{\mbox{sky}}$ approximation to treat the finite DES-Y3 survey footprint (although taking into account the exact survey geometry when computing sampling noise contributions to the covariance). In order to assess the accuracy of that model, we study the impact of several approximations and assumptions that go into it (and into 2-point function covariance models in general):
\begin{itemize}
     \item the Gaussian likelihood assumption, \ie\ whether knowledge of the covariance is sufficient to calculate the likelihood;
     \item robustness with respect to the modelling of the non-Gaussian covariance contributions, \ie\ contributions from the trispectrum and super sample covariance;
     \item treatment of the fact that 2-point functions are measured in finite angular bins;
     \item cosmology dependence of the covariance model;
     \item random point shot-noise;
     \item the assumption of Poissonian shot-noise;
     \item survey geometry and the $f_{\mathrm{sky}}$ approximation;
     \item other covariance modelling details such as flat sky vs. curved sky calculations, Limber approximation and redshift space distortions.
\end{itemize}
We generate different types of mock data and/or analytical estimates to determine how each of these effects impacts the quality of the fit between measurements of the 3x2pt data vector and maximum posterior models (quantified by the distribution of $\chi^2$ between the two). We also show how they impact cosmological parameter constraints derived from measurements of the 3x2pt data vector. For most of these tests we employ a linearized Gaussian likelihood framework which allows us to analytically quantify the impact of covariance errors on the $\chi^2$ distribution and parameter constraints. This is complemented by a set of lognormal simulations and importance sampling techniques to quickly assess large numbers of mock (non-linear) likelihood analyses.

This paper is part of a larger release of scientific results from year-3 data of the Dark Energy Survey and our analysis is informed by the (in some cases preliminary) analysis choices of the other DES-Y3 studies. In addition to carving out the most stringent constraints on cosmological parameters from late-time 2-point statistics of galaxy density and cosmic shear yet, the year-3 analysis of the DES collaboration is introducing and testing numerous methodological innovations that pave the way for future experiments. Details of the DESY3 galaxy catalogs and the photometric estimation of their redshift distribution are presented by \citet{y3-gold, y3-deepfields, y3-balrog, y3-sompz, y3-sourcewz, y3-lenswz, y3-sompzbuzzard, y3-hyperrank}. The measurements of galaxy shapes and the calibration of these measurements for the purpose of cosmic weak gravitational lensing analyses are detailed by \citet{y3-shapecatalog, y3-piff, y3-imagesims}. \citet{y3-generalmethods} develop and test the theoretical modelling pipeline of the DES-Y3 3x2pt analysis, \citet{y3-2x2ptbiasmodelling} outline how galaxy bias is incorporated in this pipeline, \citet{y3-simvalidation} validate this pipeline with the help of simulated data and \citet{y3-blinding} describe how we have blinded our analysis to focus our efforts on model independent validation criteria and reduce the chance for confirmation bias. The DESY3 methodology to sample high-dimensional likelihoods and to characterize external and internal tensions is outlined by \citet{y3-tensions, y3-inttensions}. Measurements of cosmic shear 2-point correlation functions and analyses thereof are presented by \citet{y3-cosmicshear1, y3-cosmicshear2}, the measurement and analysis of galaxy clustering 2-point statistics is carried out by \citet{y3-galaxyclustering} and 2-point cross-correlations between galaxy density and cosmic shear (galaxy-galaxy lensing) are measured and analysed by \citet{y3-gglensing}, with additional analyses of lensing magnification and shear ratios carried out by \citet{y3-2x2ptmagnification, y3-shearratio} and results for an alternative lens galaxy sample presented by \citet{y3-2x2maglimforecast, y3-2x2ptaltlensresults}. Finally, in \citet{y3-3x2ptkp} we present our cosmological analysis of the full 3x2pt data vector.

Our paper is structured as follows. We start by presenting a discussion of our validation strategy in \secref{validation}, where we also summarize our main findings before plunging into the details in the remaining of the paper. In \secref{model} we review the modelling and structure of the 3x2pt data vector. \Secref{covariance_modelling} describes our fiducial covariance model as well as two alternatives to it that are used to validate several modelling assumptions. In \secref{strategy} we describe our linearized likelihood formalism and derive analytically how different covariance matrices impact parameter constraints and maximum posterior $\chi^2$ within that formalism (including the presence of nuisance parameters and allowing for Gaussian priors on these parameters). In \secref{impact_of_modelling} we present the details of each step in our validation strategy followed by a short \secref{simplechi2} presenting a simple test to corroborate some of the results from the linearized framework. We conclude with a discussion of our results in \secref{summary}.
Seven appendices describe in more detail some results used in this work.

\section{Covariance validation strategy and summary of the results}
\label{sec:validation}

How should one validate the quality of a covariance model (and the associated likelihood model) for the purpose of constraining cosmological model parameters from a measured statistic? A straightforward answer seems to be that one should run a large number of accurate cosmological simulations, then measure and analyse the statistic at hand in each of the simulated data sets and test whether the true parameters of the simulations are located within the, say, $68.3\%$ quantile of the inferred parameter constraints in $68.3\%$ of the simulations. There are however at least 2 problems with such an approach.

The first one is a conceptual problem. Consider a Bayesian analysis of a measured statistic $\boldsymbol{\hat\xi}$ with a model $\boldsymbol{\xi}[\boldsymbol{\pi}]$ that is parametrised by model parameters $\boldsymbol{\pi}$. For each value of $\boldsymbol{\pi}$ the statistical uncertainties in the measurement $\boldsymbol{\hat\xi}$ will have some distribution
\begin{equation}
    \mathcal{L}(\boldsymbol{\pi}|\boldsymbol{\hat\xi}) \equiv p(\boldsymbol{\hat\xi}|\boldsymbol{\pi})
\end{equation}
which is also called the likelihood of the parameters given the data. If this function is known, then a Bayesian analysis will assign a posterior probability distribution to the parameters as
\begin{equation}
    p(\boldsymbol{\pi}|\boldsymbol{\hat\xi}) = \frac{1}{\mathcal{N}}\ \mathcal{L}(\boldsymbol{\pi}|\boldsymbol{\hat\xi})\ \mathrm{pr}(\boldsymbol{\pi})\ .
\end{equation}
Here $\mathrm{pr}(\boldsymbol{\pi})$ is a prior probability distribution that parametrises prior knowledge from other experiments (or theoretical constraints) and the normalisation constant $\mathcal{N}$ is fixed by demanding that $p(\boldsymbol{\pi}|\boldsymbol{\hat\xi})$ be a probability distribution. The $68.3\%$ confidence region for the parameters $\boldsymbol{\pi}$ would then \eg\ be stated as a volume $V_{68.3\%}$ in parameter space that contains $68.3\%$ of the probability. To unambiguously define that volume one can \eg\ impose the additional condition that
\begin{equation}
\min_{\boldsymbol{\pi} \in V_{68.3\%}} p(\boldsymbol{\pi}|\boldsymbol{\hat\xi}) \geq \max_{\boldsymbol{\pi} \notin V_{68.3\%}} p(\boldsymbol{\pi}|\boldsymbol{\hat\xi})
\end{equation}
or, more frequently, one would directly define one dimensional intervals that satisfy the above conditions for the marginalised posterior distributions on the individual parameter axes. Unfortunately, if one performs such an analysis many times one is not guaranteed that the true parameters (\eg\ of a simulation) are located within $V_{68.3\%}$ in $68.3\%$ of the times. This has recently been referred to as \emph{prior volume effect} (this issue is discussed in, e.g. \citet{Raveri:2018wln} and \citet{Abbott:2018xao}). One may argue that a Bayesian posterior should not be interpreted in terms of frequencies but that doesn't help for the task of validating this posterior on the basis of a large number of simulated data sets.
\footnote{In order to deal with the prior volume effect \citet{Joachimi2020} proposed to report parameter constraints through what they call projected joint highest posterior density. This topic will be addressed in a separate DES paper  (Raveri et al. - in prep.).
}

Another, more practical problem is the fact that it is not (yet) feasible to generate enough sufficiently accurate mock data sets to validate covariance matrices of large data vectors with high precision. We recall that for the DES- Y1 analysis, a total of 18 realistic simulated data sets were available to validate the inference pipeline \citep{MacCrann2018}. At the same time, the main reason why N-body simulations would be required to test the accuracy of covariance (and likelihood) models is to capture contributions to the covariance coming from the trispectrum (connected 4-point function) of the cosmic density field. But for DES-like analyses it has been shown that this contribution is negligible  (see \eg\ \citet{Krause2017, Barreira:2018jgd}). The reason for this is twofold: first,  very small scales (where the trispectrum contribution to the covariance would matter most) are often cut off from analyses because on these scales already the modelling of the data vector, $\boldsymbol{\xi}[\boldsymbol{\pi}]$, is inaccurate. And secondly, on small scales the covariance matrix is often dominated by effects coming from sparse sampling such as shot noise and shape noise. These covariance contributions are typically easy to model (although one has to be careful when estimating effective number densities and shape-noise dispersions or when estimating the number of galaxy pairs in the presence of complex survey footprints, see \citealt{Troxel:2017xyo, Troxel2018b}).

As a result of the considerations above we base our covariance validation strategy mostly on the use of a linearized likelihood (where the model $\boldsymbol{\xi}[\boldsymbol{\pi}]$ is linear in the parameters $\boldsymbol{\pi}$). In this framework the Bayesian likelihood allows for an interpretation in terms of frequencies - both for total and marginalised constraints. Also, this allows us to perform large numbers of simulated likelihood analyses very efficiently, without the need to run computationally expensive Markov Chain Monte Carlo (MCMC) codes. In addition, any leading order deviation from a linearized likelihood will be next-to-leading order for the purpose of studying the impact of covariance errors (\ie\ errors on errors) on our analysis.

Within the linearized likelihood formalism we confirm the findings of \citet{Krause2017, Barreira:2018jgd} for the DES-Y3 setup: both super-sample covariance and trispectrum have a negligible impact on our analysis. This allows us to estimate the impact of other assumptions in our covariance and likelihood model either analytically or by the means of simplified mock data such as lognormal simulations (as opposed to full N-body simulations, \cf\ \secrefnospace{FLASK}). 

\begin{figure*}
  \includegraphics[width=0.48\textwidth]{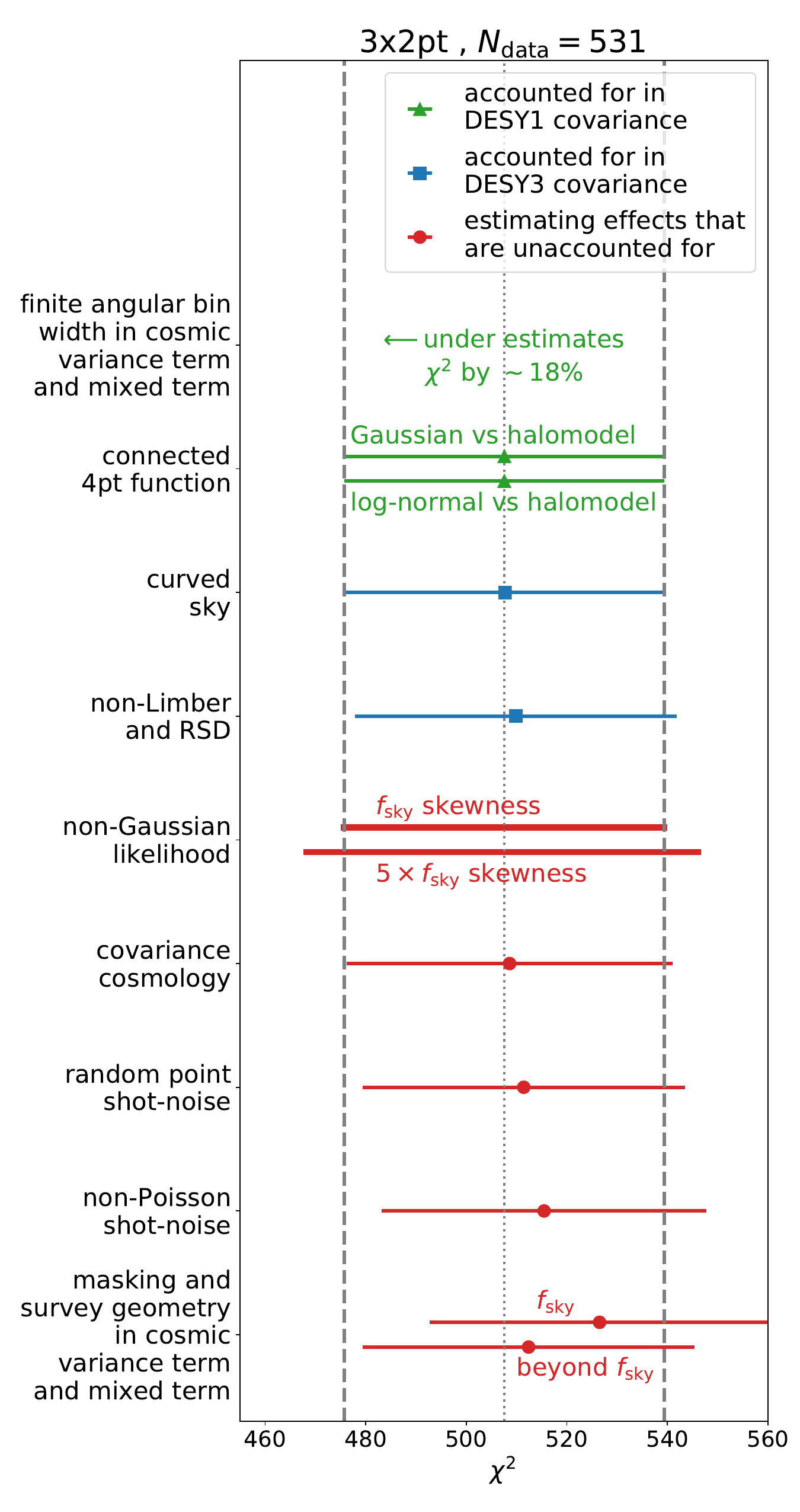} \hspace{0.1cm} \includegraphics[width=0.48\textwidth]{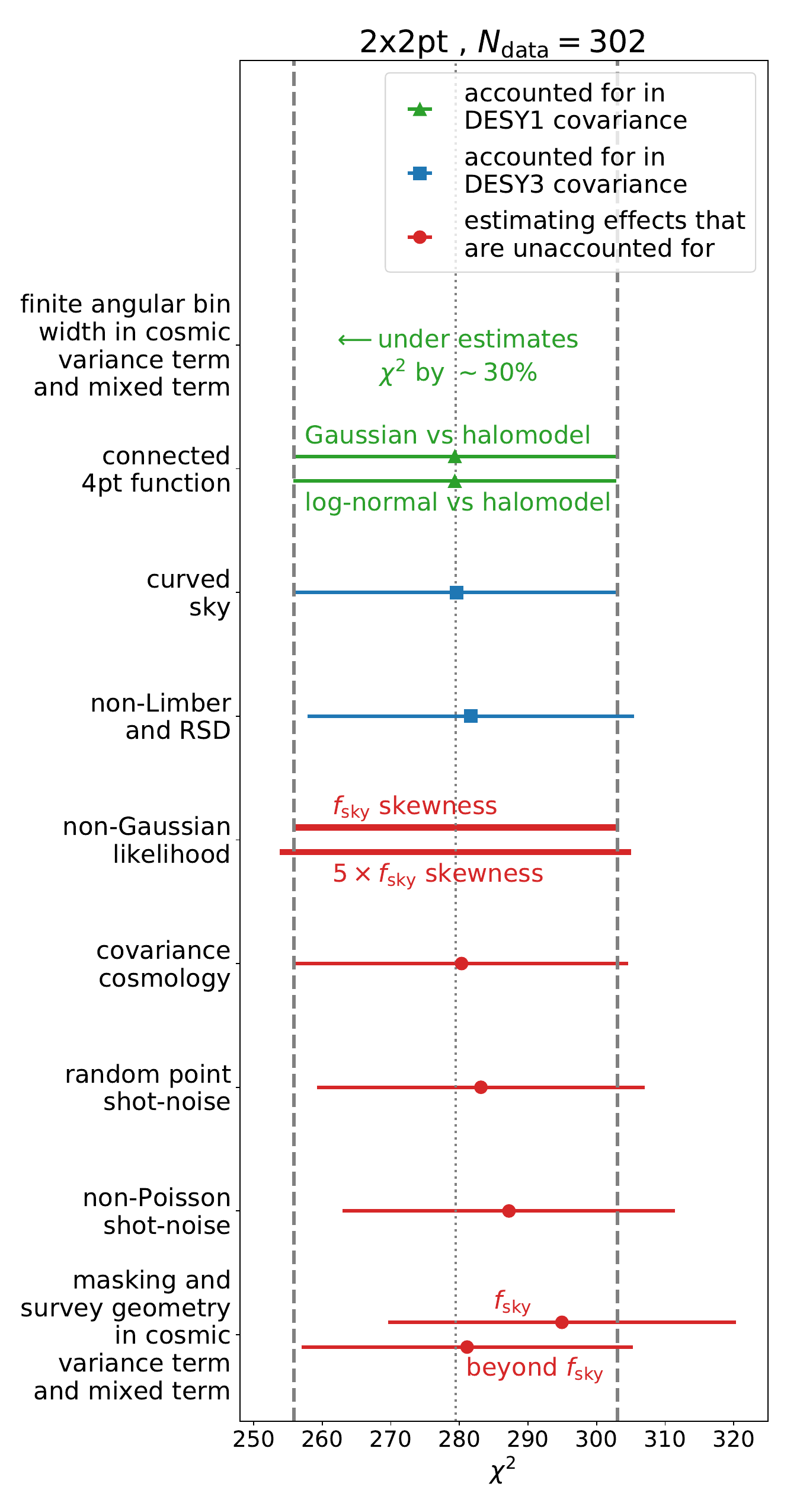}
   \caption{Impact of different covariance modelling choices on $\chi^2$ between measured 3x2pt (left panel) and 2x2pt (right panel) data vectors  and maximum posterior models.
   The dashed vertical lines and error bars indicate the $1\sigma$ fluctuations expected in $\chi^2$. See main text for details.
   }
  \label{fig:chiSq_offsets}
\end{figure*}

We summarize our main findings in \Figref{chiSq_offsets} and \tabrefnospace{summary} for the busy reader. For the combined data vector of the DES-Y3 two-point function analysis (the 3x2pt data vector, see details in \secrefnospace{model}) the left panel of \Figref{chiSq_offsets} shows the impact of different assumptions in our likelihood model on the mean and scatter of $\chi^2$ between maximum posterior model and measurements. To obtain the maximum posterior model we are fitting for all the 28 parameters listed in \tabref{params} within the linearized likelihood framework described in \secrefnospace{linearized_likelihoods}. Since we assume Gaussian priors on 13 nuisance parameters, the effective number of parameters in that fit will be between $28$ and $15$. Within the linearized likelihood approach we find that with a perfect covariance model the average $\chi^2$ is expected to be about $507.6$, \ie\ the effective number of degrees of freedom in the fit is $N_{\mathrm{param,eff}} \approx 23.4$. The right panel of \Figref{chiSq_offsets} shows the equivalent results when cosmic shear correlation functions are excluded from the data vector (the 2x2pt data vector). The green points in both panels denote effects that have been already accounted for in the previous year-1 analysis of DES.

What stands out in our analysis is the large effect of finite angular bin sizes on the cosmic variance and mixed terms of our covariance model (\cf\ \secref{covariance_modelling} for this terminology, where we also show that it is unavoidable to take into account finite bin width in the pure shot noise and shape noise terms of the covariance). In DES-Y1 this has been dealt with in an approximate manner, by computing the covariance model for a very fine angular binning and than re-summing the matrix to obtain a coarser binning \citep{Krause2017}. This time we incorporate the exact treatment of finite angular bin size for all the three two-point functions into our fiducial covariance model (\cf\ \secrefnospace{covariance_modelling}). 
The blue points in \Figref{chiSq_offsets} denote improvements that have been made in the year-3 analysis compared to the year-1 covariance model. And the red points are estimates of effects that are not taken into account in the fiducial DES-Y3 likelihood - either because they are negligible, or because an exact treatment is unfeasible (\cf\ \secref{impact_of_modelling} for details). Adding these effects in quadrature, our results suggest that the maximum posterior $\chi^2$ of the DES-Y3 3x2pt analysis should be on average $\approx 4\%$ ($\Delta \chi^2 \approx 20.3$) higher than expected if the exact covariance matrix of our data vector was known.

\Tabref{summary} summarizes the offsets in $\chi^2$ displayed in the left panel of \Figref{chiSq_offsets} and also shows how parameter constraints based on the 3x2pt data vector are impacted by assumptions of our covariance and likelihood model. We distinguish two effects here: the scatter of a maximum posterior parameter $\pi$ (which we denote by $\sigma[\hat \pi]$) and the width of posterior constraints inferred from our likelihood model (which we denote by $\sigma_{\pi}$).  For our tests of likelihood non-Gaussianity we state the changes in the difference between the fiducial parameter values and the upper (high) and lower (low) boundaries of the $68.3$\% quantile with respect to\ the standard deviation of the Gaussian likelihood. For our tests of the impact of covariance cosmology, we show the mean of all $\sigma_\pi$ obtained from our $100$ different covariances and also indicate the scatter of these $\sigma_\pi$ values. 

The effect that has the dominant impact on parameter constraints is that of evaluating the covariance model at a
17
 set of parameters that do not represent the exact cosmology of the Universe. When computing the covariance at 100 different cosmologies that were randomly drawn from a Monte Carlo Markov chain (run around a fiducial model data vector, see \secref{impact_of_cov_cosmology} for details) we find that the differences between these covariances introduce an additional scatter in maximum posterior parameter values. This scatter increases by about $3\%$ for $\Omega_m$ and $\sigma_8$ and by about $5\%$ for the dark energy equation of state parameter $w$. This increased scatter is in fact the dominant effect, since the width of the derived parameter constraints hardly changes between the different covariance matrices. Note especially that re-running the analysis with a covariance updated to the best-fit parameters does not mitigate this effect.

\begin{figure*}
  \includegraphics[width=0.95\textwidth]{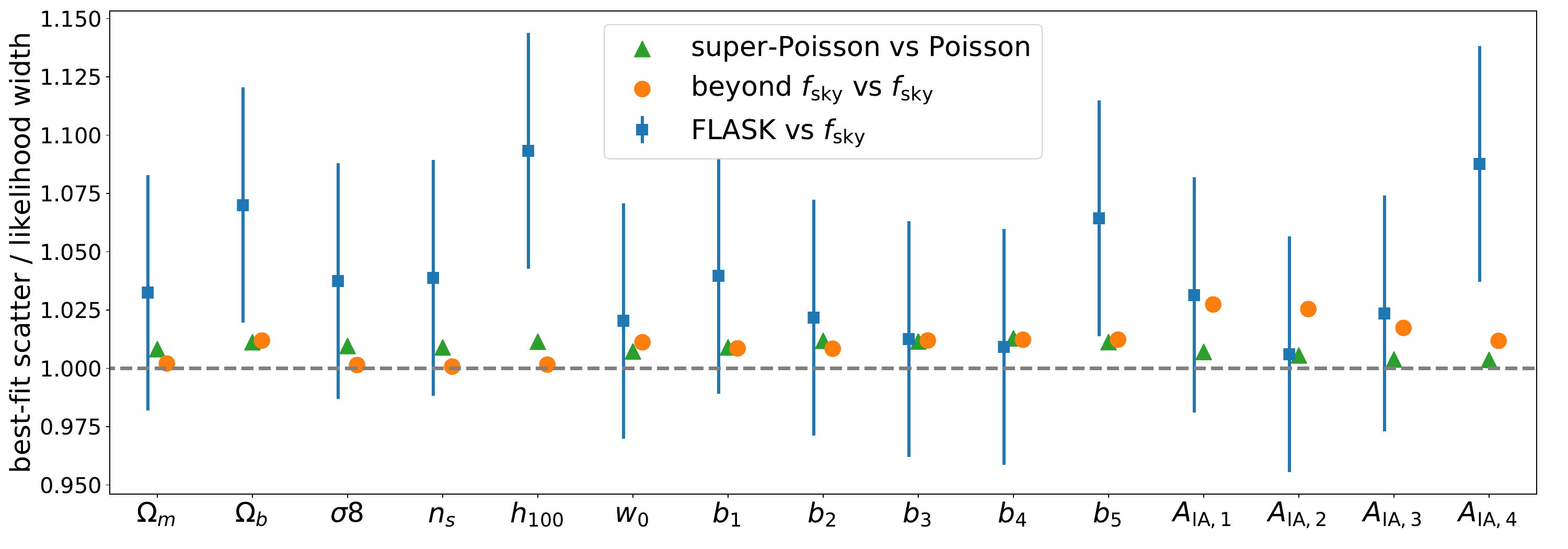}
   \caption{Impact of covariance errors on the ratio of the standard deviation of maximum posterior parameters to the width of the posterior derived from the erroneous covariance. Green triangles shot the effect caused by non-Poissonian shot-noise and orange circles show the effect caused by the $f_{\mathrm{sky}}$ approximation (\cf\ \appref{masking} for our beyond-$f_{\mathrm{sky}}$ treatment). These ratios have been calculated purely on the base of different analytic covariance models and within the linearized likelihood framework discussed in \secrefnospace{linearized_likelihoods}. We also show the ratio of maximum posterior parameter scatter observed from the 197 FLASK simulations to the statistical uncertainties expected from a log-normal covariance matrix matching the FLASK configuration. Within the statistical uncertainties, these ratios are consistent with $1$.}
  \label{fig:parameter_offsets_masking}
\end{figure*}

In \figref{parameter_offsets_masking} we take the two effects that had the largest impacts on $\chi^2$ and show the resulting mismatch between scatter of maximum posterior values and width of the inferred contours for a wider range of parameters. All of our results take into account marginalisation over nuisance parameters (and all other parameters).

Our reason for exclusively investigating the impact of covariance errors on $\chi^2$ and parameter constraints is that those are the two measures by which our final (on-shot) data analysis will be interpreted and judged\footnote{Alternatively, one could investigate the distribution of $p$-values \citep[or \emph{probability to exceed}, \cf][]{Hall2019} as opposed to the distribution of $\chi^2$.}. In the remainder of this paper we detail how the above results were obtained.

\begin{table*}
{
\renewcommand{\arraystretch}{2.0}
\begin{tabular}{p{1.4cm}|cc|p{1.7cm}c|p{1.7cm}c|p{1.9cm}c}
Effect & $\langle\chi^2\rangle$ & $\sigma[\chi^2]$ & $\sigma[\hat\Omega_m]$ & $\sigma_{\Omega_m}$ & $\sigma[\hat\sigma_8]$ & $\sigma_{\sigma_8}$ & $\sigma[\hat w]$ & $\sigma_{w}$  \\
\hline
Fiducial & 507.6 & 31.8 & 0.0509 & 0.0509 & 0.0975 & 0.0975 & 0.244 & 0.244 \\
\hline
angular bin width & 402.1 & 26.0 & +0.8\% & +7.4\% & +0.8\% & +8.3\% & +1.0\% & +7.4\% \\
connected 4-point function & 507.6 & 31.8 & +0.1\% & -0.8\% & +0.1\% & -0.9\% & +0.1\% & -0.8\% \\
\hline
curved sky & 507.7 & 31.8 & +0.0\% & -0.0\% & +0.0\% & -0.0\% & +0.0\% & -0.0\% \\
non-Limber \& RSD & 511.4 & 32.1 & +0.1\% & -0.6\% & +0.1\% & -0.6\% & +0.3\% & -1.4\% \\
\hline
non-Gauss. likelihood & - & 32.6 & +0.8\% (low) \newline -0.9\% (high) & - & +0.4\% (low) \newline -0.4\% (high) & - & +0.5\% (low) \newline +0.05\% (high) & - \\
covariance cosmology & 508.6 & 32.4 & +2.9\% & +$(0.1\pm0.06)$\% & +2.8\% & +$(0.1\pm0.05)$\% & +4.7\% & +$(0.1\pm0.06)$\% \\
random point shot-noise & 511.3 & 32.0 & +0.0\% & -0.5\% & +0.0\% & -0.6\% & +0.0\% & -0.2\% \\
non-Poisson shot-noise & 515.0 & 32.3 & +0.0\% & -0.7\% & +0.0\% & -0.8\% & +0.0\% & -0.6\% \\
masking and survey geometry & 526.5 & 33.8 & +0.6\% & -0.8\% & +0.7\% & -0.3\% & +0.3\% & -1.3\% \\
\hline
\end{tabular}
}
\caption{Summary of the impact of the different effects tested here on the distribution of $\chi^2$ between measurement and maximum posterior model, on the scatter $\sigma[\hat \pi]$ of maximum posterior parameters $\hat \pi$ and on the standard deviations $\sigma_\pi$ on these parameters inferred from the likelihood. See text for details.}
\label{tab:summary}
\end{table*}

\section{The 3x2-point data vector}
\label{sec:model}

The combined 3x2pt data vector of the DES-Y3 analysis consists of measurements of the following 2-point correlations:
\begin{itemize}
\item the angular 2-point correlation function $w(\theta)$ of galaxy density contrast measured for luminous red galaxies in 5 different redshift bins \citep[see \eg][as well as other relevant references given in \secref{intro}]{y3-galaxyclustering, y3-lenswz},
\item the auto- and cross-correlation functions $\xi_+(\theta)$ and $\xi_-(\theta)$ between the galaxy shapes of 4 redshift bins of source galaxies \citep[see \eg][]{y3-cosmicshear1, y3-cosmicshear2, y3-sompz, y3-sourcewz} ,
\item the tangential shear $\gamma(\theta)$ imprinted on source galaxy shapes around positions of foreground redMaGiC galaxies \citep[see \eg][]{y3-gglensing}.
\end{itemize}
At the time of writing this paper the exact choices for redshift intervals and angular bins considered for each 2-point function are still being determined by a careful study of their impact on the robustness of DES-Y3 parameter constraints \citep{y3-generalmethods, y3-3x2ptkp}. For the purposes of testing the modelling of the covariance matrix we will use the most recent but possibly not final DES-Y3 analysis choices. We do not expect that our tests and conclusions will change in a significant manner with further updated analysis choices.
We assume that each of the correlation functions are measured in 20 logarithmically spaced angular bins between $\theta_{\min} = 2.5'$ and  $\theta_{\max} = 250'$. Some of these bins in some of the measured 2-point functions are being cut from the analysis to ensure unbiased cosmological results, resulting in a total of 531 data points when using the preliminary DES-Y3 scale cuts. 

Our starting point of modelling the different 2-point functions in the 3x2pt data vector is the 3D nonlinear matter power spectrum $P(k,z)$ at a given wavenumber $k$ and redshift $z$. We obtain it by using either of the Boltzmann solvers CLASS \footnote{\tt www.class-code.net} or CAMB \footnote{\tt camb.info} to calculate the linear power spectrum and the HALOFIT fitting formula \citep{Smith2003} in its updated version \citep{Takahashi2012} to turn this into the late time nonlinear power spectrum. From this 3D power spectrum the angular power spectra required for our three 2-point functions (cosmic shear ($\kappa \kappa$), galaxy-galaxy lensing ($\delta_g \kappa$) and galaxy-galaxy clustering ($\delta_g \delta_g$)) in the Limber approximation are given by \citep[\eg][]{Krause2017, Limber1953}:
\begin{equation}
\label{eq:C_ell_cosmic_shear}
    C^{ij}_{\kappa \kappa}(\ell) = \int d\chi \frac{q^i_\kappa(\chi) q^j_\kappa(\chi)}{\chi^2} P\left(\frac{\ell+\frac{1}{2}}{\chi}, z(\chi)\right),
\end{equation}
\begin{equation}
\label{eq:C_ell_gamma_t}
    C^{ij}_{\delta_g \kappa}(\ell) = \int d\chi \frac{q^i_\delta\left(\frac{\ell+\frac{1}{2}}{\chi},\chi \right) q^j_\kappa(\chi)}{\chi^2} P\left(\frac{\ell+\frac{1}{2}}{\chi}, z(\chi)\right), 
\end{equation}
\begin{equation}
\label{eq:C_ell_delta_g}
    C^{ij}_{\delta_g \delta_g}(\ell) = \int d\chi \frac{q^i_\delta\left(\frac{\ell+\frac{1}{2}}{\chi},\chi \right) q^j_\delta\left(\frac{\ell+\frac{1}{2}}{\chi},\chi \right)}{\chi^2} P\left(\frac{\ell+\frac{1}{2}}{\chi}, z(\chi)\right)\ ,
\end{equation}
where $\chi$ is the comoving radial distance, $i$ and $j$ denote different combinations of pairs of redshift bins and the lensing efficiency
$q^i_\kappa$ and the radial weight function for clustering $q^i_\delta$ are given by
\begin{eqnarray}
 q^i_\kappa (\chi) &=& \frac{3H_0^2 \Omega_m\chi}{2a(\chi)} \int_{\chi}^{\chi_h} d\chi^\prime \left(\frac{\chi^\prime - \chi}{\chi}\right) n^i_\kappa(z(\chi^\prime)) \frac{d z}{d\chi^\prime} , \nonumber \\
 q^i_\delta (k,\chi) &=& b^i(k, z(\chi))\  n^i_\delta(z(\chi)) \frac{d z}{d\chi}\ .
\end{eqnarray}
Here $H_0$ is the Hubble parameter today, $\Omega_m$ the ratio of today's matter density to today's critical density of the Universe, $a(\chi)$ is the Universe's scale factor at comoving distance $\chi$ and $b^i(k, z)$ is a scale and redshift dependent galaxy bias. Furthermore, $n^i_{\kappa, g}(z)$ denote the redshift distributions of the different DES-Y3 redshift bins of source and lens galaxies respectively, normalised such that
\begin{equation}
    \int dz \; n^i_{\kappa, g}(z) = 1\ .
\end{equation}
Note that on large angular scales the DES-Y3 analysis does not make use of the Limber approximation for galaxy clustering but instead employs the method derived in \cite{Fang:2020vhc}.

The above angular power spectra are now related to the real space correlation functions $w(\theta)$, $\gamma_t(\theta)$ and $\xi_\pm(\theta)$ as
\begin{eqnarray}
\label{eq:xi_in_terms_of_Cell}
w^i(\theta) &=& \sum_\ell \frac{2 \ell +1}{4\pi} P_\ell(\cos \theta) C^{ii}_{\delta_g \delta_g}(\ell)\ ,  \nonumber \\
\gamma^{ij}_t(\theta) &=& \sum_\ell \frac{2\ell + 1}{4\pi} \frac{P_\ell^2\left( \cos \theta \right)}{\ell(\ell + 1)} C_{\delta_g\kappa}^{ij}(\ell)\ ,\nonumber \\
\xi_{\pm}^{ij}(\theta) &=& \sum_{\ell \geq 2} \frac{2\ell + 1}{4\pi}\ \frac{2(G_{\ell, 2}^+(x) \pm G_{\ell, 2}^-(x))}{\ell^2(\ell + 1)^2}\ C^{ij}_{\kappa \kappa}(\ell)\ .\nonumber \\
\end{eqnarray}
Here $P_\ell$ are the Legendre polynomials of order $\ell$, $P_\ell^m$ are the associated Legendre polynomials, $x = \cos \theta$ and the functions $G_{\ell, 2}^{+,-}(x)$ are given in \appref{curvedsky} \citep[see also][]{Stebbins1996}. Note that we only consider the auto-correlations $w^i(\theta)$ for each tomographic bin since in the Y1 analysis it was shown that the cross correlations do not carry significant information \citep{Elvin-Poole:2017xsf}.


The above relations between angular power spectra and real space correlation functions can all be written in the form
\begin{equation}
\label{eq:xi_in_terms_of_Cell_general}
    \xi^{\mathrm{AB}}(\theta) = \sum_{\ell = 0}^\infty \frac{2 \ell +1}{4 \pi}  F^{AB}_\ell( \theta) \, C_\ell^{AB}\ .
\end{equation}
This is particularly useful when deriving covariance expressions and when performing averages over finite bins in the angular scale $\theta$. To achieve the latter, one can simply derive analytic averages of the functions $F^{AB}_\ell( \theta)$. Both of these points will be considered in the next sections.

For most of our tests we consider the 3x2pt data vector and its covariance matrix at the fiducial cosmology described in section \ref{sec:strategy}, where we also show the Gaussian priors assumed on some of these parameters when assessing the impact of covariance modelling on parameter constraints and maximum posterior $\chi^2$.

\section{Covariance matrices for the 3$\times$2pt data vector}
\label{sec:covariance_modelling}

The covariance matrix of measurements of cosmological 2-point statistics typically contains three contributions \citep[\cf][]{Krause:2016jvl, Krause2017},
\begin{equation}
    \mathbf{C} = \mathbf{C}_{\mathrm{G}} + \mathbf{C}_{\mathrm{nG}} + \mathbf{C}_{\mathrm{SSC}}\ .
\end{equation}
Here, $\mathbf{C}_{\mathrm{G}}$ is the contribution to the covariance that would be present if the cosmic matter density and cosmic shear fields where pure Gaussian random fields \citep[see also][]{Schneider2002, Crocce2011}, $\mathbf{C}_{\mathrm{nG}}$ are contributions involving the connected 4-point function of these fields (the trispectrum) and $\mathbf{C}_{\mathrm{SSC}}$ is the so-called super-sample covariance contribution resulting from the fact that any survey only observes a 
finite volume of the Universe and that the mean density in that volume is subject to fluctuations due to long wavelenght modes \citep{Takada:2013wfa,Schaan:2014cpa}.

In the fiducial DES-Y3 analysis we model all of these covariance contributions analytically. This fiducial model is described in \secref{fiducial_covariance}. In \secref{lognormal_covariance} we describe an alternative model for the non-Gaussian covariance contributions that is used to test the robustness of our analysis with respect to the modelling of the trispectrum contribution. Finally, \secref{FLASK} describes a set of log-normal simulations \citep{Xavier2016} and the covariance matrix of the 3x2pt data vector estimated from them. These simulations also allow us to test the accuracy of our Gaussian likelihood assumption and the treatment of masking and finite survey area in our fiducial covariance model.

\begin{figure*}
    \centering
    \includegraphics[width=0.5\textwidth]{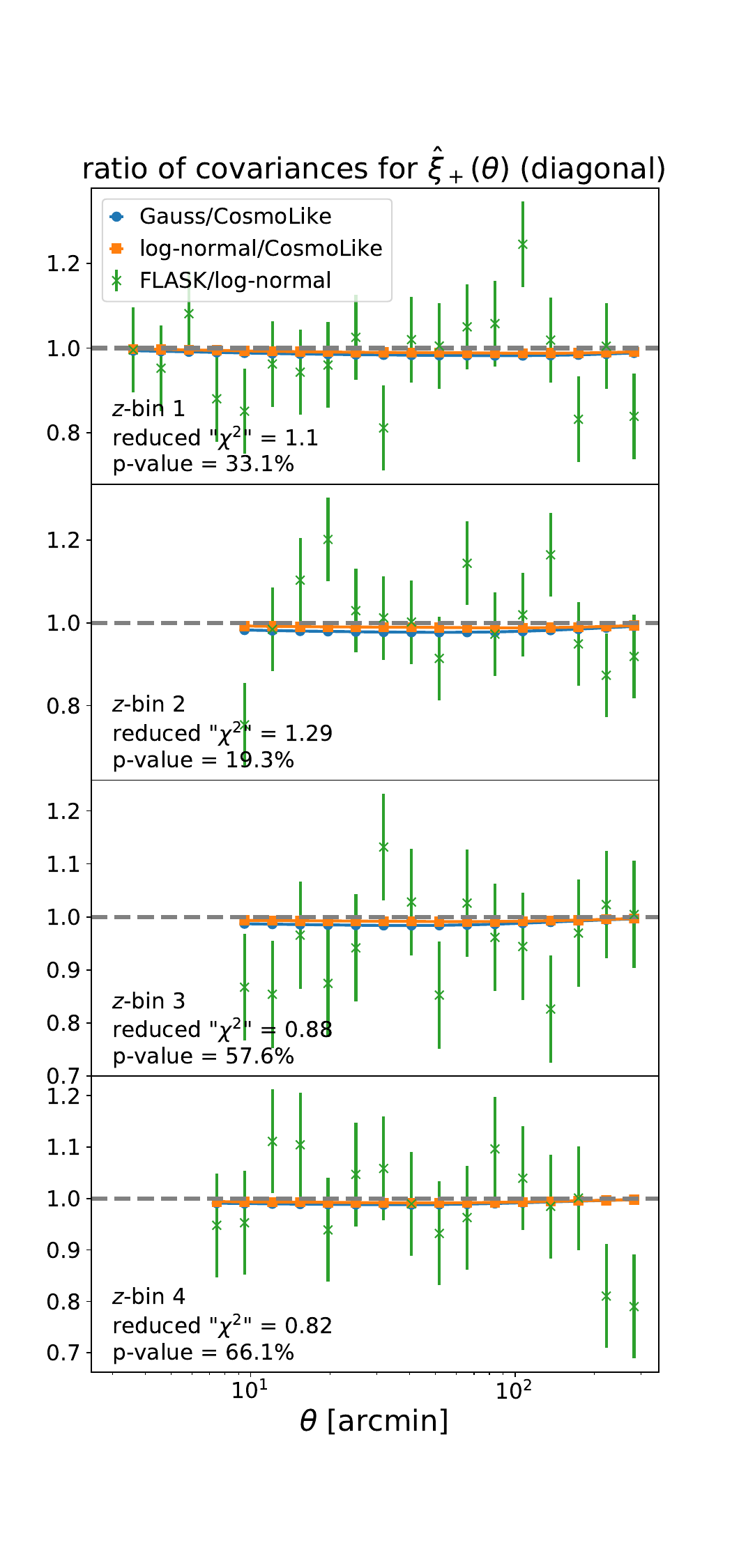}\includegraphics[width=0.5\textwidth]{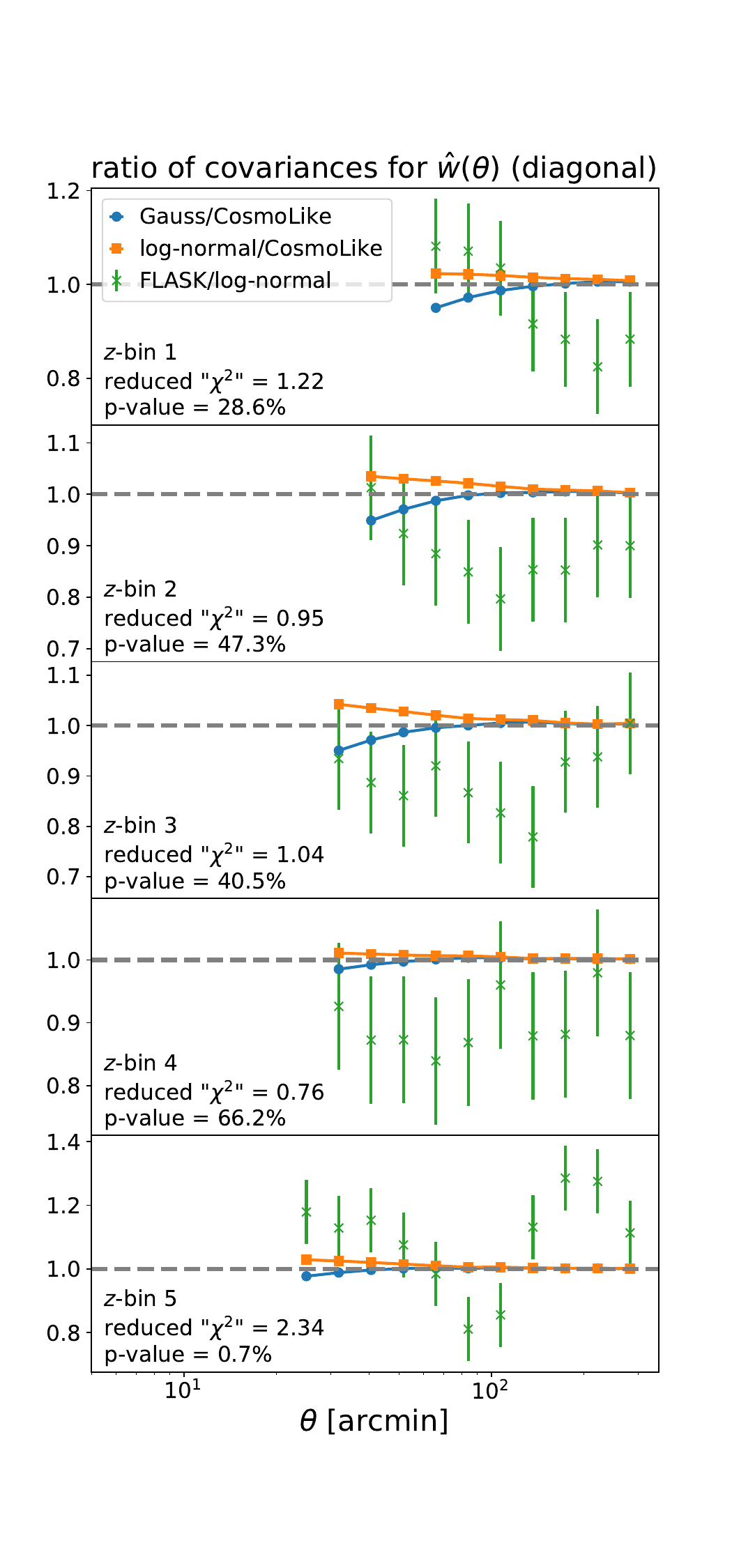}
    \caption{Ratio of the diagonal elements of the different covariance matrices introduced in this section with respect to each other. The left panel compares the variances of measurements of $\xi_+(\theta)$ while the right panel compares the variances of measurements of $w(\theta)$. To give a sense of the goodness of fit between the covariance estimated from FLASK and our fiducial analytic matrix, we treat the diagonal elements of the FLASK covariance as a multivariate Gaussian whose covariance can be inferred from the properties of the Wishart distribution \citep{Taylor2013}. The low p-value for the highest redshift bin of $w(\theta)$ most likely results from our incomplete treatment of the survey mask (\cf\ discussion in \secref{impact_of_modelling} and \apprefnospace{masking}).}
    \label{fig:comparison_diagonal}
\end{figure*}

\subsection{Fiducial DES-Y3 Covariance}
\label{sec:fiducial_covariance}

In our fiducial covariance matrix, we model the non-Gaussian covariance contributions $\mathbf{C}_{\mathrm{nG}}$ and $\mathbf{C}_{\mathrm{SSC}}$ using a halo model combined with leading-order perturbation theory to approximate the trispectrum of the cosmic density field and to compute the mode coupling between scales larger than the considered survey volume with scales inside that volume. These calculations are carried out using the \verb|CosmoCov| code package \citep{Fang:2020vhc} based on the \verb|CosmoLike| framework \citep{Krause:2016jvl}. Our modelling of these contributions has not changed with respect to the year-1 analysis of DES and we refer the reader to \citet{Krause2017} as well as to the \verb|CosmoLike| papers for details. However, the modelling of the Gaussian contribution has changed as described in the following.

\subsubsection{Gaussian covariance}

${}$

\noindent Our modelling of the Gaussian covariance part has changed with respect to the year-1 analysis in the following ways:

${}$

\begin{itemize}
\item we use (and present for the first time\footnote{We have shared our results with \citet{Fang:2020vhc} who have used them for their covariance calculations.}) analytic expression for the angular bin averaging of the functions $F^{AB}_\ell( \theta)$ (\cf\ Equation  \ref{eq:xi_in_terms_of_Cell_general}) for all 4 types of two point functions present in our data vector (see \secref{impact_of_bin_average}, this is especially relevant for the sampling-noise contribution to the covariance, \cf\ \citealt{Troxel2018b});

${}$

\item we account for redshift space distortions (RSD) and also use a non-Limber calculation to obtain the galaxy-galaxy clustering power spectrum $C_{\delta_g \delta_g}(\ell)$ (see \secref{impact_of_Limber_and_RSD});

${}$

\item we do not make use of the flat-sky approximation anymore (see \secref{impact_of_curved_sky}).
\end{itemize}

${}$

\noindent To derive expressions for the Gaussian covariance part, let us first consider an all-sky survey. If a 2-point function measurement $\hat \xi^{\mathrm{AB}}(\theta)$ could be obtained from data on the entire sky, then for most types of 2-point correlations it would be related to power spectrum measurements $C_\ell^{AB}$ from a spherical harmonics decomposition of the same all-sky data through \eqnrefnospace{xi_in_terms_of_Cell_general}, \ie
\begin{equation}
\label{eq:xi_hat_in_terms_of_Cell_hat_general}
    \hat \xi^{\mathrm{AB}}(\theta) = \sum_{\ell = 0}^\infty \frac{2 \ell +1}{4 \pi}  F^{AB}_\ell( \theta) \, \hat C_\ell^{AB}\ .
\end{equation}
A notable exception to this are the cosmic shear 2-point functions $\hat \xi_\pm$ which obtain contributions from both the so-called E-mode and B-mode power spectra \citep{Schneider2002}. For these functions equation (\ref{eq:xi_hat_in_terms_of_Cell_hat_general})  in the curved sky formalism becomes
\begin{align}
\label{eq:xi_hat_plus_minus}
    &\ \hat\xi_{\pm}^{ij}(\theta) \nonumber \\
    =&\ \sum_{\ell \geq 2} \frac{2\ell + 1}{4\pi}\ \frac{2(G_{\ell, 2}^+(x) \pm G_{\ell, 2}^-(x))}{\ell^2(\ell + 1)^2}\ \left(\hat C^{E, ij}_{\gamma \gamma}(\ell) \pm \hat C^{B, ij}_{\gamma \gamma}(\ell)\right)\ ,
\end{align}
where in the absence of shape-measurement systematics (and ignoring post-Born corrections) $\langle \hat C^{E, ij}_{\gamma \gamma}(\ell) \rangle = C^{ij}_{\kappa \kappa}(\ell)$ and $\langle \hat C^{B, ij}_{\gamma \gamma}(\ell) \rangle = 0$.

${}$

Since this is a linear equation in $C(\ell)$'s, the covariance of two different 2-point function measurements $\hat \xi^{AB}$ and $\hat \xi^{CD}$ at two different angular scales $\theta_1$ and $\theta_2$ would be given in terms of the covariance of the corresponding power spectrum measurements by
\begin{align}
\label{eq:realspace_cov_in_terms_of_harmonic_space_cov}
\mathrm{Cov}\left[\hat \xi^{AB}(\theta_1), \hat \xi^{CD}(\theta_2)\right] = \sum_{\ell_1, \ell_2} \frac{(2\ell_1 +1)(2\ell_2 +1)}{(4\pi)^2} \times
\nonumber \\
\hspace{1.5cm}  F_{\ell_1}^{AB}\left( \cos \theta_1 \right) F_{\ell_2}^{CD}\left( \cos \theta_2 \right) \mathrm{Cov}\left[\hat C_{\ell_1}^{AB}, \hat C_{\ell_2}^{CD}\right]\ .
\end{align}
Again, for $\hat\xi^{AB}(\theta) = \hat \xi_\pm(\theta)$ one would have to use $C_{\ell}^{AB} = \hat C^{E}_{\gamma \gamma}(\ell) \pm \hat C^{B}_{\gamma \gamma}(\ell)$ in this sum.

${}$

For the auto-power spectrum of galaxy density contrast in one of our redshift bins the harmonic space covariance would be \citep{Crocce2011}
\begin{align}
\mathrm{Cov}[\hat C_{\delta_g \delta_g}^{ii}(\ell_1), \hat C_{\delta_g \delta_g}^{ii}(\ell_2)] = \frac{2\delta_{\ell_1\ell_2}}{(2\ell_1 + 1)} \left( C_{\delta_g \delta_g}^{ii}(\ell_1) + \frac{1}{n_g} \right)^2\ .\nonumber \\
\end{align}

${}$

\noindent Here $n_g$ is the number density of the galaxies and $\delta_{\ell_1\ell_2}$ is the Kronecker symbol. 
To account for partial-sky surveys (such as DES) we simply divide this expression (and similar ones for the other 2-point functions) by the observed 
sky fraction $f_{\mathrm{sky}}$. This so-called $f_{\mathrm{sky}}$ approximations leads to the following harmonic space Gaussian covariances of \citep[ see also][]{Krause2017}:

\newpage

\begin{widetext}
\begin{align}
\label{eq:harmonic_covariances}
\mathrm{Cov}[\hat C_{g g}^{ij}(\ell_1), \hat C_{g g}^{kl}(\ell_2)] =&\  \frac{\delta_{\ell_1\ell_2} \left[\left(C_{g g}^{ik}(\ell_1) + \frac{\delta_{ik}}{n_g^i} \right)\left(C_{g g}^{jl}(\ell_1) + \frac{\delta_{jl}}{n_g^j} \right) + \left(C_{g g}^{il}(\ell_1) + \frac{\delta_{il}}{n_g^i} \right)\left(C_{g g}^{jk}(\ell_1) + \frac{\delta_{jk}}{n_g^j} \right)\right]}{(2\ell_1 + 1) f_{\mathrm{sky}}}
\end{align}
\begin{align}
\mathrm{Cov}[\hat C_{\gamma \gamma}^{E, ij}(\ell_1), \hat C_{\gamma \gamma}^{E, kl}(\ell_2)] =&\  \frac{\delta_{\ell_1\ell_2} \left[\left(C_{\kappa \kappa}^{ik}(\ell_1) + \frac{\delta_{ik}\sigma_{\epsilon, i}^2}{n_s^i} \right)\left(C_{\kappa \kappa}^{jl}(\ell_1) + \frac{\delta_{jl} \sigma_{\epsilon, j}^2}{n_s^j} \right) + \left(C_{\kappa \kappa}^{il}(\ell_1) + \frac{\delta_{il}\sigma_{\epsilon, i}^2}{n_s^i} \right)\left(C_{\kappa \kappa}^{jk}(\ell_1) + \frac{\delta_{jk} \sigma_{\epsilon, j}^2}{n_s^j} \right)\right]}{(2\ell_1 + 1) f_{\mathrm{sky}}}
\end{align}
\begin{align}
\mathrm{Cov}[\hat C_{\gamma \gamma}^{B, ij}(\ell_1), \hat C_{\gamma \gamma}^{B, kl}(\ell_2)] =&\  \frac{\delta_{\ell_1\ell_2} \left[\frac{\delta_{ik}\sigma_{\epsilon, i}^2}{n_s^i} \frac{\delta_{jl} \sigma_{\epsilon, j}^2}{n_s^j} + \frac{\delta_{il}\sigma_{\epsilon, i}^2}{n_s^i}\frac{\delta_{jk} \sigma_{\epsilon, j}^2}{n_s^j} \right]}{(2\ell_1 + 1) f_{\mathrm{sky}}}
\end{align}
\begin{align}
\mathrm{Cov}[\hat C_{g \kappa}^{ij}(\ell_1), \hat C_{g \kappa}^{kl}(\ell_2)] =&\  \frac{\delta_{\ell_1\ell_2} \left[\left(C_{g g}^{ik}(\ell_1) + \frac{\delta_{ik}}{n_g^i} \right)\left(C_{\kappa \kappa}^{jl}(\ell_1) + \frac{\delta_{jl} \sigma_{\epsilon, j}^2}{n_s^j} \right) + C_{g \kappa}^{il}(\ell_1) C_{g \kappa}^{kj}(\ell_1)\right]}{(2\ell_1 + 1) f_{\mathrm{sky}}}
\end{align}
\begin{align}
\mathrm{Cov}[\hat C_{g g}^{ij}(\ell_1), \hat C_{\gamma \gamma}^{E, kl}(\ell_2)] =&\  \frac{\delta_{\ell_1\ell_2} \left[C_{g \kappa}^{ik}(\ell_1)C_{g \kappa}^{jl}(\ell_1) + C_{g \kappa}^{il}(\ell_1)C_{g \kappa}^{jk}(\ell_1) \right]}{(2\ell_1 + 1) f_{\mathrm{sky}}}
\end{align}
\begin{align}
\mathrm{Cov}[\hat C_{gg}^{ij}(\ell_1), \hat C_{g \kappa}^{kl}(\ell_2)] =&\  \frac{\delta_{\ell_1\ell_2} \left[\left(C_{gg}^{ik}(\ell_1) + \frac{\delta_{ik}\sigma_{\epsilon, i}^2}{n_s^i} \right)C_{g \kappa}^{jl}(\ell_1) + C_{g \kappa}^{il}(\ell_1) \left(C_{g g}^{jk}(\ell_1) + \frac{\delta_{jk}}{n_g^j} \right)\right]}{(2\ell_1 + 1) f_{\mathrm{sky}}}
\end{align}
\begin{align}
\mathrm{Cov}[\hat C_{g \kappa}^{ij}(\ell_1), \hat C_{\gamma \gamma}^{E, kl}(\ell_2)] =&\  \frac{\delta_{\ell_1\ell_2} \left[C_{g \kappa}^{ik}(\ell_1)\left(C_{\kappa \kappa}^{jl}(\ell_1) + \frac{\delta_{jl} \sigma_{\epsilon, j}^2}{n_s^j} \right) + C_{g \kappa}^{il}(\ell_1)\left(C_{\kappa \kappa}^{jk}(\ell_1) + \frac{\delta_{jk} \sigma_{\epsilon, j}^2}{n_s^j} \right)\right]}{(2\ell_1 + 1) f_{\mathrm{sky}}}
\end{align}
\begin{align}
\label{eq:harmonic_covariances_end}
\mathrm{Cov}[\hat C_{g g}^{ij}(\ell_1), \hat C_{\gamma \gamma}^{B, kl}(\ell_2)] =&\  0\ (\mathrm{as\ are\ all\ other\ covariances\ with\ only\ one}\ \hat C_{\gamma \gamma}^{B})\ .
\end{align}
\end{widetext} 

\noindent At this point let us introduce the following nomenclature: 
we will denote the terms that contain two power spectra as \emph{cosmic variance} contribution to the covariance, the terms that contain no power spectrum at all as the \emph{sampling noise} contributions (or \emph{shape noise} and \emph{shot noise} contributions) and the terms that contain contribution from one power spectrum and a sampling noise  as the \emph{mixed terms}. We test the accuracy of the $f_{\mathrm{sky}}$-approximation that results in Equations (\ref{eq:harmonic_covariances}-\ref{eq:harmonic_covariances_end}) in \secref{masking_tests} by comparing it to more accurate expressions. 

\subsection{Analytic lognormal covariance model}
\label{sec:lognormal_covariance}

To test the robustness of the CosmoLike covariance we also employ an alternative model for the connected 4-point function part of the covariance - the lognormal model. \citet{Hilbert2011} originally derived this as a model for the covariance of cosmic shear correlation function, assuming that that the lensing convergence $\kappa$ can be written in terms of a Gaussian random field $n$ as \citep[see also][]{Xavier2016}
\begin{equation}
\label{eq:definition_lognormal_variable}
    \kappa = \lambda \left(e^{n+\mu} - 1\right)\ ,
\end{equation}
where it is assumed that $\langle n \rangle = 0$. For given values $\lambda > 0$ and $\mu$ the power spectrum of $n$ can be chosen such as to reproduce a desired 2-point correlation function $\xi_\kappa$ (see \citet{Xavier2016} for caveats). Furthermore, for any given value $\lambda > 0$ one can choose $\mu$ such that $\langle \kappa \rangle = 0$. This makes $\lambda$ the only free parameter of the lognormal covariance model. \citet{Hilbert2011} show that this model leads to a number of correction terms to the Gaussian covariance model, and identify the most dominant of these terms to be
\begin{align}
    &\ C_{\mathrm{LN}}[\hat\xi_\kappa(\theta_1), \hat\xi_\kappa(\theta_2)]\nonumber \\
    \approx&\ C_{\mathrm{G}}[\hat\xi_\kappa(\theta_1), \hat\xi_\kappa(\theta_2)] + \frac{4\ \xi_\kappa(\theta_1)  \xi_\kappa(\theta_2)}{A_{\mathrm{S}}\lambda^2} \mathrm{Var}_{\mathrm{S}}(\kappa)\ .
\end{align}
Here $A_{\mathrm{S}}$ is the area of the considered survey footprint and $\mathrm{Var}_{\mathrm{S}}(\kappa)$ is the variance of $\kappa$ when averaged over the footprint. We generalise this to the covariance of 2-point correlations $\hat\xi_{AB}$ and $\hat\xi_{CD}$ between arbitrary scalar fields $\delta_A, \delta_B, \delta_C, \delta_D$ as
\begin{align}
\label{eq:extension_of_Hilbert_lognormal}
    &\ C_{\mathrm{LN}}[\hat\xi_{AB}(\theta_1), \hat\xi_{CD}(\theta_2)] - C_{\mathrm{G}}[\hat\xi_{AB}(\theta_1), \hat\xi_{CD}(\theta_2)]\nonumber \\
    \approx&\ \frac{\xi_{AB}(\theta_1) \xi_{CD}(\theta_2)}{A_{\mathrm{S}}} \left\lbrace \frac{\mathrm{Cov}_{\mathrm{S}}(\delta_A, \delta_C)}{\lambda_A \lambda_C} + \frac{\mathrm{Cov}_{\mathrm{S}}(\delta_A, \delta_D)}{\lambda_A \lambda_D} + \right. \nonumber \\
    &\ \ \ \ \ \ \ \ \ \ \ \ \ \ \ \ + \left. \frac{\mathrm{Cov}_{\mathrm{S}}(\delta_B, \delta_C)}{\lambda_B \lambda_C} + \frac{\mathrm{Cov}_{\mathrm{S}}(\delta_B, \delta_D)}{\lambda_B \lambda_D} \right\rbrace\ .
\end{align}
Here, $\mathrm{Cov}_{\mathrm{S}}(\delta_A, \delta_C)$ is the covariance of $\delta_A$ and $\delta_C$ after the two fields have been averaged over the entire survey footprint (and likewise for the other terms appearing above). Following \citet{Hilbert2011} we use this expression even when considering non-scalar fields (\ie\ the shear field) by replacing $\xi_{XY}(\theta)$ by the appropriate 2-point functions $\xi_+(\theta), \xi_-(\theta), \gamma_t(\theta)$ (or $w(\theta)$, for the scalar galaxy density contrast).

To choose the parameters $\lambda_X$ (also called the lognormal shift parameters, \cf\ \citealt{Xavier2016}) we follow a procedure similar to the one outlined in \citet{Friedrich2018}. There it is shown how the value of $\lambda_X$ can be adjusted in order to match the re-scaled cumulant
\begin{equation}
    S_3(\vartheta) \equiv \frac{\langle \delta_X(\vartheta)^3 \rangle}{\langle \delta_X(\vartheta)^2 \rangle^2}
\end{equation}
of the random field $\delta_X$ smoothed with a top-hat filter of angular radius $\vartheta$ to the value of $S_3$ predicted by leading-order perturbation theory for that same smoothing scale. Since the focus in our paper is the covariance matrix of 2-point statistics (hence a 4-point function), we modify their method to match instead the value of reduced fourth order cumulant
\begin{equation}
    S_4(\vartheta) \equiv \frac{\langle \delta_X(\vartheta)^4 \rangle - 3\langle \delta_X(\vartheta)^2 \rangle^2}{\langle \delta_X(\vartheta)^2 \rangle^3}\ .
\end{equation}
The field $\delta_X$ here will be either projections of the 3D matter density contrast along the line-of-sight distribution of our lens galaxies or the lensing convergence fields corresponding to our 4 source redshift bins. The smoothing scale $\vartheta$ at which we use the $\lambda_X$ to match $S_4$ to its perturbation theory value is chosen such that it corresponds to about $10$Mpc$/h$ at the mean redshift of the line-of-sight projection kernels corresponding to the different $\delta_X$. This is approximately the scale at which \citet{Friedrich2018} found the shifted log-normal model to be a good approximation of the overall PDF of density fluctuations in N-body simulations (\cf\ their figure 5). 

Our results are shown in \tabref{lognormal}, where
we present the number density, galaxy bias (relevant for lenses only), shape-noise dispersion (per shear component; relevant for sources only) and the lognormal shift parameters obtained from the procedure described above. Note that for the source galaxy samples, the relevant line-of-sight projection kernel used to derive the shift parameter is the lensing kernel (and not the redshift distribution of the source galaxies). For the lens galaxies, all shift parameters come out to be $> 1$. As a consequence there will be pixels with negative density in our lognormal simulations. However, the fraction of such pixels is $<0.01$ for all runs and all bins and setting $\delta_g=-1$ in these bins has an unnoticeable effect on the statistics measured in these maps (\eg\ for bin 4, which is affected most, the standard deviation of $\delta_g$ changes by $0.053\%$).
Note further that at the time of completing the simulation runs presented in Section \ref{sec:FLASK}, the DES Y3 shear catalog and redshift distribution were not finalized. As a consequence, the shape noise dispersion values used for simulations differ from the values in this table.

\begin{figure}
    \centering
    \includegraphics[width=0.5\textwidth]{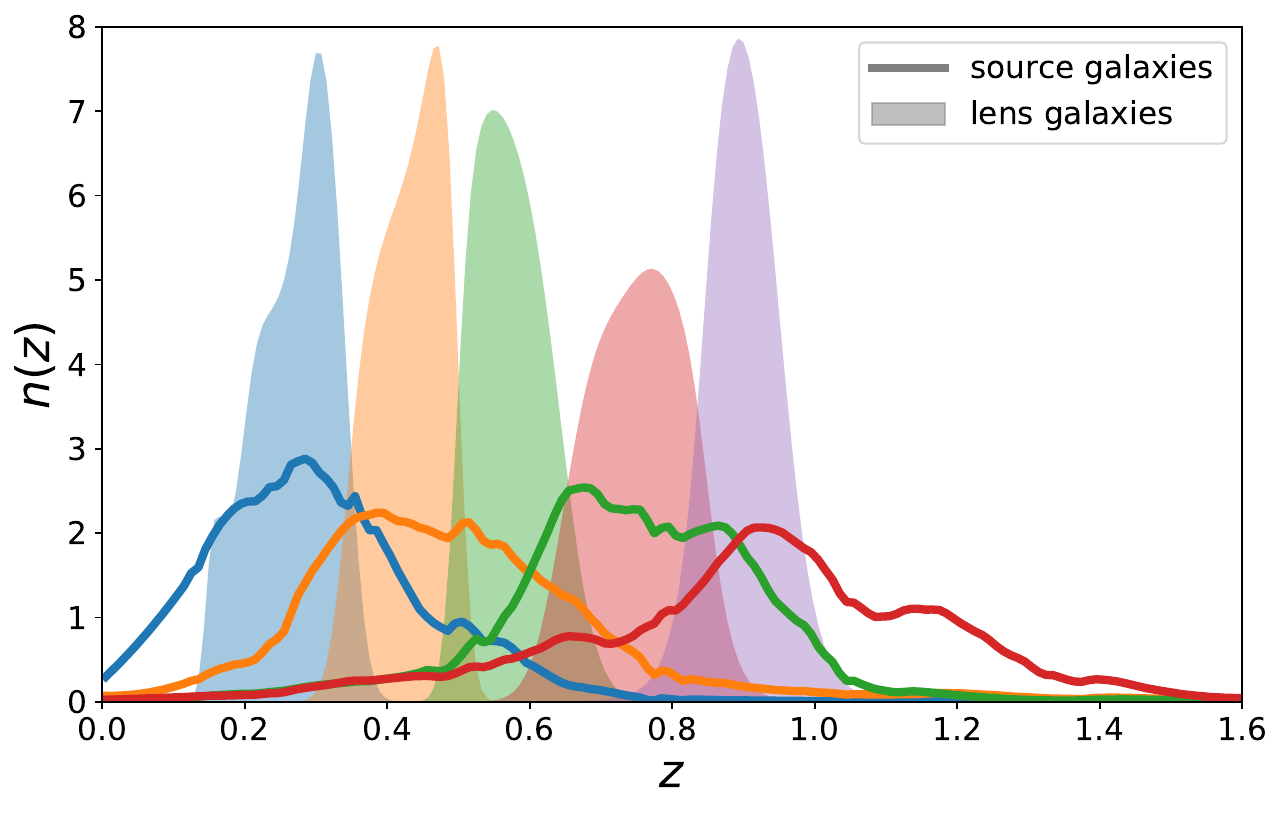}
\caption{Redshift distributions of lens galaxies (shaded regions) and source galaxies (solid lines) in our fiducial test configuration.}
  \label{fig:n_of_z}
\end{figure}   
    
\begin{table}
\begin{tabular}{|c|c|c|c|c|}
\hline
z-bin & $n_g$ [arcmin$^{-2}$] & bias & $\sigma_\epsilon$ & log-normal shift \\
\hline
lenses 1 & $0.0221$ & $1.7$ & $-$ & $1.089$ \\
lenses 2 & $0.0381$ & $1.7$ & $-$ & $1.106$ \\
lenses 3 & $0.0583$ & $1.7$ & $-$ & $1.047$ \\
lenses 4 & $0.0295$ & $2.0$ & $-$ & $1.252$ \\
lenses 5 & $0.0251$ & $2.0$ & $-$ & $1.177$ \\
\hline
sources 1 & $1.7971$ & $-$ & $0.2724$ & $0.00453$ \\
sources 2 & $1.5521$ & $-$ & $0.2724$ & $0.00885$ \\
sources 3 & $1.5967$ & $-$ & $0.2724$ & $0.01918$ \\
sources 4 & $1.0979$ & $-$ & $0.2724$ & $0.03287$ \\
\hline
\end{tabular}
    \caption{ Number density, galaxy bias (relevant for lenses only), shape-noise dispersion (per shear component; relevant for sources only) and the lognormal shift parameters obtained from the procedure described in \secref{lognormal_covariance}.}
\label{tab:lognormal}
\end{table}

\subsection{Lognormal covariance from simulations}
\label{sec:FLASK}

\begin{figure}
\includegraphics[width=0.5\textwidth]{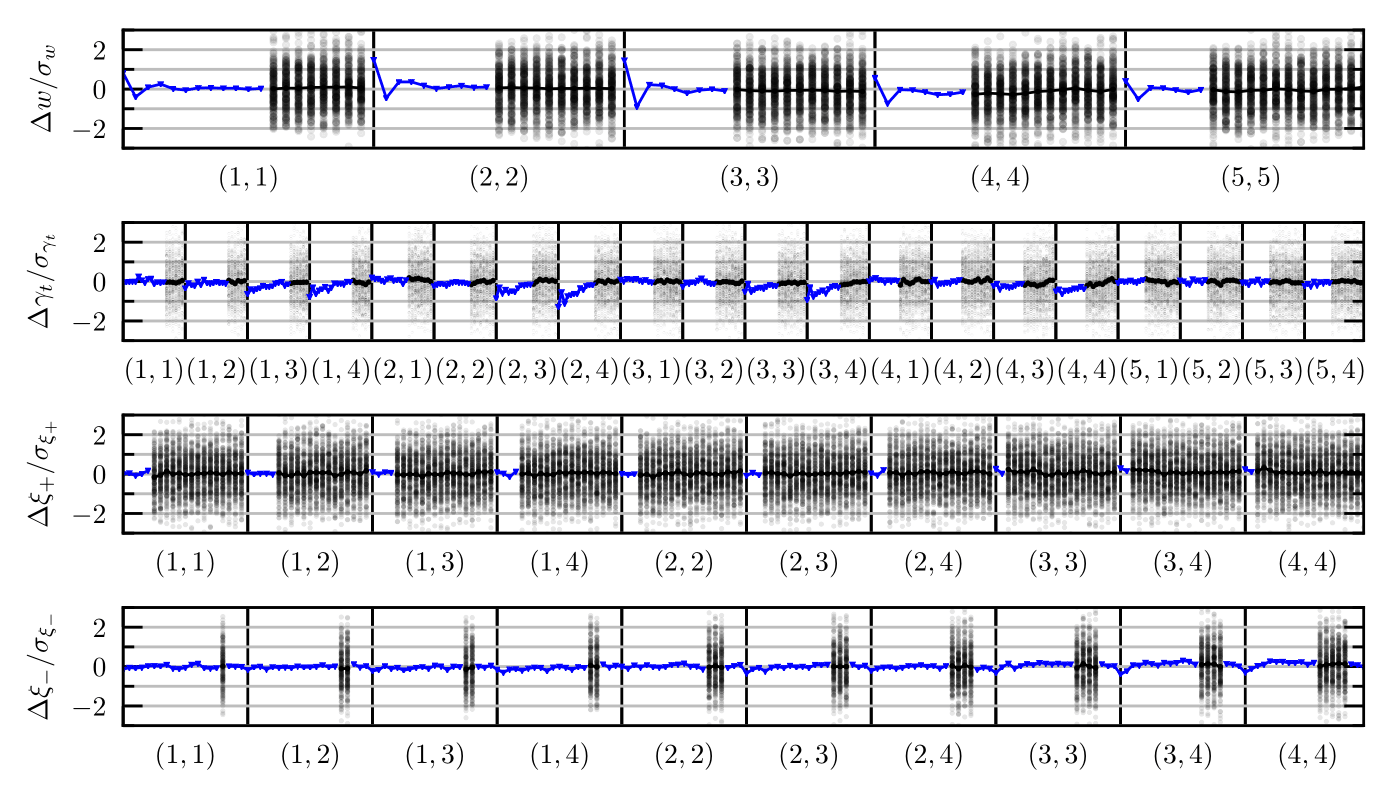}
\caption{Validation of FLASK simulations. Each panel shows the absolute difference of three 2-point correlations measured on FLASK realizations and the predicted correlation functions from input $C(\ell)$s normalized to the statistical error given by the standard deviation along FLASK realizations ($\Delta X /\sigma_X$, where $X=w, \gamma_t, \xi_+, \xi_{-}$).
Gray dots are single realizations and blue dots its mean.}
\label{fi:flask_validation}
\end{figure}

\begin{figure}
\includegraphics[width=0.5\textwidth]{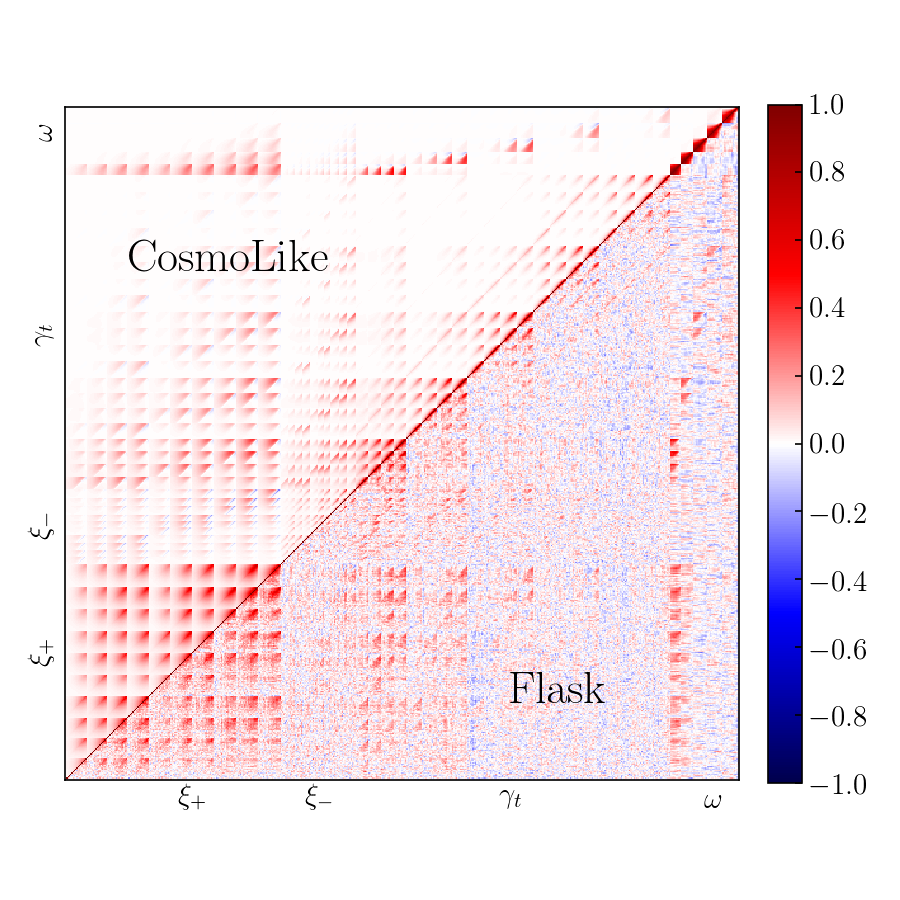}
   \caption{FLASK (lower diagonal) vs. CosmoLike halo model (upper diagonal) correlation matrix.}
  \label{fig:halo_vs_lognormal_correlation}
\end{figure}

We also produce a test DES-Y3 covariance matrix from a set of simulations. We use the publicly available code FLASK (Full sky Lognormal Astro fields Simulation Kit) \citep{Xavier2016} \footnote{http://www.astro.iag.usp.br/~flask/} to generate 800 DES-Y3 footprint sky maps of density, convergence and shear healpix  maps \citep{Gorski2005} with NSIDE=8192, as well as galaxy positions catalogs, used to reproduce the DES-Y3 properties. FLASK is able to  quickly  produce tomographic correlated simulations of clustering and weak lensing lognormal fields based on the DES-Y3 lens and sources samples.  
The lognormal distribution of cosmological fields has been shown to be a good approximation \citep{Coles:1991if,Wild:2004me,Clerkin:2016kyr}  but much less computationally expensive to generate than full N-body simulations. 

As input for the simulations, we used a set of auto and cross correlated power spectrum and the lognormal field shift parameters. 
The theoretical input power spectrum was generated using CosmoLike, 
and the lognormal shifts are the ones listed in \tabrefnospace{lognormal}.  
In order to reproduce the properties of shear fields, we added the shape-noise term by sampling each pixel of the simulated maps to match the correspondent shape-noise dispersion $\sigma_\epsilon$ and number density $n_g$  of the tomographic bin.
At the time of completing the simulation runs, the DES Y3 shear catalog and redshift distribution were not finalized. For this reason, the values used in the simulations are slightly diffeent from the values in \tabrefnospace{lognormal}.  
For the simulations, we set the number density for the five tomographic lens bins as $0.0227, 0.0392, 0.0583, 0.0451, 0.0278$ (arcmin${}^{-2}$).
The shape-noise dispersion values for the four tomographic bins of sources were set to $0.27049, 0.33212, 0.32537, 0.35037$.
The cosmology adopted for the theoretical power spectra is set as 
$\Omega_m   = 0.3$, $\sigma_8   = 0.82355$, $n_s= 0.97$, $\Omega_b=0.048$, $h_0=0.69$, and $\Omega_\nu h_0^2  = 0.00083$. 

We use the publicly available code \texttt{TreeCorr}\footnote{\url{https://github.com/rmjarvis/TreeCorr}} \citep{Jarvis2004} to measure the 3x2 point correlation measurements for 200 DES-Y3 realizations.
For all measurements, we used 20 log-spaced angular separation bins on scales between 2.5 and 250 arcmin.
We set the \texttt{bin\_slop} \texttt{TreeCorr} parameter to zero, essentially setting all estimators to brute-force computation.
In Figure \ref{fi:flask_validation} we show the validation of the measurements comparing with the theoretical input.

We will use the FLASK covariance mainly to estimate the impact of the survey geometry.

\subsection{Comparisons among covariances}

Here we present some comparisons between the different covariance matrices.
In Figure \ref{fig:comparison_diagonal} we show the
ratio of the diagonal elements of the different covariance matrices introduced in this section displaying both the variances of the 
measurements of $\xi_+(\theta)$  of $w(\theta)$.

In Figure \ref{fig:halo_vs_lognormal_correlation} we compare the covariance matrices obtained from the FLASK simulations and the analytical halo model covariance.


\section{Impact of covariance errors on a linearized Gaussian likelihood}
\label{sec:strategy}

As discussed above a full assessment of the impact of using different covariance matrices to parameter estimation becomes unfeasible due to the computational demand of running a large number of MCMC chains.
Since the covariance matrices studied in this work differ by subdominant effects we do not expect large modifications in the results of the estimation of the parameters.
Therefore we will bypass this difficulty by using a linearized approximation of the model data vector as a function of the parameters. The measured data is assumed to be a Gaussian multivariate variable characterized by a covariance matrix and a given prior matrix. This approach is called the Gaussian linear model \citep{Seehars:2014ora,Seehars:2015qza,Raveri:2018wln}. 

Within this approach we study the following impacts of different covariances:
\begin{itemize}
    \item error in the parameter estimation, characterized by the width of the contours;
    \item the scatter of the best fit (maximum posteriors) parameters;
    \item change in the maximum posterior $\chi^2$ value;
    \item error in the  maximum posterior $\chi^2$ value.
\end{itemize}
In the remainder of this section we detail this method. 

\subsection{Linearized likelihoods}
\label{sec:linearized_likelihoods}

\begin{table}
\caption{Fiducial cosmology and standard deviation of Gaussian parameter priors used in our mock likelihood analyses. $A_{\mathrm{IA}, i}$ is the intrinsic alignment amplitude in the $i$th source redshift bin, $m_i$ is the multiplicative shear bias and $\Delta z_{s, i}$ parametrizes systematic shifts in the photometric redshift distribution of that bin. $\Delta z_{l, i}$ parametrizes systematic shifts in the photometric redshift distribution of the $i$th lens redshift bin. The Gaussian priors we choose for the parameters follow the analysis choices of \citet{DES2018}  and we assume infinite flat priors for all other parameters.}
\label{tab:params}
\begin{center}
\begin{tabular}{| c  c  c |}
\hline
\hline
Parameter & Fiducial value & $\sigma_{\mathrm{prior}}$ \\  
\hline 
\multicolumn{2}{|c|}{{\bf Cosmology}} \\
$\Omega_m$  &  0.3 & -  \\ 
$\sigma_8$ &  0.82355 & - \\
$h_{100}$ & 0.69& -\\
$n_s$ & 0.97& - \\
$w_0$ & -1& - \\
$\Omega_b$ &  0.048& -  \\
$\Omega_\nu$ &  0.001743& -  \\
$\Omega_\Lambda$ &  $1 - \Omega_m - \Omega_\nu$ \\
\hline
$b_1$ & 1.7& - \\
$b_2$ & 1.7& - \\
$b_3$ & 1.7& - \\
$b_4$ & 2.0& - \\
$b_5$ & 2.0& - \\
\hline
$\Delta z_{l, 1}$ & 0.0& $0.04$ \\
$\Delta z_{l, 2}$ & 0.0& $0.04$ \\
$\Delta z_{l, 3}$ & 0.0& $0.04$ \\
$\Delta z_{l, 4}$ & 0.0& $0.04$ \\
$\Delta z_{l, 5}$ & 0.0& $0.04$ \\
\hline
$\Delta z_{s, 1}$ & 0.0& $0.08$ \\
$\Delta z_{s, 2}$ & 0.0& $0.08$ \\
$\Delta z_{s, 3}$ & 0.0& $0.08$ \\
$\Delta z_{s, 4}$ & 0.0& $0.08$ \\
\hline
$A_{\mathrm{IA}, 1}$ & 0.0& - \\
$A_{\mathrm{IA}, 2}$ & 0.0& - \\
$A_{\mathrm{IA}, 3}$ & 0.0& - \\
$A_{\mathrm{IA}, 4}$ & 0.0& - \\
\hline
$m_1$ & 0.0& $0.03$ \\
$m_2$ & 0.0& $0.03$ \\
$m_3$ & 0.0& $0.03$ \\
$m_4$ & 0.0& $0.03$ \\
\hline
\end{tabular}
\end{center}
\end{table}

To speed up our simulated likelihood analyses, we employ a linearized model of the data vector  $\boldsymbol{\xi}$ (\eg\ the DES-Y3 3x2-point function data vector). 
This can be considered a linear Taylor expansion of our full model around a fiducial set of parameters $\boldsymbol{\pi}^0$ which is summarized in \tabrefnospace{params}. In this approximation our model data vector becomes
\begin{align}
\label{eq:linearized_model}
    \boldsymbol{\xi}(\boldsymbol{\pi}) =&\ \boldsymbol{\xi}(\boldsymbol{\pi}^0) + \sum_\alpha (\pi_\alpha - \pi_\alpha^0)\left.\frac{\partial \boldsymbol{\xi}(\boldsymbol{\pi})}{\partial \pi_\alpha}\right|_{\boldsymbol{\pi} = \boldsymbol{\pi}^0}
\end{align}
where the sum is over all components $\pi_\alpha$ of the parameter vector $\boldsymbol{\pi}$ (we will use latin indices for the components of the data vector and greek indices for the components of the parameter vector). Given a 2-point function measurement $\boldsymbol{\hat \xi}$ and abbreviating 
\begin{align}
     \boldsymbol{\xi}^0 =&\ \boldsymbol{\xi}(\boldsymbol{\pi}^0) \nonumber \\
    \delta \boldsymbol{\xi} =&\ \boldsymbol{\hat \xi} - \boldsymbol{\xi}^0 \nonumber \\
    \delta \boldsymbol{\pi} =&\ \boldsymbol{\pi} - \boldsymbol{\pi}^0 \nonumber \\
    \partial_\alpha \boldsymbol{\xi} =&\ \left.\frac{\partial \boldsymbol{\xi}(\boldsymbol{\pi})}{\partial \pi_\alpha}\right|_{\boldsymbol{\pi} = \boldsymbol{\pi}_0} \nonumber
\end{align}
our figure of merit $\chi^2$ as a function of the parameters becomes in the linearized approximation
\begin{align}
\label{eq:fom_with_pior}
    \chi^2[\delta\boldsymbol{\pi}] =&\ \left(\delta \boldsymbol{\xi} - \sum_\alpha \delta \pi_\alpha \partial_\alpha \boldsymbol{\xi} \right)^T \mathbf{C}^{-1} \left(\delta \boldsymbol{\xi} - \sum_\alpha \delta \pi_\alpha \partial_\alpha \boldsymbol{\xi} \right)\nonumber \\
    &\ + \left(\boldsymbol{\pi} - \boldsymbol{\pi}^{\mathrm{prior}}\right)^T\ \mathbf{P}\ \left(\boldsymbol{\pi} - \boldsymbol{\pi}^{\mathrm{prior}}\right)\ .
\end{align}
Here we have allowed for a Gaussian prior with covariance matrix $\mathbf{P}^{-1}$ and central value $\boldsymbol{\pi}^{\mathrm{prior}}$.
To find the deviation $\delta\boldsymbol{\pi}^{\mathrm{MP}} = \boldsymbol{\pi}^{\mathrm{MP}} - \boldsymbol{\pi}^0$ from our fiducial parameters that minimizes this function (the maximum posterior value of the parameters is denoted by $\boldsymbol{\pi}^{\mathrm{MP}}$) we have to solve
\begin{equation}
 \left. \frac{\partial \chi^2}{\partial (\delta\pi_\beta)} \right|_{\delta\boldsymbol{\pi}=\delta\boldsymbol{\pi}^{\mathrm{MP}}}\ = 0 \ .
\end{equation}
Defining a vector $\mathbf{x}$ such that $x_\beta = \delta \boldsymbol{\xi}^T \mathbf{C}^{-1} \partial_\beta \boldsymbol{\xi}$ as well as the Fisher matrix $F_{\alpha\beta} = \partial_\beta \boldsymbol{\xi}^T \mathbf{C}^{-1} \partial_\alpha \boldsymbol{\xi}$ this becomes
\begin{align}
    (\mathbf{F}+\mathbf{P})\ \delta\boldsymbol{\pi}^{\mathrm{MP}} = \mathbf{x} + \mathbf{P}\ (\boldsymbol{\pi}^{\mathrm{prior}} - \boldsymbol{\pi}^{0}) \nonumber
\end{align}
\begin{align}
\label{eq:ML_parameter_residuals}
    \Rightarrow\  \boldsymbol{\pi}^{\mathrm{MP}} = \boldsymbol{\pi}^0 + (\mathbf{F}+\mathbf{P})^{-1} \mathbf{x} + (\mathbf{F}+\mathbf{P})^{-1}\mathbf{P}\ (\boldsymbol{\pi}^{\mathrm{prior}} - \boldsymbol{\pi}^{0})\ .
\end{align}

We now want to consider the situation when a model covariance matrix $\mathbf{C}_{\mathrm{mod}}$ is used to calculate the likelihood in equation (\ref{eq:fom_with_pior} which is different from the true covariance matrix $\mathbf{C}_{\mathrm{true}}$ of the statistical uncertainties in the data vector $\boldsymbol{\hat\xi}$. In that case our linearized likelihood will be a Gaussian centered around $\boldsymbol{\pi}^{\mathrm{MP}}$ and with parameter covariance matrix
\begin{equation}
    \mathbf{C}_{\boldsymbol{\pi}, \mathrm{like}} = (\mathbf{F}_{\mathrm{mod}}+\mathbf{P})^{-1}\ ,
\end{equation}
where $F_{\mathrm{mod},\alpha\beta} = \partial_\beta \boldsymbol{\xi}^T \mathbf{C}_{\mathrm{mod}}^{-1} \partial_\alpha \boldsymbol{\xi}$ is the Fisher matrix calculated from the model covariance.

The actual covariance matrix of $\boldsymbol{\pi}^{\mathrm{MP}}$ includes two sources of noise. First, statistical uncertainties in the measurement $\boldsymbol{\hat\xi}$ which are described by the covariance matrix $\mathbf{C}_{\mathrm{true}}$ and are represented by the first term in equation (\ref{eq:ML_parameter_residuals}) that is proportional to $\mathbf{x}$. And secondly, statistical uncertainties in our choice of the prior center which are described by the prior covariance matrix $\mathbf{P}^{-1}$ and are represented by the second term in equation (\ref{eq:ML_parameter_residuals}) that is proportional to $\boldsymbol{\pi}^{\mathrm{prior}}$. The latter term has the covariance matrix $(\mathbf{F}_{\mathrm{mod}}+\mathbf{P})^{-1}\mathbf{P}(\mathbf{F}_{\mathrm{mod}}+\mathbf{P})^{-1}$ (because the covariance matrix of $\boldsymbol{\pi}^{\mathrm{prior}}$ is $\mathbf{P}^{-1}$). Hence, the total covariance matrix of $\boldsymbol{\pi}^{\mathrm{MP}}$ can be written as
\begin{align}
    &\ \left(\mathbf{C}_{\boldsymbol{\pi}, \mathrm{MP}}\right)_{\alpha\beta}\ \equiv\  \mathrm{Cov}[\pi_{\alpha}^{\mathrm{MP}}, \pi_{\beta}^{\mathrm{MP}}]  = \nonumber \\
    =&\ (\mathbf{F}_{\mathrm{mod}}+\mathbf{P})^{-1}\mathbf{P}(\mathbf{F}_{\mathrm{mod}}+\mathbf{P})^{-1}\ +\nonumber \\
    &\ +\ \sum_{\kappa, \lambda} (\mathbf{F}_{\mathrm{mod}} + \mathbf{P})_{\alpha\kappa}^{-1}\ (\mathbf{F}_{\mathrm{mod}} + \mathbf{P})_{\lambda\beta}^{-1}\ \times \nonumber \\
    &\ \times \ \sum_{i,k} \partial_\kappa \xi_i\  (\mathbf{C}_{\mathrm{mod}}^{-1}\mathbf{C}_{\mathrm{true}}\mathbf{C}_{\mathrm{mod}}^{-1})_{ik}\ \partial_\lambda \xi_k\ .
\end{align}
For $\mathbf{C}_{\mathrm{mod}} = \mathbf{C}_{\mathrm{true}}$ it is easy to see that this parameter covariance $\mathbf{C}_{\boldsymbol{\pi}, \mathrm{MP}}$ equals the covariance $\mathbf{C}_{\boldsymbol{\pi}, \mathrm{like}}$ that describes the shape of our likelihood (as it should).

\subsection{Impact on the width of the likelihood and scatter of best fit parameters}
We can use the above findings to study the impact of different effects in covariance modelling on parameter constraints. If a covariance matrix $\mathbf{C}_1$ contains a noise contribution that is missing in another covariance matrix $\mathbf{C}_2$, then we quantify the difference between these matrices by considering two effects:
\begin{itemize}
    \item \textbf{Width of likelihood contours:}
    \\
    
    Denoting the Fisher matrices obtained from $\mathbf{C}_1$ or $\mathbf{C}_2$ as $\mathbf{F}_1$ and $\mathbf{F}_2$ respectively, the width of likelihood contours drawn from the different covariances are given by
    \begin{align}
        \mathbf{C}_{\boldsymbol{\pi}, \mathrm{like},\ 1} =&\ (\mathbf{F}_1+\mathbf{P})^{-1}\nonumber \\
        \mathbf{C}_{\boldsymbol{\pi}, \mathrm{like},\ 2} =&\ (\mathbf{F}_2+\mathbf{P})^{-1}\ .
    \end{align}
    Hence, if the difference $\mathbf{C}_1-\mathbf{C}_2 = \mathbf{E}$ represents noise contributions missing from (or miss-estimated in $\mathbf{C}_2$), then a comparison of $\mathbf{C}_{\boldsymbol{\pi}, \mathrm{like},\ 1}$ and $\mathbf{C}_{\boldsymbol{\pi}, \mathrm{like},\ 2}$ quantifies the impact of this on the width of parameter contours.
    \\
    
    \item\textbf{Scatter in the center of likelihood contours:}
    \\
    
    If the data vector $\boldsymbol{\hat\xi}$ had $\mathbf{C}_1$ as its true covariance matrix but $\mathbf{C}_2$ would be used to derive the maximum posterior parameters $\boldsymbol{\pi}^{\mathrm{MP}}$ from it, then the maximum posterior parameter covariance would be given by
    \begin{align}\left(\mathbf{C}_{\boldsymbol{\pi}, \mathrm{MP},\ 2}\right)_{\alpha\beta} =&\ (\mathbf{F}_{\mathrm{2}}+\mathbf{P})^{-1}\ \mathbf{P}\ (\mathbf{F}_{\mathrm{2}}+\mathbf{P})^{-1}\ +\nonumber \\
    &\ +\ \sum_{\kappa, \lambda} (\mathbf{F}_{\mathrm{2}} + \mathbf{P})_{\alpha\kappa}^{-1}\ (\mathbf{F}_{\mathrm{2}} + \mathbf{P})_{\lambda\beta}^{-1}\ \times \nonumber \\
    &\ \times \sum_{i,k} \partial_\kappa \xi_i\  (\mathbf{C}_{\mathrm{2}}^{-1\ }\mathbf{C}_{\mathrm{1}}\ \mathbf{C}_{\mathrm{2}}^{-1})_{ik}\ \partial_\lambda \xi_k\ .
    \end{align}
    If the difference $\mathbf{C}_1-\mathbf{C}_2 = \mathbf{E}$ represents noise contributions missing from (or miss-estimated in $\mathbf{C}_2$), then a comparison of $\mathbf{C}_{\boldsymbol{\pi}, \mathrm{MP},\ 2}$ and $\mathbf{C}_{\boldsymbol{\pi}, \mathrm{MP},\ 1} \equiv \mathbf{C}_{\boldsymbol{\pi}, \mathrm{like},\ 1}$ quantifies the impact of this on the scatter in the location of parameter contours.
\end{itemize}

An inaccurate covariance model will in general have a different impact on the width and the location of parameter contours. Hence, in order to quantify the importance of different effects in covariance modelling for parameter estimation, we compare both the pair $\mathbf{C}_{\boldsymbol{\pi}, \mathrm{like},\ 1}\ /\ \mathbf{C}_{\boldsymbol{\pi}, \mathrm{like},\ 2}$ and the pair $\mathbf{C}_{\boldsymbol{\pi}, \mathrm{MP},\ 1}\ /\ \mathbf{C}_{\boldsymbol{\pi}, \mathrm{MP},\ 2}$.

\subsection{Distribution of $\chi^2$ when fitting for parameters}
\label{sec:chiSq_in_linearized_likelihood}

Within the linearized likelihood model developed in the previous section we now investigate how errors in the covariance model impact the distribution of $\chi_{\mathrm{MP}}^2$ between measured data vector $\boldsymbol{\hat\xi}$ and a maximum posterior model $\boldsymbol{\xi}_{\mathrm{MP}} = \boldsymbol{\xi}(\boldsymbol{\pi}^{\mathrm{MP}})$,
\begin{equation}
    \hat{\chi}_{\mathrm{MP}}^2 = (\boldsymbol{\hat\xi} - \boldsymbol{\xi}_{\mathrm{MP}})^T \mathbf{C}^{-1} (\boldsymbol{\hat\xi} - \boldsymbol{\xi}_{\mathrm{MP}})\ .
\end{equation}
We start with the case that
\begin{enumerate}
    \item the true covariance $\mathbf{C}$ of $\boldsymbol{\hat\xi}$ is known
    \item no parameter priors are used when determining the best fitting model $\boldsymbol{\xi}_{\mathrm{MP}}$
    \item the true expectation value $\boldsymbol{\bar\xi} \equiv\langle\boldsymbol{\hat\xi}\rangle$ lies within our parameter space. I.e. there are parameters $\boldsymbol{\pi}^{\mathrm{true}}$ such that $\boldsymbol{\xi}(\boldsymbol{\pi}^{\mathrm{true}}) = \boldsymbol{\bar\xi}$ .
\end{enumerate}
We will show that, as expected, in this case $\hat{\chi}_{\mathrm{MP}}^2$ should follow a $\chi^2$-distribution with $N_{\mathrm{data}}-N_{\mathrm{param}}$ degrees of freedom.

Using  equations (\ref{eq:linearized_model}) and (\ref{eq:ML_parameter_residuals}) (and setting again $\delta \boldsymbol{\xi} \equiv \boldsymbol{\hat\xi} - \boldsymbol{\xi}^0$) one can see that the maximum posterior data vector is given by
\begin{align}
    \boldsymbol{\xi}_{\mathrm{MP}} =&\ \boldsymbol{\xi}^0 +  \sum_{\alpha \beta} \partial_\alpha \boldsymbol{\xi}\ (\mathbf{F}^{-1})_{\alpha \beta}\ \left(\delta \boldsymbol{\xi}^T \mathbf{C}^{-1} \partial_\beta \boldsymbol{\xi}\right)\nonumber \\
    =&\ \boldsymbol{\bar\xi} +  \sum_{\alpha \beta} \partial_\alpha \boldsymbol{\xi}\ (\mathbf{F}^{-1})_{\alpha \beta}\ \left((\boldsymbol{\hat\xi}-\boldsymbol{\bar\xi})^T \mathbf{C}^{-1} \partial_\beta \boldsymbol{\xi}\right)\nonumber \\
    =&\ \boldsymbol{\bar\xi} +  \sum_{\alpha \beta} \sum_{kl} \partial_\alpha \boldsymbol{\xi}\ (\mathbf{F}^{-1})_{\alpha \beta}\ (\hat\xi_k-\bar\xi_k) \left(\mathbf{C}^{-1}\right)_{kl} \partial_\beta \xi_l\nonumber \\
    \equiv&\ \boldsymbol{\bar\xi} + \boldsymbol{\mathcal{P}} \cdot (\boldsymbol{\hat\xi} - \boldsymbol{\bar\xi}) \ .
\label{eq:xiML}
\end{align}
Here, the second line follows from the fact that $\boldsymbol{\bar\xi} =\langle\boldsymbol{\hat\xi}\rangle=\langle\boldsymbol{\xi}_{\mathrm{MP}}\rangle$ and we have defined the matrix
\begin{equation}
    \mathcal{P}_{ij} = \sum_{\alpha \beta} \partial_\alpha \xi_i \sum_{l} (\mathbf{F}^{-1})_{\alpha \beta}\ \left(\mathbf{C}^{-1}\right)_{lj} \partial_\beta \xi_l\ .
\label{eq:P}
\end{equation}
It can be shown that $\boldsymbol{\mathcal{P}}$ is an idempotent matrix ($\boldsymbol{\mathcal{P}}^2=\boldsymbol{\mathcal{P}}$) and furthermore that
\begin{align}
    \mathrm{Trace}\left(\boldsymbol{\mathcal{P}}\right) &= N_{\mathrm{param}} \nonumber \\
    \boldsymbol{C}^{-1} \boldsymbol{\mathcal{P}} \boldsymbol{C} &=
 \boldsymbol{\mathcal{P}}^T   \ .
\end{align}
The residual between the measurement $\boldsymbol{\hat\xi}$ and the best fitting model $\boldsymbol{\xi}_{\mathrm{MP}}$ can be written in terms of $\boldsymbol{\mathcal{P}}$ as
\begin{align}
    \boldsymbol{\hat\xi} - \boldsymbol{\xi}_{\mathrm{MP}} =&\ (\boldsymbol{\hat\xi} - \boldsymbol{\bar\xi}) - (\boldsymbol{\xi}_{\mathrm{MP}} - \boldsymbol{\bar\xi}) \nonumber \\
    =&\ (\mathbb{1} - \boldsymbol{\mathcal{P}})\cdot (\boldsymbol{\hat\xi} - \boldsymbol{\bar\xi})\ .
\end{align}
Hence, the covariance matrix of $\boldsymbol{\hat\xi} - \boldsymbol{\xi}_{\mathrm{MP}}$ is given by 
\begin{equation}
    \mathbf{C}_{\mathcal{P}} \equiv 
    \langle (\boldsymbol{\hat\xi} - \boldsymbol{\xi}_{\mathrm{MP}})^T (\boldsymbol{\hat\xi} - \boldsymbol{\xi}_{\mathrm{MP}})   \rangle = 
    (\mathbb{1} - \boldsymbol{\mathcal{P}}) \mathbf{C}\ (\mathbb{1} - \boldsymbol{\mathcal{P}})^T
\end{equation}
This makes it straightforward to find the expectation value
\begin{align}
\label{eq:chiSq_ML_with_correct_cov}
    \langle \chi_{\mathrm{MP}}^2 \rangle =&
    \langle (\boldsymbol{\hat\xi} - \boldsymbol{\xi}_{\mathrm{MP}})^T \boldsymbol{C}^{-1}(\boldsymbol{\hat\xi} - \boldsymbol{\xi}_{\mathrm{MP}})   \rangle \nonumber \\
    =&\ \mathrm{Trace}\left( \mathbf{C}_{\mathcal{P}}\ \mathbf{C}^{-1}\right) \nonumber \\
    =&\ \sum_{jk} C_{kj}\ \left(C^{-1}\right)_{jk}\ -\ \sum_{k} \mathcal{P}_{kk} \nonumber \\
    =&\ N_{\mathrm{data}} - N_{\mathrm{param}}\ .
\end{align}
Similarly, the variance of $\chi_{\mathrm{MP}}^2$ can be shown to be
\begin{align}
\label{eq:VarchiSq_ML_with_correct_cov}
    \mathrm{Var}(\chi_{\mathrm{MP}}^2) =&\ \langle (\chi_{\mathrm{MP}}^2)^2 \rangle - \langle \chi_{\mathrm{MP}}^2 \rangle^2 \nonumber \\
    =&\ 2\ \mathrm{Trace}\left( \left[ \mathbf{C}_{\mathcal{P}}\ \mathbf{C}^{-1}\right]^2 \right) \nonumber \\
    =&\ 2(N_{\mathrm{data}} - N_{\mathrm{param}})\ .
\end{align}
So far, we have only re-derived textbook results \citep{Anderson2003}. Now how do $\langle \chi_{\mathrm{MP}}^2 \rangle$ and $\mathrm{Var}(\chi_{\mathrm{MP}}^2)$ change if the covariance model $\mathbf{C}_{\mathrm{mod}}$ we use to find the best fitting model $\boldsymbol{\xi}_{\mathrm{MP}}$ and to compute $\chi_{\mathrm{MP}}^2$ is different from the true covariance matrix $\mathbf{C}$ of $\boldsymbol{\hat\xi}$?

Following similar steps as Eqs. (\ref{eq:xiML}) and (\ref{eq:P}) one can show that
\begin{equation}
\label{eq:xi_ML_with_wrong_cov}
    \boldsymbol{\xi}_{\mathrm{MP}} = \boldsymbol{\bar\xi} + \boldsymbol{\mathcal{P}}_{\mathrm{mod}} \cdot (\boldsymbol{\hat\xi}-\boldsymbol{\bar\xi})
\end{equation}
where
\begin{equation}
    (\mathcal{P}_{\mathrm{mod}})_{ij} = \sum_{\alpha \beta} \partial_\alpha \xi_i \sum_{l} (\mathbf{F}_{\mathrm{mod}}^{-1})_{\alpha \beta}\ \left(\mathbf{C}_{\mathrm{mod}}^{-1}\right)_{lj} \partial_\beta \xi_l
\end{equation}
and where the Fisher matrix $\mathbf{F}_{\mathrm{mod}}$ is computed from the model covariance $\mathbf{C}_{\mathrm{mod}}$. \Eqnref{xi_ML_with_wrong_cov} especially shows that $\boldsymbol{\xi}_{\mathrm{MP}}$ is still an unbiased estimator of $\boldsymbol{\bar\xi}$ even when $\mathbf{C}_{\mathrm{mod}} \neq \mathbf{C}$. When deriving the moments of $\chi_{\mathrm{MP}}^2$ we will still come across expectation values like (\cf\ \eqnrefnospace{chiSq_ML_with_correct_cov})
\begin{equation}
    \langle (\hat\xi_i - \bar\xi_i)(\hat\xi_j - \bar\xi_j) \rangle \equiv (\mathbf{C})_{ij} \neq (\mathbf{C}_{\mathrm{mod}})_{ij}\ .
\end{equation}
Hence the expectation value and variance of $\chi_{\mathrm{MP}}^2$ are given by
\begin{align}
\label{eq:expectation_chiSq_ML_wrong_cov}
    \langle \chi_{\mathrm{MP}}^2 \rangle =&\ \mathrm{Trace}\left( \mathbf{C}_{\boldsymbol{\mathcal{P}}_{\mathrm{mod}}} \mathbf{C}_{\mathrm{mod}}^{-1} \right) \\
\label{eq:variance_chiSq_ML_wrong_cov}
    \mathrm{Var}(\chi_{\mathrm{MP}}^2) =&\ 2\ \mathrm{Trace}\left( \left[ \mathbf{C}_{\boldsymbol{\mathcal{P}}_{\mathrm{mod}}} \mathbf{C}_{\mathrm{mod}}^{-1} \right]^2 \right),
\end{align}
where
\begin{equation}
    \mathbf{C}_{\boldsymbol{\mathcal{P}}_{\mathrm{mod}}} = 
 (\mathbb{1} - \boldsymbol{\mathcal{P}}_{\mathrm{mod}}) \mathbf{C}\ (\mathbb{1} - \boldsymbol{\mathcal{P}}_{\mathrm{mod}})^T    
\end{equation}
Now we are left to investigate how Equations \ref{eq:expectation_chiSq_ML_wrong_cov} and \ref{eq:variance_chiSq_ML_wrong_cov} change when a Gaussian parameter prior $\mathbf{P}$ is included in the likelihood function (\cf\ \eqnrefnospace{fom_with_pior}). A complication in this case is, that now $\boldsymbol{\xi}_{\mathrm{MP}}$ is not necessarily an unbiased estimate of $\boldsymbol{\bar\xi}$ anymore. This is because in \eqnrefnospace{fom_with_pior} we have centered our prior around the model parameters $\boldsymbol{\pi}^{\mathrm{prior}}$ which may be different from the true parameters $\boldsymbol{\pi}^{\mathrm{true}}$. Inserting the full expression for the maximum posterior parameters (\eqnrefnospace{ML_parameter_residuals}) into our linearized model we now get
\begin{align}
\label{eq:xi_ML_with_wrong_cov_and_prior}
    \boldsymbol{\xi}_{\mathrm{MP}} =&\ \boldsymbol{\xi}^0 + \boldsymbol{\mathcal{P}}_{\mathrm{mod}} \cdot (\boldsymbol{\hat\xi}-\boldsymbol{\xi}^0) + \boldsymbol{\zeta}
\end{align}
with
\begin{align}
(\mathcal{P}_{\mathrm{mod}})_{ij} =&\ \sum_{\alpha \beta} \partial_\alpha \xi_i \sum_{l} (\mathbf{F}_{\mathrm{mod}} + \mathbf{P})_{\alpha \beta}^{-1}\ \left(\mathbf{C}_{\mathrm{mod}}^{-1}\right)_{lj} \partial_\beta \xi_l\nonumber \\
\boldsymbol{\zeta} =&\ \sum_\alpha \left[(\mathbf{F}_{\mathrm{mod}}+\mathbf{P})^{-1}\mathbf{P}\ (\boldsymbol{\pi}^{\mathrm{prior}} - \boldsymbol{\pi}^{0})\right]_{\alpha}\partial_\alpha \boldsymbol{\xi}\nonumber \\
\end{align}
The residual between $\boldsymbol{\hat\xi}$ and $\boldsymbol{\xi}_{\mathrm{MP}}$ hence becomes
\begin{align}
    \boldsymbol{\hat\xi} - \boldsymbol{\xi}_{\mathrm{MP}} =&\ (\mathbb{1} - \boldsymbol{\mathcal{P}}_{\mathrm{mod}})\cdot (\boldsymbol{\hat\xi} - \boldsymbol{\xi}^0) - \boldsymbol{\zeta}\ .
\end{align}
Treating the prior center $\boldsymbol{\pi}^{\mathrm{prior}}$ again as a random vector centered around $\boldsymbol{\pi}^{\mathrm{true}}$, $\boldsymbol{\zeta}$ also becomes a random vector with covariance
\begin{align}
 \left(\mathbf{C}_\zeta\right)_{ij} \equiv&\    \mathrm{Cov}[\zeta_i , \zeta_j]\nonumber \\
 =&\ \sum_{\alpha \beta \gamma \delta} \partial_{\alpha} \xi_i\ (\mathbf{F}_{\mathrm{mod}} + \mathbf{P})_{\alpha \beta}^{-1}\ \mathbf{P}_{\beta\gamma}\ (\mathbf{F}_{\mathrm{mod}} + \mathbf{P})_{\gamma \delta}^{-1}\ \partial_{\delta} \xi_j\ .
\end{align}
Hence, along lines similar to the case without a prior, we can write the moments of $\chi_{\mathrm{MP}}^2$ for a given model covariance as
\begin{align}
\label{eq:expectation_chiSq_ML_wrong_cov_and_pior}
    \langle \chi_{\mathrm{MP}}^2 \rangle =&\ \mathrm{Trace}\left( \lbrace\mathbf{C}_{{\mathcal{P}}_{\mathrm{mod}}}+\mathbf{C}_\zeta\rbrace\ \mathbf{C}_{\mathrm{mod}}^{-1}\right) \\
\label{eq:variance_chiSq_ML_wrong_cov_and_pior}
    \mathrm{Var}(\chi_{\mathrm{MP}}^2) =&\ 2\ \mathrm{Trace}\left( \left[ \lbrace\mathbf{C}_{{\mathcal{P}}_{\mathrm{mod}}}+\mathbf{C}_\zeta\rbrace\ \mathbf{C}_{\mathrm{mod}}^{-1}\right]^2 \right) \ .
\end{align}
Notice that in the absence of priors $\mathbf{C}_\zeta = \mathbb{0}$ and for the true covariance $\mathbf{C}$ we recover equations
(\ref{eq:chiSq_ML_with_correct_cov}) and (\ref{eq:VarchiSq_ML_with_correct_cov}) as expected. 
Equations (\ref{eq:expectation_chiSq_ML_wrong_cov_and_pior}) and (\ref{eq:variance_chiSq_ML_wrong_cov_and_pior}) are used to produce our main result shown in Figure \ref{fig:chiSq_offsets} for different covariance matrices.


\section{Exploring different effects in the covariance modelling}
\label{sec:impact_of_modelling}


Our main goal is to study the impact of including different effects in the covariance modelling on the estimation of parameters.
Several covariance matrices were generated and tested under different assumptions and approximations. The main results were already shown in  \secref{validation}.
We now present the details of each step in the validation strategy that was outlined in \secref{strategy}.

\subsection{Gaussian likelihood assumption}
\label{sec:Gauss}

\begin{figure}
\includegraphics[width=0.49\textwidth]{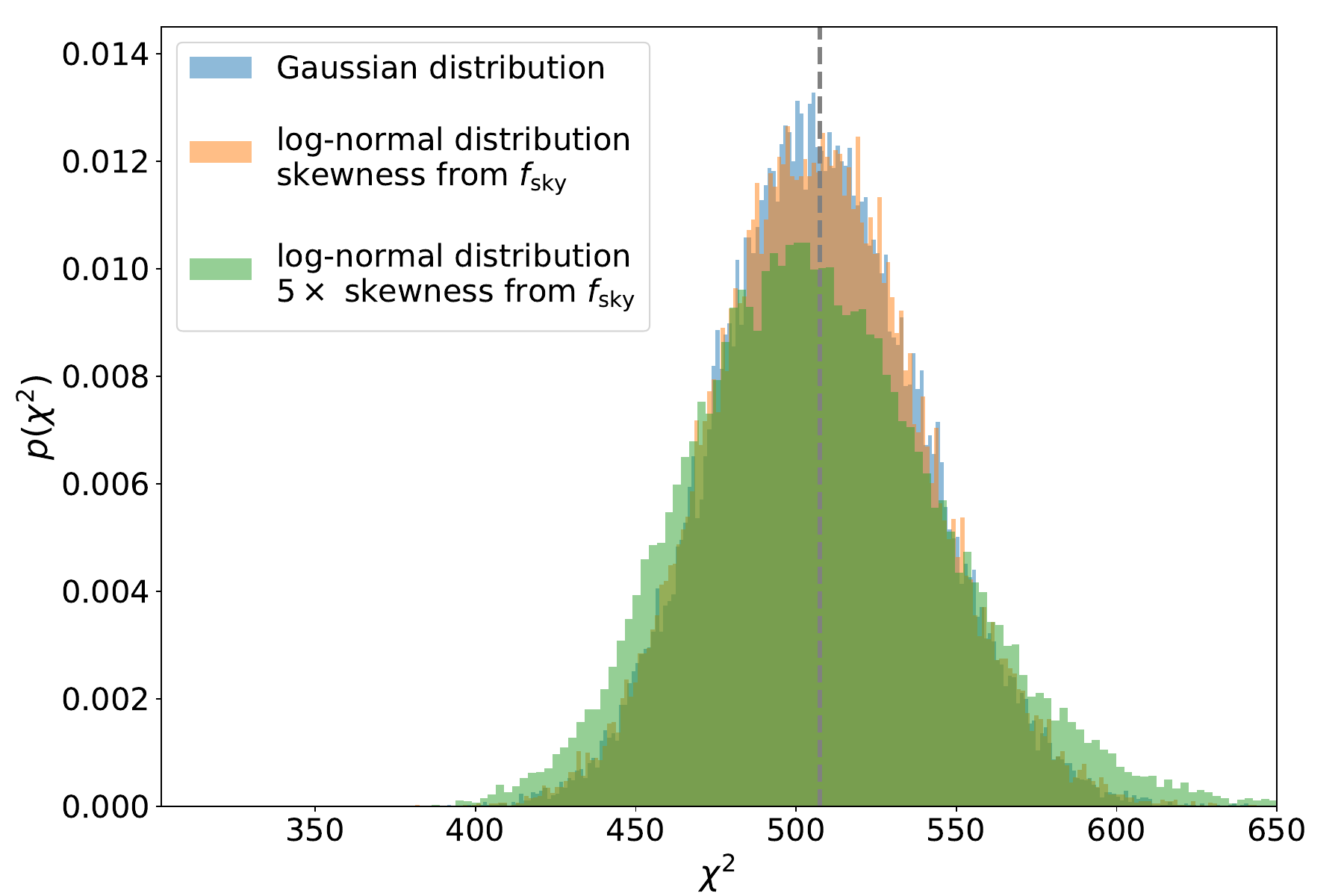}

\includegraphics[width=0.49\textwidth]{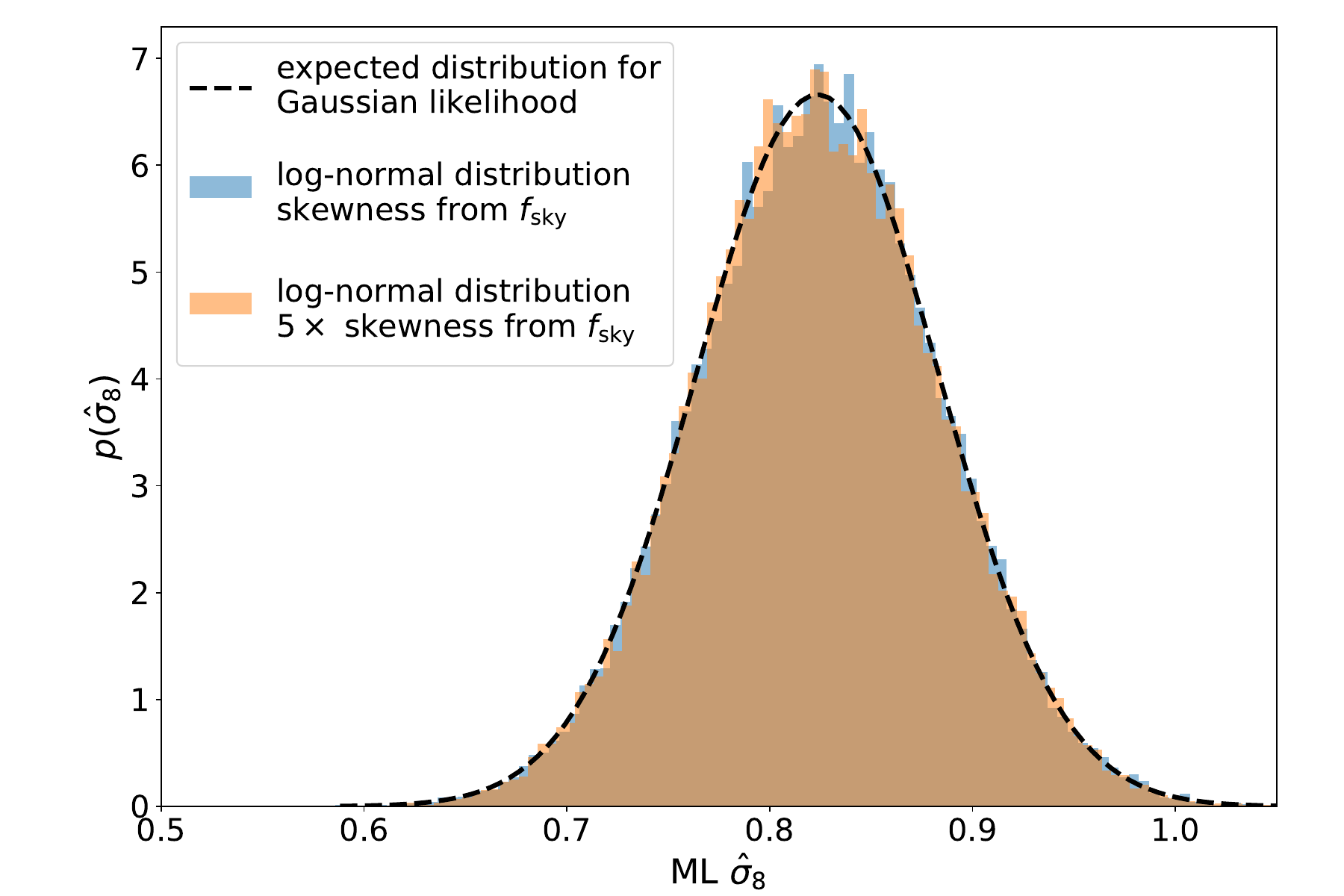}
\caption{Top panel: Distribution of $\chi^2$ when drawing 3x2pt data vectors from a Gaussian distribution (blue histogram), from a shifted log-normal distribution where the skewness of each data point was computed in the $f_{\mathrm{sky}}$ approximation (red histogram) and when assuming that the skewness of the data points is $5$ times that of the $f_{\mathrm{sky}}$ approximation (green histogram). Bottom panel: Distribution of maximum posterior $\sigma_8$ when fitting the linearized model to \secref{linearized_likelihoods} Gaussian realisations of our fiducial data vector , to lognormal realisations of our fiducial data vector (blue histogram)  and to lognormal realisations with 5 times the skewness of the $f_{\mathrm{sky}}$ approximation employed on \secref{Gauss} (orange histogram). 
}
\label{fig:nG_likelihood}
\end{figure}

A basic assumption of our framework of testing different covariance matrices is that the likelihood function of the data is Gaussian.
One simple reason of why the sampling distribution of the correlation functions can not be an exact multivariate Gaussian is that this violates the positivity constraint of the power spectrum \citep{Schneider:2009kd}. There are also other reasons described below. The purpose of this Subsection is to assess the impact of non-Gaussianity of the likelihood of 2-point functions in the parameter estimation. In this sense checking this basic assumption is a test of the whole framework and is different from the robustness tests for the covariance matrix modelling described in the remaining Subsections of this Section.

The impact of a non-Gaussian likelihood in parameter estimation of weak lensing correlation functions has been recently studied in \citet{Lin:2019omj} where no significant biases were found in one-dimensional posteriors of $\Omega_m$ and
$\sigma_8$ between the multivariate Gaussian likelihood model and more complex non-Gaussian likelihood models.
Also in \citet{Sellentin2018b} the skewed distributions of weak lensing shear correlation functions are used to derive an analytical expression for a non-Gaussian likelihood.

We first consider a full-sky survey such that each of our 2-point function estimators 
$\hat\xi^{\mathrm{AB}}(\theta)$ is a harmonic transform of a harmonic space estimator $\hat C_\ell^{AB}$ (\cf\ \eqnrefnospace{xi_hat_in_terms_of_Cell_hat_general}), \ie
\begin{equation}
    \hat \xi^{\mathrm{AB}}(\theta) = \sum_{\ell = 0}^\infty \frac{2 \ell +1}{4 \pi}  F^{AB}_\ell( \theta) \, \hat C_\ell^{AB}\ .
\end{equation}
Each $\hat C_\ell^{AB}$ is given in terms of the spherical harmonics coefficients $a_{\ell m}$, $b_{\ell m}$ of two Gaussian random fields as
\begin{equation}
\label{eq:C_ell_estimator}
    \hat C_\ell^{AB} = \frac{1}{2\ell+1} \sum_{m=-\ell}^{\ell} a_{\ell m} b_{\ell m}^*\ .
\end{equation}
The product of two Gaussian random variables does not follow a Gaussian distribution. Therefore, in principle one would not expect $\hat C_\ell^{AB}$ (and consequently $\hat \xi^{\mathrm{AB}}(\theta)$)  to have a Gaussian likelihood. 
However, at small scales, \ie\ at high multipoles $\ell$ , the sum of the random variables $a_{\ell m} b_{\ell m}^*$ in \eqnref{C_ell_estimator} will approach a Gaussian distribution by means of the central limit theorem, since there is a large number ($2 \ell + 1$) of independent modes. 
It should be pointed out that at these small scales the galaxy density and shear fields  characterized by $a_{\ell m}$ and $b_{\ell m}$ are  themselves non-Gaussian due to the non-linear evolution of gravity.

It is hence our working hypothesis that non-Gaussianity of $\hat C_\ell^{AB}$ only matters at the largest scales (small $\ell$'s) 
where both $a_{\ell m}$ and $b_{\ell m}$ 
can be considered Gaussian random variables but not their product.
In the full-sky case it can then be shown that the second and third central moments of $\hat C_\ell^{AB}$ are given by
\begin{align}
    \langle \left(\hat C_\ell^{AB} - C_\ell^{AB}\right)^2\rangle =&\  \frac{\left[\left(C_\ell^{AB}\right)^2+ C_\ell^{AA}C_\ell^{BB}\right]}{2\ell+1}\\
    \langle \left(\hat C_\ell^{AB} - C_\ell^{AB}\right)^3\rangle =&\ \frac{2 \left[ \left(C_\ell^{AB}\right)^3 + 3 C_\ell^{AA}C_\ell^{BB}C_\ell^{AB}\right]}{(2\ell+1)^2}\ .
\end{align}
If only a fraction $f_{\mathrm{sky}}$ of the sky is being observed, these moments get divided by   $f_{\mathrm{sky}}$ and $f_{\mathrm{sky}}^2$ respectively.

Assuming different multipoles to be uncorrelated, the corresponding moments of $\hat\xi^{\mathrm{AB}}(\theta)$ can be computed as
\begin{align}
\label{eq:variance_of_xiAB}
    &\ \langle \left(\hat\xi^{\mathrm{AB}}(\theta)-\xi^{\mathrm{AB}}(\theta)\right)^2\rangle\nonumber \\
    =&\  \sum_{\ell = 0}^\infty \left(\frac{2 \ell +1}{4 \pi}  F^{AB}_\ell( \theta)\right)^2 \ \langle \left(\hat C_\ell^{AB} - C_\ell^{AB}\right)^2\rangle \\
\label{eq:skewness_of_xiAB}
    &\ \langle \left(\hat\xi^{\mathrm{AB}}(\theta)-\xi^{\mathrm{AB}}(\theta)\right)^3\rangle\nonumber \\
    =&\  \sum_{\ell = 0}^\infty \left(\frac{2 \ell +1}{4 \pi}  F^{AB}_\ell( \theta)\right)^3 \ \langle \left(\hat C_\ell^{AB} - C_\ell^{AB}\right)^3\rangle \ .
\end{align}
\Eqnref{variance_of_xiAB} is of course nothing but the diagonal of the covariance matrix (\cf\ \eqnref{realspace_cov_in_terms_of_harmonic_space_cov}).

The dominant effect of the non-Gaussianity of the $C_\ell$'s is a positive skewness in the distribution of our data vectors \citep{Sellentin2018b}. To estimate its impact on our parameter constraints, we approximate the entire distribution of our 3x2pt data vector  by a multivariate lognormal distribution. The covariance of our data vector and the skewness of each data point as given by \eqnref{skewness_of_xiAB} are sufficient to fix the parameters of a shifted log-normal distribution. We have already discussed this in Sections~\ref{sec:lognormal_covariance} and \ref{sec:FLASK}, though with a conceptual difference: in that section we describe how to configure log-normal simulations of the cosmic density field, while here we assume measurements of the 3x2-point functions to have a multivariate log-normal distribution. To be explicit, we fix the shift parameters $\lambda(\theta)$ that enter the log-normal PDF of the measurements $\hat\xi^{\mathrm{AB}}(\theta)$ in the different angular bins (\cf\ \eqnrefnospace{definition_lognormal_variable} for the definition of $\lambda$) via the equation
\begin{align}
    &\ \langle \left(\hat\xi^{\mathrm{AB}}(\theta)-\xi^{\mathrm{AB}}(\theta)\right)^3\rangle = \nonumber \\
    &\ \frac{3\langle \left(\hat\xi^{\mathrm{AB}}(\theta)-\xi^{\mathrm{AB}}(\theta)\right)^2\rangle^2}{\lambda(\theta)} + \frac{\langle \left(\hat\xi^{\mathrm{AB}}(\theta)-\xi^{\mathrm{AB}}(\theta)\right)^2\rangle^3}{\lambda(\theta)^3}\ ,
\end{align}
which relates the 2nd and 3rd central moments of log-normal random variables \citep{Hilbert2011}.

In the top panel of \figref{nG_likelihood} we show the impact of this non-Gaussianity on the distribution of maximum posterior $\chi^2$. For that figure we generated $300,000$ random realisations of our fiducial data vector from a multi-variate Gaussian distribution, $300,000$ random realisations of that data vector from a multi-variate lognormal distribution and $300,000$ random realisations from another lognormal distribution, whose skewness in each data point was increased by a factor of $5$. For each of these random realisations we analytically determined the maximum posterior model within the linearized likelihood formalism of \secref{linearized_likelihoods} and then computed the $\chi^2$ between that model and the random realisation. The blue histogram in the top panel of \figref{nG_likelihood} shows the distribution of these $\chi^2$ values for the Gaussian random realisations and the red histogram corresponds to the log-normal random realisations. The two histograms are almost identical. Hence, within the $f_{\mathrm{sky}}$-approximation employed above non-Gaussianity in the likelihood does not seem to affect our analysis. And even in the extreme scenario of enhancing the skewness of the data vector by a factor of $5$ (green histogram) the increase in the scatter of $\chi^2$ remains smaller than about $3\%$ of the average $\chi^2$ - which still wouldn't dominate over the other effects discussed in subsequent sections (\cf\ \figrefnospace{chiSq_offsets}).

The impact of non-Gaussianity on the likelihood becomes even more negligible when directly considering the distribution of maximum posterior parameters. We demonstrate this in the bottom panel of \figref{nG_likelihood} for the best-fit values of $\sigma_8$ but find similar results for our other key cosmological parameters $\Omega_m$ and $w_0$.
Therefore, we conclude that it is safe to assume a Gaussian distribution for the statistical uncertainties of the DES-Y3 2-point function measurements. 

\subsection{Modelling of connected 4-point function in covariance}
\label{sec:impact_of_4point}
    
The connected 4-point contribution to the covariance is the part that is most challenging to model analytically
\citep{Schneider2002,Hilbert2011,Sato2011,Takada:2013wfa}. This contribution is most relevant at small scales and turns out to be a small one for current large-scale structure analyses \citep{Krause2017,Barreira:2018jgd}. This is for two reasons: 1) such analyses typically cut away their smallest scales because of uncertainties in the modelling of their data vectors  and 2) at small scales the covariance matrix is often dominated by shape noise and shot noise which are believed to be well understood. 

We test whether the non-Gaussian covariance parts (by which we mean both the connected 4-point function and super-sample covariance) are a relevant contribution to our error budget by either
\begin{itemize}
    \item replacing the non-Gaussian contributions from the fiducial halo model with the lognormal covariance described in \secref{lognormal_covariance}
    \item or setting it to zero, \ie\ using a Gaussian covariance matrix only.
\end{itemize}
\Figref{chiSq_offsets} and \tabref{summary} show that neither of these changes has a significant impact on the distribution of $\chi^2$ and our parameter constraints. Assuming that our halo model and lognormal recipes do not underestimate the non-Gaussian covariance parts by orders of magnitude
(see \eg\ \citet{Sato2009, Hilbert2011} for justifications of this assumption)
this demonstrates that we are insensitive to the exact modelling of these contributions. At the same time, we want to stress that this finding holds for the specific scale cuts, redshift distributions and tracer densities of the DESY3 3x2pt analysis and cannot necessarily be generalized to other analysis setups.

\subsection{Exact angular bin averaging}
\label{sec:impact_of_bin_average}

\Eqnref{realspace_cov_in_terms_of_harmonic_space_cov} holds when measuring the 2-point correlation functions in infinitesimally small bins around the angular scales $\theta_1$ and $\theta_2$. This is unfeasible in practice and in fact also leads to divergent covariance matrices. This can for example be seen for the galaxy clustering correlation functions, where the constant term proportional to $1/n_g^2$ in the harmonic space covariance gives a contribution to the real space covariance of
\begin{eqnarray}
&&\frac{1}{4\pi^2 n_g^2f_{\mathrm{sky}}}\ \underset{N\rightarrow \infty} \lim \sum_{\ell = 1}^N \frac{(2\ell +1)}{2} P_{\ell}\left( \cos \theta \right)^2\nonumber \\
&\rightarrow& \frac{1}{4\pi^2 n_g^2f_{\mathrm{sky}}}\ \delta_D(\cos \theta - \cos \theta) \nonumber \\
&& (= \infty)\ .\nonumber
\end{eqnarray}
The reason for this divergence is simply the fact that the number of galaxy pairs found in an infinitesimal bin vanishes, leading to infinite shot-noise. This problem disappears when considering finite angular bins.

To analytically average over a finite angular bin $[\theta_{\min}, \theta_{\max}]$, we assume that the number of galaxy pairs with angular separation $\theta$ is proportional to $\sin \theta$ (corresponding to a uniform distribution of galaxies on the sky). We then replace the functions $F^{AB}_\ell( \theta)$ in Equations \ref{eq:xi_in_terms_of_Cell} and \ref{eq:xi_in_terms_of_Cell_general} by
\begin{equation}
    F^{AB}_\ell( \theta) \rightarrow \frac{1}{\cos \theta_{\min} - \cos \theta_{\max}}\ \int_{\theta_{\min}}^{\theta_{\max}} \dd \theta\ \sin \theta\ F^{AB}_\ell( \theta)\ .
\end{equation}
For the galaxy clustering correlation function $w(\theta)$ this leads to
\begin{eqnarray}
P_\ell(\cos \theta) \rightarrow \frac{\left[P_{\ell+1}(x) - P_{\ell-1}(x) \right]_{\cos\theta_{\max}}^{\cos\theta_{\min}}}{(2\ell+1)(\cos\theta_{\min } - \cos\theta_{\max })} .
\end{eqnarray}

The corresponding expressions for the galaxy-galaxy lensing correlation function $\gamma_t(\theta)$ and for the cosmic shear correlation functions $\xi_\pm$ are presented  (together with derivations of all the bin averaged expressions) in \appref{bin_averaging}.

We show below how the bin averaging solves the problem of diverging diagonal values of the covariance for $w(\theta)$. This can be seen from
\begin{eqnarray}
\label{eq:shotnoise}
&&\sum_{\ell} \frac{\left(\left[ P_{\ell+1}\left( x \right) - P_{\ell-1}\left( x \right)\right]_{\cos\theta_{\max}}^{\cos\theta_{\min}}\right)^2}{2(2\ell +1)f_{\mathrm{sky}}(n_g A_{\mathrm{bin}})^2}\nonumber \\
&=& \int_{\cos \theta_{\max}}^{\cos \theta_{\min}}\mathrm d x_1\int_{\cos \theta_{\max}}^{\cos \theta_{\min}}\mathrm d x_2 \sum_{\ell} \frac{2\ell +1}{2} \frac{P_{\ell}\left( x_1 \right)P_{\ell}\left( x_2 \right)}{f_{\mathrm{sky}}(n_g A_{\mathrm{bin}})^2} \nonumber \\
&=& \int_{\cos \theta_{\max}}^{\cos \theta_{\min}}\mathrm d x_1\int_{\cos \theta_{\max}}^{\cos \theta_{\min}}\mathrm d x_2 \ \frac{\delta_D(x_1 - x_2)}{f_{\mathrm{sky}}(n_g A_{\mathrm{bin}})^2} \nonumber \\
&=& \frac{\cos \theta_{\min} - \cos \theta_{\max}}{f_{\mathrm{sky}}(n_g A_{\mathrm{bin}})^2} \nonumber \\
&=& \frac{2}{A_{\mathrm{survey}} A_{\mathrm{bin}} n_g^2}\nonumber \\
&=& \frac{1}{N_{\mathrm{pair}}}\ ,
\end{eqnarray}
where $A_{\mathrm{survey}} = 4\pi f_{\mathrm{sky}}$ is the total survey area, $A_{\mathrm{bin}} = 2 \pi \left(\cos \theta_{\min} - \cos \theta_{\max}\right) $ is the bin area and $N_{\mathrm{pair}}$ the total number of galaxy pairs used to estimate $\hat w$. The above expression is the usual formula for the shot-noise part of the real
space covariance. 

The impact of the exact angular bin averaging for the noise and mixed terms in the Gaussian part of the covariance matrix is included for all 4 types of two point functions present in the DES-Y3 data vector and the DES-Y3 fiducial covariance.

\subsection{Flat vs. Curved sky}
\label{sec:impact_of_curved_sky}

For the Y1 analysis it was shown that the flat-sky approximation was valid for
the galaxy-galaxy shear and shear-shear 2-point correlation function \citep{Krause2017}. In Y3 the fiducial covariance computes the full sky correlations, see equations (\ref{eq:xi_hat_in_terms_of_Cell_hat_general}) and (\ref{eq:xi_hat_plus_minus}). We show in Fig. (\ref{fig:chiSq_offsets}) that the effect of including curved sky results has negligible impact on the $\chi^2$ distribution. \Tabref{summary} shows that this is also true for parameter constraints.

\subsection{RSD and Limber approximation and redshift space distortion effects}
\label{sec:impact_of_Limber_and_RSD}
The modelling of the angular power spectrum of two tracers involves a projection from the three dimensional power spectrum that requires integrals with integrands containing the product of two spherical Bessel functions, which are highly oscillatory. The inclusion of redshift space distortion (RSD) effects in a simple linear modelling \citep{1987MNRAS.227....1K} involves the computation of those integrals with derivatives of the Bessel functions. 
These integrals are notoriously difficult to perform numerically and it is usual to apply the so-called Limber approximation \citep{Limber1953,LoVerde:2008re}.
An efficient computation of these integrals without resorting to the Limber approximation was recently implemented in the case of the angular power spectrum for galaxy clustering in \citet{Fang20nonlimber}. We use their approach to study the impact of taking into account both non-Limber computations and RSD effects in the covariance matrix. Figure \ref{fig:chiSq_offsets} and \tabref{summary} show that not taking these effects into account leads to an increase in average $\chi^2$ of about $0.5\%$ and an underestimation of uncertainties in key cosmological parameters by $0.6\%$ to $1.4\%$.
\\

\subsection{Effect of the mask geometry}
\label{sec:masking_tests}

The analytical covariance models described in \secref{covariance_modelling} make use of the so called $f_{\mathrm{sky}}$ approximation, \ie\ they take the covariance of an all-sky survey and divide this by the sky fraction of DES-Y3 to approximate the covariance of our partial sky data. In \appref{masking} we show how to go beyond this approximation. First, we note there that the covariance of the 2-point function measurements between pairs of scalar random fields ($\delta_a,\delta_b$) and ($\delta_c,\delta_d$) within angular bins $[\theta_{-}^{ab},\theta_{+}^{ab}]$ and $[\theta_{-}^{cd},\theta_{+}^{cd}]$ is given by
\begin{align}
    &\ \mathrm{Cov}\left\lbrace\hat\xi^{ab}[\theta_{-}^{ab},\theta_{+}^{ab}],\hat\xi^{cd}[\theta_{-}^{cd},\theta_{+}^{cd}]\right\rbrace\ \frac{N_{\mathrm{pair}}^{ab}[\theta_{-}^{ab},\theta_{+}^{ab}] \ N_{\mathrm{pair}}^{cd}[\theta_{-}^{cd},\theta_{+}^{cd}]}{n_a n_b n_c n_d}\nonumber \\
    =&\ (2\pi)^2\sum_{\ell_1\ \ell_2} \left[P_{\ell_1+1}(x) - P_{\ell_1-1}(x) \right]_{\theta_{+}^{ab}}^{\theta_{-}^{ab}}\ \left[P_{\ell_2+1}(x) - P_{\ell_2-1}(x) \right]_{\theta_{+}^{cd}}^{\theta_{-}^{cd}}\cdot\nonumber\\
    &\ \mathrm{Cov}\left\lbrace\tilde C_{\ell_1}^{ab}, \tilde C_{\ell_2}^{cd} \right\rbrace\ .
\end{align}
Here $P_\ell$ are again the Legendre polynomials and the angular bin averaging was already evaluated. The factor $N_{\mathrm{pair}}^{ab}[\theta_{-}^{ab},\theta_{+}^{ab}]$ (\resp\ $N_{\mathrm{pair}}^{cd}[\theta_{-}^{cd},\theta_{+}^{cd}]$) is the average number of pairs of random points that uniformly sample the footprint with densities $n_a$, $n_b$ (\resp\ $n_c$, $n_d$) and whose separation falls into the angular bin $[\theta_{-}^{ab},\theta_{+}^{ab}]$ (\resp\ $[\theta_{-}^{cd},\theta_{+}^{cd}]$). Hence, these factors describe how the exact survey geometry suppresses the number of pairs of positions in the bins $[\theta_{-}^{ab},\theta_{+}^{ab}]$ and $[\theta_{-}^{cd},\theta_{+}^{cd}]$ with respect to\ the $f_{\mathrm{sky}}$ approximation. And finally, \smash{$\mathrm{Cov}\lbrace\tilde C_{\ell_1}^{ab}, \tilde C_{\ell_2}^{cd} \rbrace$} is the covariance of pseudo-$C_\ell$ estimates of the power spectra between the fields ($\delta_a,\delta_b$) and ($\delta_c,\delta_d$) (see \appref{masking} for more details). Note that the full survey footprint modifies the covariance with respect to\ the $f_{\mathrm{sky}}$ approximation used in \secref{model} both through the factors $N_{\mathrm{pair}}^{ab}[\theta_{-}^{ab},\theta_{+}^{ab}]/n_an_b$, $N_{\mathrm{pair}}^{cd}[\theta_{-}^{cd},\theta_{+}^{cd}]/n_c n_d$ and by changing \smash{$\mathrm{Cov}\lbrace\hat C_{\ell_1}^{ab}, \hat C_{\ell_2}^{cd} \rbrace$} compared to Equations \ref{eq:harmonic_covariances}-\ref{eq:harmonic_covariances_end}.

One can determine the factors $N_{\mathrm{pair}}^{ab}[\theta_{-}^{ab},\theta_{+}^{ab}]/n_an_b$ and $N_{\mathrm{pair}}^{cd}[\theta_{-}^{cd},\theta_{+}^{cd}]/n_c n_d$ either by counting pairs in a set of random points that trace the survey footprint homogeneously or they can be calculated analytically from the power spectrum of the survey mask (see our \appref{masking} as well as \citealt{Troxel2018b}). This will generally lead to an enhancement of statistical uncertainties (\ie\ of the covariance matrix) with respect to\ the $f_{\mathrm{sky}}$ approximation. To calculate \smash{$\mathrm{Cov}\lbrace\tilde C_{\ell_1}^{ab}, \tilde C_{\ell_2}^{cd} \rbrace$} one could \eg\ follow approximations made by \citet{Efstathiou2004}. We slightly modify their arguments in \appref{masking} to arrive at
\begin{align}
    \label{eq:Efstathiou_2.0_main_text}
    &\ \mathrm{Cov}\left\lbrace\tilde C_{\ell_1}^{ab}, \tilde C_{\ell_2}^{cd} \right\rbrace \approx \nonumber \\
    &\ \ \frac{1}{4}\left(\frac{C_{\ell_1}^{ac} C_{\ell_2}^{bd} + C_{\ell_2}^{ac} C_{\ell_1}^{bd} + C_{\ell_1}^{ac} C_{\ell_1}^{bd} + C_{\ell_2}^{ac} C_{\ell_2}^{bd}}{(2\ell_1+1)(2\ell_2+1)}\right.\ +\nonumber \\
    &\ \ +\left.\frac{C_{\ell_1}^{ad} C_{\ell_2}^{bc} + C_{\ell_2}^{ad} C_{\ell_1}^{bc} + C_{\ell_1}^{ad} C_{\ell_1}^{bc} + C_{\ell_2}^{ad} C_{\ell_2}^{bc}}{(2\ell_1+1)(2\ell_2+1)}\right)\ \mathcal{M}_{\ell_1 \ell_2}\ .
\end{align}
Here the mode coupling matrix $\mathcal{M}_{\ell_1 \ell_2}$ again depends on the power spectrum of the survey mask and is also detailed in \apprefnospace{masking}. Note that in order to keep our notation brief, we have assumed that the power spectra $C_{\ell_1}^{ac}$ \etc\ in the above equation include both the underlying cosmological power spectra and contributions to the power spectra from sampling noise, such as shape-noise and shot-noise.

\begin{figure*}
    \centering
    \includegraphics[width=0.48\textwidth]{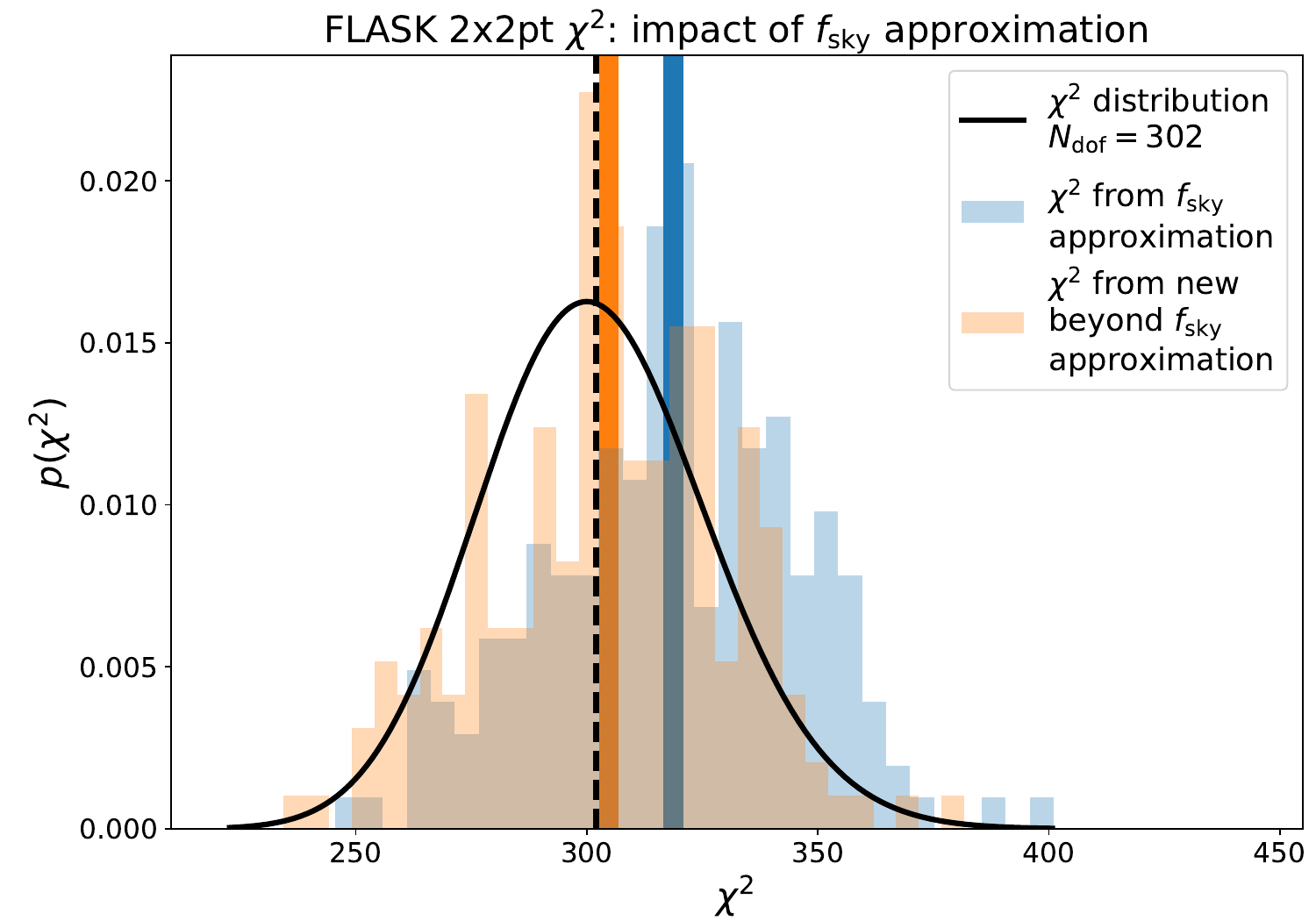}\vspace{0.04\textwidth}\includegraphics[width=0.48\textwidth]{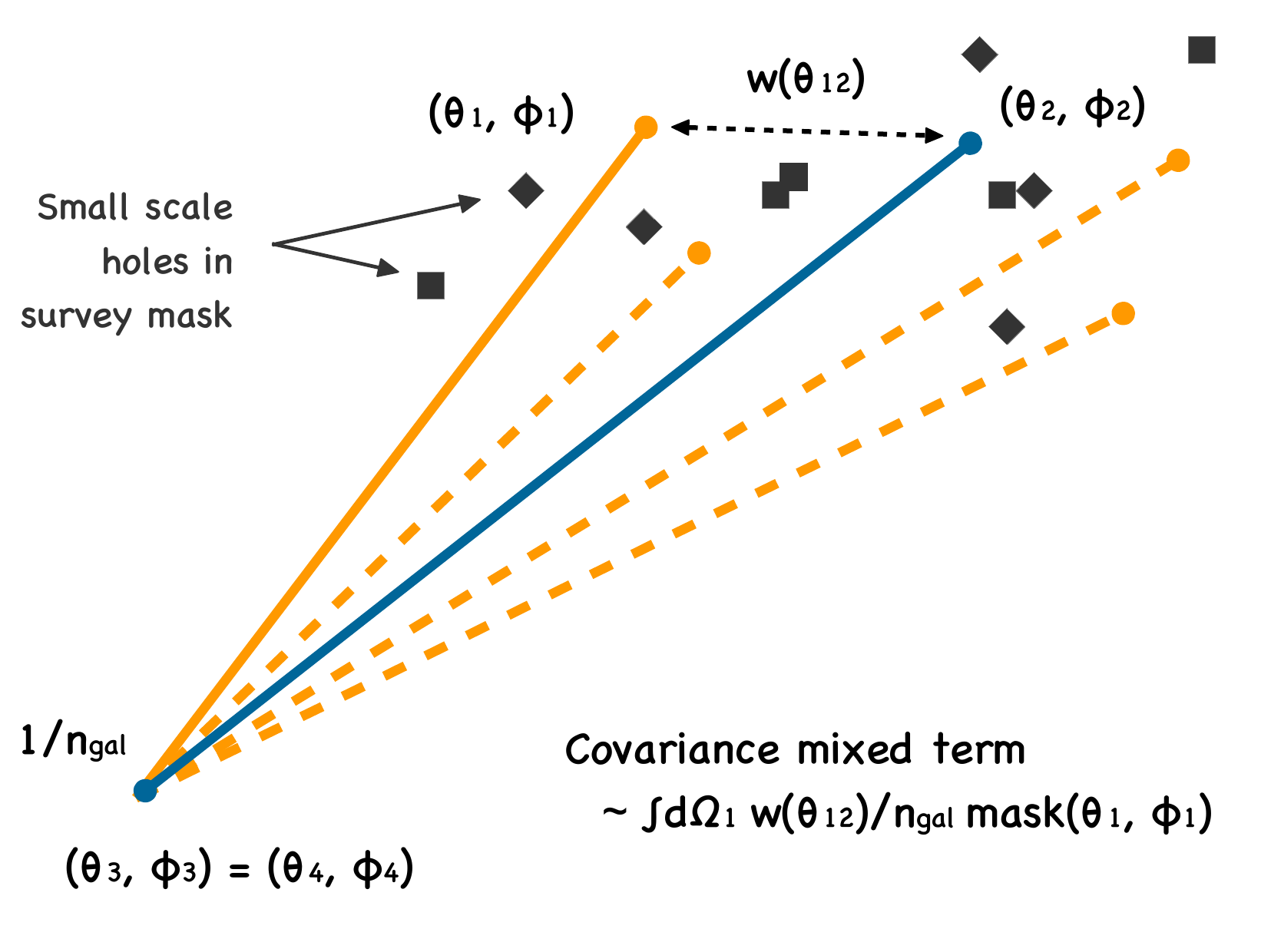}
\caption{The impact of masking on the DES-Y3 covariance. The blue histogram in the right panel shows the distribution of $\chi^2$ obtained from our FLASK data vectors when using the $f_{\mathrm{sky}}$ approximation. We restrict this figure to the 2x2pt function part of the data vector since it is this part that suffers the most from masking effects (\cf\ \figref{chiSq_offsets}). Ansatzes in the CMB literature (\eg\ \citealt{Efstathiou2004}) are not sufficient to correct for this, because the DES footprint has features down to very small scales. In the main text we have motivated a possible way to correct for these small scale masking features and the orange histogram in the left panel shows that this ansatz indeed significantly improves the $\chi^2$ obtained from our FLASK measurements. The sketch in the right panel visualises how small scale features in the mask lead to an overestimation in the covariance when using common ways to treat the impact of survey geometry on the 2-point function covariance (see main text for explanation).}
  \label{fig:masking_and_fsky}
\end{figure*}

In practice, \eqnref{Efstathiou_2.0_main_text} and the approximations proposed by \citet{Efstathiou2004} yield very similar results and they are both valid on scales $\ell_1, \ell_2$ which are much smaller than the typical scales of the mask $W$. Unfortunately, the DES-Y3 analysis mask has features and holes over a large range of scales. Hence, the angular scales of interest in the 3x2pt analysis are never strictly smaller than the scales of our mask. Hence, \eqnref{Efstathiou_2.0_main_text} is not sufficiently accurate in our case and in fact significantly overestimates our statistical uncertainties. In \figref{masking_and_fsky} we explain a simple scheme that can be used to correct for this. To motivate this procedure, consider how one would compute the Gaussian covariance model directly from the real space 2-point correlation functions, \ie\ without taking the detour to Fourier space that was used in \secrefnospace{model}. For the covariance of $\hat\xi^{ab}[\theta_{-}^{ab},\theta_{+}^{ab}]$ and $\hat\xi^{cd}[\theta_{-}^{cd},\theta_{+}^{cd}]$ this would amount to integration over all pairs of locations within our survey mask that fall into the angular bins $[\theta_{-}^{ab},\theta_{+}^{ab}]$ and $[\theta_{-}^{cd},\theta_{+}^{cd}]$. Schematically, this leads to terms of the form
\begin{align}
\label{eq:covariance_real_space_integrals}
    &\mathrm{Cov} \propto \nonumber \\
    &\underset{(ab)\in\mathrm{mask,bin}}{\int} \dd \Omega^{a}\dd \Omega^{b} \underset{(cd)\in\mathrm{mask,bin}}{\int} \dd \Omega^{c}\dd \Omega^{d} \xi^{ac}(\theta^{ac})\xi^{bd}(\theta^{bd}) + \dots\ .
\end{align}
Here $\Omega^a \dots \Omega^d$ are fours locations inside the survey mask such that the distance between $\Omega^a$ and $\Omega^b$ lies inside the angular bin $[\theta_{-}^{ab},\theta_{+}^{ab}]$ and the distance between $\Omega^c$ and $\Omega^d$ lies inside the angular bin $[\theta_{-}^{cd},\theta_{+}^{cd}]$. Now the approximation of \citet{Efstathiou2004} assumes that the 2-point functions $\xi^{ac}(\theta),\xi^{bd}(\theta)$ are negligible on scales $\theta$ comparable to the smalles features in the mask. Schematically this amounts to making the approximation
\begin{align}
\label{eq:Efstathiou_approx_1}
    &\ \int \dd \Omega^{a}\ \dots\ W(\Omega^{a})\xi^{ac}(\theta^{ac})\nonumber \\
    \approx&\ \bar{\xi}^{ac}\int \dd \Omega^{a}\ \dots\ W(\Omega^{a})\delta_{\mathrm{Dirac}}^2(\Omega^{a}-\Omega^{c})
\end{align}
where $\bar{\xi}^{ac}$ is a suitable average of the 2-point function over different scales. Our understanding of the approximation of \citet{Efstathiou2004} via \eqnref{Efstathiou_approx_1} is based on findings that we present in \apprefnospace{motivation_for_rescaling}. This approximation fails when the mask contains features (\eg\ holes) on scales where the 2-point function has not yet decayed. Assuming that such small scales holes cover a fraction of $f_{\mathrm{mask}}$ of a more coarse version of the footprint, then this can roughly be corrected for with a multiplicative factor, \ie\ by instead using the approximation
\begin{align}
\label{eq:Efstathiou_approx_2}
    &\ \int \dd \Omega^{a}\ \dots\ W(\Omega^{a})\xi^{ac}(\theta^{ac})\nonumber \\
    \approx&\ f_{\mathrm{mask}}\ \bar{\xi}^{ac}\int \dd \Omega^{a}\ \dots\ W(\Omega^{a})\delta_{\mathrm{Dirac}}^2(\Omega^{a}-\Omega^{c})\ .
\end{align}

The right panel of \figref{masking_and_fsky} visualizes this for the mixed terms in the covariance, where one of the correlation functions $\xi^{ac}$ or $\xi^{bd}$ is due to sampling noise such as shape-noise or shot-noise and is hence exactly proportional to a Dirac delta function. In that case, the integration is over pairs that share one end point. Now the approximation made \eg\ in \citet{Efstathiou2004} or by our \eqnref{Efstathiou_2.0_main_text} assumes that also the correlation function between the other two end points effectively acts as a delta function with respect to the smallest scale features in the survey mask (\cf\ \eqnrefnospace{Efstathiou_approx_1}). We find that this is not the case for the DES-Y3 mask and that it contains features on all scales relevant to our analysis. But as indicated in \eqnref{Efstathiou_approx_2}, one can approximately correct for this by multiplying the mixed terms in the covariance by the fraction $f_{\mathrm{mask}}$ of the coarser survey geometry that is covered by small scale holes in the mask. This can be considered a next-to-leading order correction to our \eqnrefnospace{Efstathiou_2.0_main_text}.

By applying \eqnref{Efstathiou_approx_2} twice one can see that the cosmic variance terms (terms where neither of the 2-point functions $\xi^{ac}$ or $\xi^{bd}$ are exactly proportional to delta functions) can be corrected by multiplication with $f_{\mathrm{mask}}^2$. To implement this correction in practice we draw circles within the DES-Y3 survey footprint with radii ranging from $5$arcmin to $20$arcmin and measure the masking fraction in these circles. We find that this fraction is $\approx 90\%$ across the considered scales. Multiplying the mixed terms in the covariance by that fraction and the cosmic variance terms by the square of that fraction (together with using \eqnrefnospace{Efstathiou_2.0_main_text}) we indeed find significant improvement of the maximum posterior $\chi^2$ obtained for the FLASK simulations - as is shown in the left panel of \figref{masking_and_fsky} (as well as in \figref{chiSq_offsets}).

\begin{figure}
    \centering
    \includegraphics[width=0.48\textwidth]{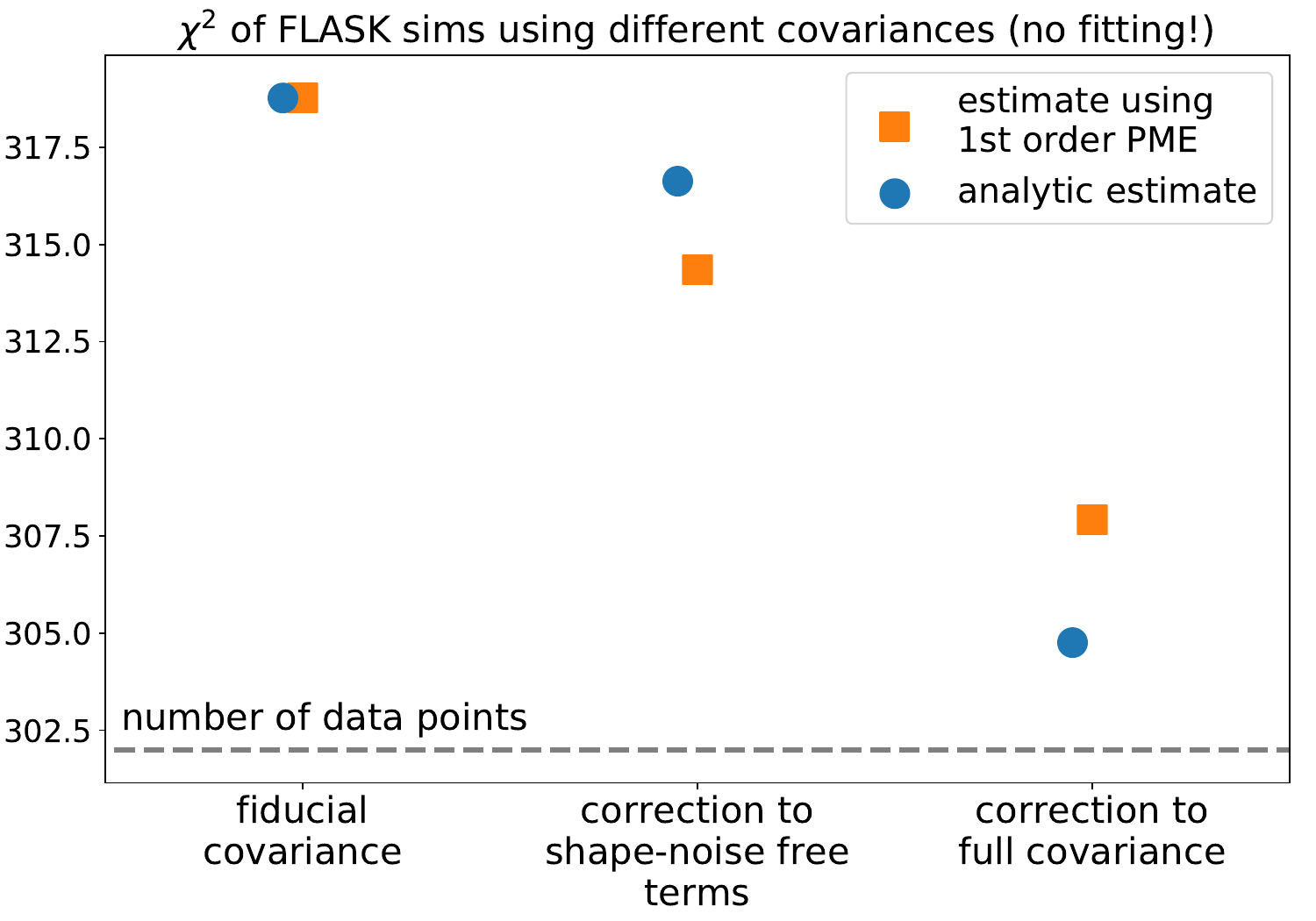}
\caption{The method of \emph{precision matrix expansion} \citep{Friedrich_Eifler} allows us to estimate the impact of individual covariance terms on $\chi^2$ even when only few simulated measurements are available. The orange squares show the average $\chi^2$ between our FLASK measurements and their mean (rescaled by a factor of $N_{\mathrm{FLASK}}/(N_{\mathrm{FLASK}}-1)$ to account for the correlation of individual measurements and mean) when using either no PME at all or when using PME estimates from shape-noise free sims or from the full sims. The blue dots show the corresponding $\chi^2$ values when using the heuristically motivated analytical treatment of masking and survey geometry presented in the main text. The grey dashed line represents the number of data points and should be the average $\chi^2$ if we had a perfect covariance model (note that for this comparison we have not performed any parameter fitting).}
  \label{fig:masking_chiSq_model_vs_PME}
\end{figure}

In \figref{masking_chiSq_model_vs_PME} we use our FLASK measurements together with the technique of \emph{precision matrix expansion} (PME, from inverse covariance $=$ precision matrix \citealt{Friedrich_Eifler}) and perform a consistency of the modelling ansatz described above by investigating the impact of masking on individual covariance terms. We find both with the PME methods and with our analytic ansatz that masking effects are most impactful in the covariance terms that depend on shape-noise of the weak lensing source galaxies (\ie\ in what we called mixed terms in \secrefnospace{model}). This also agrees with the findings of \citet{Joachimi2020} and it further motivates the modelling of masking effects that we have described here. Nevertheless, we do not elevate this modelling ansatz to our fiducial covariance model because its motivation remains rather heuristic. But we consider it a realistic estimate for the error made by the $f_{\mathrm{sky}}$ approximation and can hence use it to estimate the impact of that approximation on parameter constraints. In \figref{parameter_offsets_masking} we have already shown that this impact below the $1\%$ level, \ie\ we underestimate the scatter of maximum posterior parameters by less than $1\%$ when making the $f_{\mathrm{sky}}$ approximation in our fiducial covariance model.

Note that \citet{Kilbinger2, Sato2011, Shirasaki2019, Philcox_covariance_and_masking} have devised and promoted an alternative method to correct for masking, which amounts to direct Monte-Carlo integration of expressions like \eqnrefnospace{covariance_real_space_integrals}. Given the large area of DES-Y3 and and its numerous combinations of redshift bins, we did not find this to be feasible.

\subsection{Non-Poissonian shot noise}
\label{sec:nonPoisson_shot_noise}

In the Poissonian limit and in a complete region of the sky the power spectrum of shot-noise is scale-independent and given by
\begin{equation}
    N^{\rm{complete}}_{\ell} = \frac{1}{\bar{n}},
\label{eq:sn1}
\end{equation}
where $\bar{n}$ is the galaxy density per steradian. As noted in the previous subsection, in galaxy surveys not every region of the sky is fully accessible, \ie, the presence of bright stars, satellite trails, etc. lead to artificial changes in the measured galaxy density. These density changes can potentially modify the observed galaxy power spectrum and bias any cosmological analyses derived from them, and thus, they are avoided by removing certain regions of the sky where artifacts may be found. These regions are usually smaller than the resolution of the (pixelated) survey mask used to determine whether a region of the sky is within the footprint or not, since it is computationally expensive to increase the resolution. This this can be described through a fractional mask $W_{i}=1/f_{i}$, where $i$ is a given pixel of the mask $f_{i}$ is the fractional area of the pixel unaffected by the presence of artifacts. If we compute the galaxy overdensity as $\delta_{g, i} = N_{i}/(\bar{N} W_{i})-1$, with $\bar{N} = \sum_{i} N_{i}/ \sum_{i} W_{i}$, the mean number of sources per pixel we can estimate the Poissonian noise power spectrum as~\citep{Nicola:2019yiw}
\begin{equation}
    N_{\ell} = \Omega_{pix}\frac{\bar{W}}{\bar{N}},
    \label{eq:sn2}
\end{equation}
where $\bar{w}$ is the mean of the weights $w_{i}$ across the footprint, and $\Omega_{pix}$ is the area of the pixels from the mask in steradians. In the case where all the pixels in the footprint are fully complete we recover Equation~(\ref{eq:sn1}) since $\bar{n} = \bar{N}/\Omega_{pix}$, and $\bar{N}=\sum_{i}N_{i}/\sum_{i}1$. However, in the case that any of the pixels of the mask are not fully complete we obtain an increased shot-noise contribution by a factor $\bar{W} \geq 1$ (since $0 \leq f_{i} \leq 1$).

\begin{figure}
    \centering
    \includegraphics[width=0.49\textwidth]{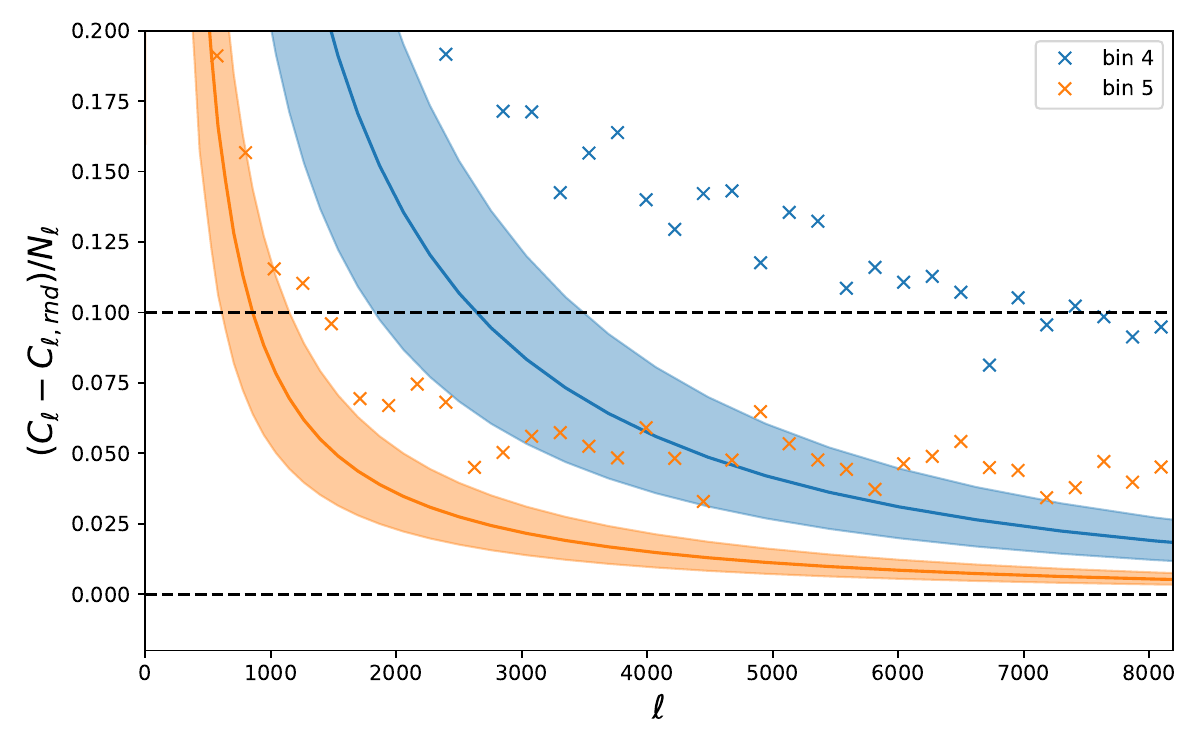}
    \caption{Measured ratio $r_{\ell} = \frac{C_{\ell}-C_{\ell,rnd}}{N_{\ell}}$ (crosses) compared to predicted contribution of the galaxy power spectra over the shot noise (solid line) for the fiducial parameters at~\citet{Elvin-Poole:2017xsf} allowing for a 20\% uncertainty in the galaxy bias (shaded regions) for 2 redshift bins (bin 4 in blue and bin 5 in orange). Horizontal dashed lines are just to guide the eye. If the shot-noise were to be completely Poissonian, the measured and predicted ratios would agree, however we find an excess between $2\%$ and $6\%$.}
    \label{fig:shot_noise_comparison}
\end{figure}

In previous studies the DES-Y1 lens galaxies were shown to prefer a super-Poissonian variance~\citep{Friedrich2018, Gruen2018} which might be a consequence of their complex selection criteria or due to the nature of their formation and evolution \citep[see \eg\ ][]{Baldauf2013, Dvornik2018}. This super-Poissonian variance leads to an enhance in shot-noise. In order to test for this effect, we proceeded to estimate the angular power spectrum, $C_{\ell} \approx C_{\ell, galaxies} + N_{\ell} + \delta C_{\ell}$, of DES-Y1 \textsc{redmagic} galaxies selected in \citet{Elvin-Poole:2017xsf} using \texttt{NaMaster}~\citep{Alonso2019}, where $N_{\ell}$ is the shot noise contribution from equation (\ref{eq:sn2}), and $\delta C_{\ell}$ is the excess power which can be due to a number of factors (variations in completeness not captured by the mask, super-Poissonian shot noise, observational systematics, etc.). We also compute the power spectrum, $C_{\ell, rnd}$ of a random field with the same number of objects as the galaxy sample considered, and the probability of populating a pixel $i$ is proportional to its weight, $W_{i}$. We find that $C_{\ell, rnd}$  is statistically consistent with $N_{\ell}$. We then compute the ratio: 
\begin{equation}
r_{\ell} = \frac{C_{\ell}-C_{\ell,rnd}}{N_{\ell}} \approx \frac{C_{\ell, galaxies}}{N_{\ell}} + \frac{\delta C_{\ell}}{N_{\ell}}.
\label{eq:ratio_power_noise}
\end{equation}
In Figure \ref{fig:shot_noise_comparison} we show $r_{\ell}$ compared to the theoretical expectation for $C_{\ell, galaxies}/N_{\ell} = C_{\ell, th}/N_{\ell}$, where $C_{\ell, th}$ is the theoretical power spectrum computed using the best-fit parameters found in~\citet{Elvin-Poole:2017xsf}. We also allow for a 20\% variation in the linear galaxy bias, which is much larger than the uncertainty found in~\citet{Elvin-Poole:2017xsf}. We find that there is an excess power at $\ell \geq 3000$ that cannot be explained by an excess galaxy clustering (i.e., a larger than measured linear bias or a non-linear bias component). We identify this excess (between $2\%$ and $6\%$) with $\frac{\delta C_{\ell}}{N_{\ell}}$ in equation (\ref{eq:ratio_power_noise}). 

This excess will translate into an extra shot-noise-like contribution to the covariance matrix \citep{Philcox2020}. The way we include this is by fitting a correction to the number density $\alpha_{n}$ such that the excess power is compatible with zero. In order to do so we minimize the following $\chi^{2}$:
\begin{equation}
\chi^{2}(\alpha_{n}) = \sum_{\ell}\left(\frac{C_{\ell}-C_{\ell, th}}{\alpha_{n}N_{\ell}}-\frac{C_{\ell,rnd}}{N_{\ell}}\right)^{2}\left(\frac{\Delta C_{\ell, th}}{\alpha_{n} N_{\ell}}\right)^{-2}
\end{equation}
where $\Delta C_{\ell, th}$ is varied in the range  $1.2^{2} C_{\ell, th} - 0.8^{2} C_{\ell, th}$ (so we are allowing for 20\% uncertainty in the bias for the fit).
The resulting values for $\alpha_{n}$ can be found in Table~\ref{tab:alpha_corr}. In \figref{chiSq_offsets}, \figref{parameter_offsets_masking} and \tabref{summary} one can see that depleting the lens galaxy densities in our fiducial covariance model by these factors has a negligible effect on both maximum posterior $\chi^2$ and parameters constraints.

\begin{table}
\centering
\begin{tabular}{|c|c|}
\hline
Bin number & $\alpha$ \\
\hline
1 & $1.042 \pm  0.002$ \\
2 & $1.069 \pm 0.003$ \\
3 & $1.072 \pm 0.003$ \\
4 & $1.057 \pm 0.003$ \\
5 & $1.021 \pm 0.001$ \\
\hline
\end{tabular}
\caption{Best-fit values of $\alpha_{n}$ to correct for the excess shot-noise with the DES-Y1 \textsc{redmagic} galaxies.}
\label{tab:alpha_corr}
\end{table}

\subsection{Cosmology dependence of the covariance model}
\label{sec:impact_of_cov_cosmology}

In order to evaluate our covariance model, we choose a particular set of cosmological parameters. We do not vary these parameters when sampling our parameter posterior and this may impact the width of our parameter constraints \citep{Hamimeche2008, Eifler2009, White2015, Kalus2016}. Our main reason for not sampling the covariance model along with the data model is that computing a covariance matrix is computationally too costly for this to be feasible.
Recently, \citet{Carron2013} have also indicated that it may indeed be incorrect to vary the covariance cosmology when running MCMC chains.

It is only after running the MCMC chains that we can recompute the covariance at our best-fit parameters and re-derive our parameter constraints - repeating this process until our constraints have converged \citep[\cf][for the application of this procedure in DES Y1 data]{DES2018}.
Therefore, the cosmology at which we compute our covariance is expected to be off from the best-fit cosmology.  In this subsection, we investigate how $\chi^2$, as well as cosmological parameter constraints, shift when computing the covariance at cosmologies that are randomly drawn from the DESY3-like posterior.

We test the robustness of our constraints against the choice of cosmological parameters at which we evaluate the covariance model by taking a set of 100 different cosmologies drawn randomly from the simulated DES Y3 3x2pt posterior and generating 100 lognormal covariance matrices. 
Using each of these covariances, we estimate posteriors for a given realization of simulated DES Y3 3x2pt data with noise drawn from a fiducial lognormal covariance. 

Since it is prohibitively expensive to perform simulated analyses running MCMC chains for each covariance matrix, we use the technique of importance sampling. That allows us to quickly evaluate how these different likelihood modeling choices impact the derived parameter constraints without repeatedly running expensive sampling algorithms. In our importance sampling pipeline, we take a fiducial analysis as a proposal distribution, re-evaluate the likelihoods using the alternative covariance matrix, and compute importance weights as:
\begin{equation}
\label{eq:Importance_weights}
w_i = \frac{\mathcal{L}(\pi_i | \hat\xi, \textbf{C}_\text{alt})}{\mathcal{L}(\pi_i | \hat\xi, \textbf{C}_\text{fid})},
\end{equation}
where $\textbf{C}_\text{fid}$ is the fiducial covariance in the analysis and $\textbf{C}_\text{alt}$ the alternative one. If the changes induced by the new covariance matrix in the posterior are not too large, the re-weighted samples represent the target distribution (i.e., the posterior for the alternative covariance matrix). So we have:
\begin{equation}
E_\textbf{p}[f(X_i)] = \sum_i p_i f(X_i) = \sum_i q_i \frac{p_i}{q_i} f(X_i) = E_\textbf{q}[w_i f(X_i)],
\end{equation}
for a function $f(X_i)$ of the posterior samples. Here, $p_i$ is the probability of $X_i$ under the target distribution $\textbf{p}$ and $q_i$ is the probability of of $X_i$ under the proposal distribution $\textbf{q}$ (see \eg\ \citet{Owen:2013} and \citet{MacKay:2002}).

To diagnose the performance of our importance sampling estimates, we use the Effective Sample Size (ESS):

\begin{equation}
\text{ESS} = \frac{\left<w_i\right>^2}{\left<w_i^2\right>} N_\text{samples}
\end{equation}
where $N_\text{samples}$ is the total number of posterior samples used in the estimation. The ESS as defined above quantifies the statistical power of the sample set after the re-weighting process (assuming uncorrelated samples). It is equal to the original sample size re-scaled by the ratio of the variances under each of the distributions \citep{Martino_2017}, such that the error of the mean of a quantity $x$ with standard deviation $\sigma_x$ under the target distribution can be estimated as $\sigma_x / \sqrt{\text{ESS}}$. Additionally, since our proposal distribution is itself a weighted sample set, we incorporate both the original and the importance weights in our ESS estimate.

\begin{figure}
        \includegraphics[width=0.50\textwidth]{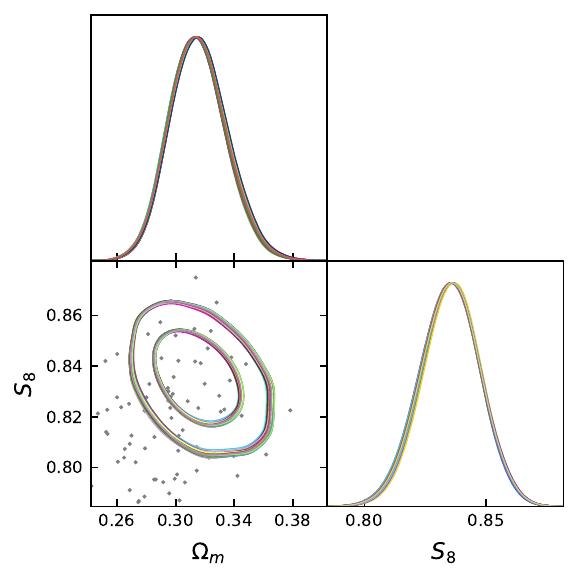}
        \caption{$(S_8, \Omega_m)$ constraints for a given noisy realization of the DES Y3 3x2pt data vector analyzed using 100 log-normal covariance matrices, each computed from a different cosmology drawn from a simulated DES Y3 3x2pt posterior. The 100 contours are superimposed in the plot, showing very small change in constraints. The points indicate the cosmologies at which the covariances were evaluated.}
        \label{fig:cosmofcov_1}
    \end{figure}

Using the fiducial lognormal covariance matrix we run the nested sampling algorithm {\tt MultiNest} \citep{Feroz_2008,Feroz_2009,Feroz_2019}, 
and perform the importance sampling procedure to estimate parameters 
using each of the 100 covariance matrices randomly sampled in parameter space. 
The ($S_8 , \Omega_m$) contours can be seen in Fig.~(\ref{fig:cosmofcov_1}). The effective sample sizes for the importance sampled estimates range from 16446 to 18329 (implying a standard error of the mean within $0.78\%$ of the standard deviation for all cases), and the contours show good statistics. As the impact of covariance cosmology is barely noticeable for this range of tested parameters, we repeat the analysis for a few more extreme (and unlikely) cosmologies in appendix~\ref{app:extreme_cosmo}.
    
 These results all confirm that we can safely neglect the impact of the choice of covariance cosmology in DES Y3 3x2pt analysis. One caveat of this conclusion is that we have indeed only varied cosmological parameters (including galaxy bias parameters) but not nuisance parameters (multiplicative shear bias, photometric redshift uncertainties) or parameters that describe intrinsic alignment. However, the DES-Y3 shear and photo-z calibration yield tight Gaussian priors on the corresponding nuisance parameters. And intrinsic alignment is relevant only on small angular scales where the covariance matrix is dominated by sampling noise contributions. Hence, we do not expect the results of this section to change significantly had all parameters been varied.



\subsection{Random point shot-noise}

We also consider the effect of additional shot-noise in the measurements of galaxy clustering resulting from the use of finite numbers of random points. The Landy-Szalay estimator \citep{Landy1993} is estimating the galaxy clustering correlation function inside an angular bin $[\theta_{1}, \theta_{2}]$ as
\begin{equation}
\label{eq:Landy_Szalay}
    \hat w[\theta_{1}, \theta_{2}] = \frac{DD[\theta_{1}, \theta_{2}] - 2DR[\theta_{1}, \theta_{2}] + RR[\theta_{1}, \theta_{2}]}{RR[\theta_{1}, \theta_{2}]}\ ,
\end{equation}
where $DD[\theta_{1}, \theta_{2}]$ is the number of galaxy pairs found within the angular bin, $RR[\theta_{1}, \theta_{2}]$ is the (normalised) number of pairs of random points that uniformly samples the survey footprint and $DR[\theta_{1}, \theta_{2}]$ the (normalised) number of galaxy-random-point pairs within the angular bin. If the number density of random points $n_r$ is much larger than the number density of the galaxies $n_g$ (as is recommended for reduce sampling noise) then both $RR$ and $DR$ must be rescaled by factors of $(n_g/n_r)^2$ and $(n_g/n_r)$ respectively.

We stress that the Landy-Szalay estimator was devised at a time of very limited computational resources, where it was prohibitively costly to measure galaxy pair in a large number of random points. Hence, it was vital to minimize random point shot-noise. Nowadays, footprint geometries of photometric surveys are typically characterised by high resolution healpix maps. The most straightforward way to calculate galaxy clustering correlation function is to simply assign a value of galaxy density contrast to each of these pixels and then measure the scalar auto-correlation function of the unmasked pixels. This way, one is avoiding random point shot-noise completely.

Nevertheless, it is still very common to measure $w(\theta)$ by means of \eqnrefnospace{Landy_Szalay}. So we also tested what impact a finite number of random points would have on our analysis. To do so we extended expressions of \citet[see their appendix A]{Cabre2009} to the case where the same random points are used to estimate $w(\theta)$ in each of our redshift bins and also to subtract shear around random points from our galaxy-galaxy lensing correlation functions. Note that this causes a noise contribution to the 2-point function measurements that is correlated among different redshift bins. We assumed a random point density of $1.36/\mathrm{arcmin}^2$, which is more that $20$ times larger than the density of our most dense lens galaxy sample. From \tabref{summary} it can be seen that not accounting for the random point shot-noise in the covariance leads to an increase in average $\chi^2$ of $\lesssim 1\%$ and to an underestimation of parameter uncertainties by $\approx 0.5\%$. Hence, this effect can be ignored for our analysis.

\subsection{Effective densities and effective shape-noise}
\label{sec:effective_quantities}

We are closing this section by spelling out an aspect of covariance modelling that may seem straightforward but which has repeatedly
came up in covariace discussions.

If the tracer galaxies used to estimate 2-point correlation functions are weighted according to some weighting scheme, then this may change the effective number densities and the effective shape noise that should be used when evaluating the covariance expressions in \secref{covariance_modelling}. In the following we will derive how this can be done for each of the 2-point functions in the DESY3 3x2pt data vector.

\subsubsection{Galaxy clustering}

We start with the galaxy clustering correlation function $w(\theta)$. We assume a weighting scheme that is aimed at correcting for non-cosmological density fluctuations resulting from spatially varying observing conditions \citep[as \eg\ in][]{Elvin-Poole:2017xsf}. This means that the weights assigned to each galaxy in fact sample a weight map that spans the entire footprint.

Instead of measuring $w(\theta)$ from the weighted galaxies by means of, say, the Landy-Szalay estimator \citep{Landy1993} it will be more convenient to think of the galaxy density contrast as a pixelized field on the sky. Further more, we will assume that the weight map has been normalised such that $\langle w \rangle = 1$ (which can always be done without changing the outcome of the weighted measurement). Consider pixel $i$ with galaxy count $N_{g,i}$ and weight $w_i$. If $n_g$ is the average galaxy density of the unweighted sample, then by taking expectation values with respect to many Poissonian shot-noise realisations (and hence ignoring fluctuations of the underlying matter density field) we get
\begin{align}
    \langle N_{g,i} \rangle =&\ \frac{A_{\mathrm{pix}} n_g}{w_i}\\
    \mathrm{Var}(N_{g,i}) =&\ \frac{A_{\mathrm{pix}} n_g}{w_i}\\
    \mathrm{Var}(w_i N_{g,i}) =&\ w_i A_{\mathrm{pix}} n_g\\
    \mathrm{Var}\left(\frac{w_i N_{g,i}}{A_{\mathrm{pix}} n_g} - 1\right) \equiv&\ \mathrm{Var}(\delta_{g,i})\nonumber\\
    =&\ \frac{w_i}{A_{\mathrm{pix}} n_g}\ ,
\end{align}
where $A_{\mathrm{pix}}$ is the area of each pixel and the second to last line serves as definition of $\delta_{g,i}$ and needs the fact that we demanded $\langle w \rangle = 1$. Note that these equation are only valid for an ensemble of observations that shares the same weight maps and differs only in their shot-noise realisations.

From the set of all pixels we can now estimate $w(\theta)$ within a finite angular bin $[\theta_1, \theta_2]$ as
\begin{equation}
    \hat w[\theta_1, \theta_2] = \frac{\sum_{\mathrm{pxls}\ i>j} \Delta_{[\theta_1, \theta_2]}^{ij}\ \delta_{g,i}\ \delta_{g,j}}{\sum_{\mathrm{pxls}\ i>j} \Delta_{[\theta_1, \theta_2]}^{ij}}\ ,
\end{equation}
where the symbol $\Delta_{[\theta_1, \theta_2]}^{ij}$ in the double sum over all pixels is $1$ when the distance of the pixel pair $i,j$ is within $[\theta_1, \theta_2]$ and $0$ otherwise. Note that we assume an enumeration of the pixels and that we demand $i>j$ in the sum in order to not count any pair of pixels twice.

If shot-noise is the only source of noise, then it is straight forward to calculate the variance of this measurement as
\begin{align}
\label{eq:var_of_wtheta_with_weighting}
    \mathrm{Var}(\hat w[\theta_1, \theta_2]) =&\ \frac{\sum_{\mathrm{pxls}\ i>j} \Delta_{[\theta_1, \theta_2]}^{ij}\ \langle\delta_{g,i}^2\rangle\ \langle\delta_{g,j}^2\rangle}{\left[\sum_{\mathrm{pxls}\ i>j} \Delta_{[\theta_1, \theta_2]}^{ij}\right]^2}\nonumber \\
    =&\ \frac{1}{(A_{\mathrm{pix}}n_g)^2}\frac{\sum_{\mathrm{pxls}\ i>j} \Delta_{[\theta_1, \theta_2]}^{ij}\ w_i\ w_j}{\left[\sum_{\mathrm{pxls}\ i>j} \Delta_{[\theta_1, \theta_2]}^{ij}\right]^2}\nonumber \\
    =&\ \frac{1}{N_{\mathrm{pair},g}[\theta_1, \theta_2]}\frac{\sum_{\mathrm{pxls}\ i>j} \Delta_{[\theta_1, \theta_2]}^{ij}\ w_i\ w_j}{\sum_{\mathrm{pxls}\ i>j} \Delta_{[\theta_1, \theta_2]}^{ij}}\ .
\end{align}
In the last line, $N_{\mathrm{pair},g}[\theta_1, \theta_2]$ is the number of unweighted galaxy pairs within the angular bin $[\theta_1, \theta_2]$\ . Note that in the presence of clustering, this should be calculated from a set of random points instead of from the actual galaxy catalog.

The first factor on the right side of \eqnref{var_of_wtheta_with_weighting} is what the shot-noise variance of $\hat w$ should be in the absence of a weighting scheme. The second term is a 2-point function of the weight map itself. If the weight map has a white-noise power spectrum, then this factor will be close to $1$ in any angular bin that doesn't include angular distances of $0$. This means that at large enough scales the last line of \eqnref{var_of_wtheta_with_weighting} looks like the covariance for plain Poissonian shot-noise without any notion of an effective number density.  This maybe surprising, but it stems from the fact that the weighting scheme we assumed does not simply multiply the galaxy density contrast field. Instead it reverses an already existing depletion of galaxy density from non-cosmological density fluctuations.

In conclusion, the effective number density that should be used to compute the covariance of $\hat w[\theta_1, \theta_2]$ is
\begin{equation}
    n_{g,\mathrm{eff}}[\theta_1, \theta_2] = n_g \sqrt{\frac{\sum_{\mathrm{pxls}\ i>j} \Delta_{[\theta_1, \theta_2]}^{ij}}{\sum_{\mathrm{pxls}\ i>j} \Delta_{[\theta_1, \theta_2]}^{ij}\ w_i\ w_j}} \ .
\end{equation}

\subsubsection{Galaxy-galaxy lensing}

We move on to consider the galaxy-galaxy lensing correlation function $\gamma_t[\theta_1, \theta_2]$. We assume that the lens galaxy sample comes with weights derived from a weight map $w^l$ as in the previous subsection while each source galaxy $j$ has a weight $w_j^s$ which does not come from an entire weight map but is instead the result of the individual quality of shape-measurement for this galaxy. A measurement of $\gamma_t$ can be constructed as
\begin{equation}
    \hat \gamma_t[\theta_1, \theta_2] = \frac{\sum_{\mathrm{pxl}\ i,\ \mathrm{source}\ j} \Delta_{[\theta_1, \theta_2]}^{ij}\ \delta_{l,i}\ \epsilon_{t,j\rightarrow i}\ w_j^s}{\sum_{\mathrm{pxl}\ i,\ \mathrm{source}\ j} \Delta_{[\theta_1, \theta_2]}^{ij}\ w_j^s}\ .
\end{equation}
Here, $\delta_{l,i}$ is the galaxy density contrast of the lenses defined in analogy to the previous subsection, $\epsilon_{t,j\rightarrow i}$ is the tangential component of the shear of source $j$ with respect to\ lens galaxy $i$ and $w_j$ is the weight of source galaxy $j$. Note that due to our definition of the lens galaxy density contrast this estimator already includes subtraction of shear around random points.

If shot-noise and shape-noise are the only sources of noise, then it can be readily shown that
\begin{align}
\label{eq:variance_of_gamma_t_shape-noise_only}
    &\ \mathrm{Var}(\hat \gamma_t[\theta_1, \theta_2])\nonumber \\
    =&\ \frac{\sum_{\mathrm{pxl}\ i,\ \mathrm{source}\ j} \Delta_{[\theta_1, \theta_2]}^{ij}\ \langle\delta_{l,i}^2\rangle\ \langle (\epsilon_{t,j\rightarrow i}\ w_j^s)^2\rangle}{\left[\sum_{\mathrm{pxl}\ i,\ \mathrm{source}\ j} \Delta_{[\theta_1, \theta_2]}^{ij}\ w_j^s\right]^2}\nonumber \\
    \approx &\ \frac{1}{N_{\mathrm{pair},ls}[\theta_1, \theta_2]}\frac{\sum_{\mathrm{pxl}\ i,\ \mathrm{source}\ j} \Delta_{[\theta_1, \theta_2]}^{ij}\ \langle (\epsilon_{t,j\rightarrow i}\ w_j^s)^2\rangle}{\sum_{\mathrm{pxl}\ i,\ \mathrm{sources}\ j} \Delta_{[\theta_1, \theta_2]}^{ij}\ w_j^s}\nonumber \\
    \approx &\ \frac{1}{2 N_{\mathrm{pair},ls}[\theta_1, \theta_2]}\frac{\sum_{\mathrm{source}\ j} \langle |\boldsymbol{\epsilon}_j|^2\ (w_j^s)^2\rangle}{N_s}\\
    &\ (\mathrm{only\ with}\ \langle w_j^s \rangle = 1 = \langle w_j^l \rangle\ !). \nonumber
\end{align}
Here $N_{\mathrm{pair},ls}[\theta_1, \theta_2]$ is the number of unweighted lens-source pairs in the angular bin $[\theta_1, \theta_2]$, $N_s$ is the total number of source galaxies and $\boldsymbol{\epsilon}_j = \epsilon_{1,j} + i \epsilon_{2,j}$ is the complex intrinsic ellipticity of source galaxy $j$.

Note that the final expression in \eqnref{variance_of_gamma_t_shape-noise_only} explicitly allows for the possibility that the source weights $w_j^s$ are correlated with the intrinsic ellipticities $\boldsymbol{\epsilon}_j$ of the source galaxies. One can interpret \eqnref{variance_of_gamma_t_shape-noise_only} as
\begin{equation}
\label{eq:gammaT_noise_variance_with_sigmaEff}
    \mathrm{Var}(\hat \gamma_t[\theta_1, \theta_2]) = \frac{\sigma_{\epsilon,\mathrm{eff}}^2}{N_{\mathrm{pair},ls}[\theta_1, \theta_2]}
\end{equation}
with the effective dispersion of intrinsic ellipticity \textbf{per shear component} given by
\begin{equation}
\label{eq:effective_shape-noise}
    \sigma_{\epsilon,\mathrm{eff}}^2 = \frac{1}{2} \frac{\sum_{\mathrm{source}\ j} |\boldsymbol{\epsilon}_j|^2\ (w_j^s)^2}{N_s}\ .
\end{equation}

One subtlety here is that the above derivation requires $\langle w_j^s \rangle = 1$. The above expressions mus be modified is this is not the case or when taking into account responses $R_j$ of a shape catalog generated with metacalibation \citep{Sheldon2017}. We detail what to do in the latter case in \apprefnospace{metacal}.

\subsubsection{cosmic shear}

For cosmic shear we follow \citet{Schneider2002} and construct a measurement of $\xi_+$ from a set of sources as
\begin{equation}
    \hat \xi_+[\theta_1, \theta_2] = \frac{\sum_{i>j} \Delta_{[\theta_1, \theta_2]}^{ij}\ w_i w_j\ (\epsilon_{1,i}\epsilon_{1,j}+\epsilon_{2,i}\epsilon_{2,j})}{\sum_{i>j} \Delta_{[\theta_1, \theta_2]}^{ij}\ w_i w_j}\ .
\end{equation}
If shape-noise is the only source of noise and if the intrinsic ellipticities of galaxies are not correlated with their weights, then the variance of $\hat\xi_+$ is given by
\begin{align}
\label{eq:cosmic_shear_noise_variance_with_sigmaEff}
    &\ \mathrm{Var}(\hat \xi_+[\theta_1, \theta_2])\nonumber \\
    =&\ \frac{\sum_{i>j} \Delta_{[\theta_1, \theta_2]}^{ij}\ \langle \epsilon_{1,i}^2 w_i^2\rangle \langle \epsilon_{1,j}^2 w_j^2\rangle+\langle \epsilon_{2,i}^2 w_i^2\rangle \langle \epsilon_{2,j}^2 w_j^2\rangle}{\left[\sum_{i>j} \Delta_{[\theta_1, \theta_2]}^{ij}\ w_i w_j\right]^2}\nonumber \\
    \approx&\ \frac{2 \sigma_{\epsilon,\mathrm{eff}}^4}{N_{\mathrm{pair}}[\theta_1, \theta_2]}\ .
\end{align}
Here $N_{\mathrm{pair}}[\theta_1, \theta_2]$ is the number of source galaxy pairs in the bin $[\theta_1, \theta_2]$ and we have replaced each expectation value $\langle \epsilon_{1/2,i}^2 w_i^2\rangle$ by $\sigma_{\epsilon,\mathrm{eff}}^2$ from \eqnrefnospace{effective_shape-noise}. Note that we again assumed $\langle w_j\rangle =1$ and that this may require re-scaling of both weights and $\sigma_\epsilon$ when using shape measurements from metacalibration.

\subsubsection{Testing validity of effective shape noise}

\begin{figure*}
  \includegraphics[width=0.99\textwidth]{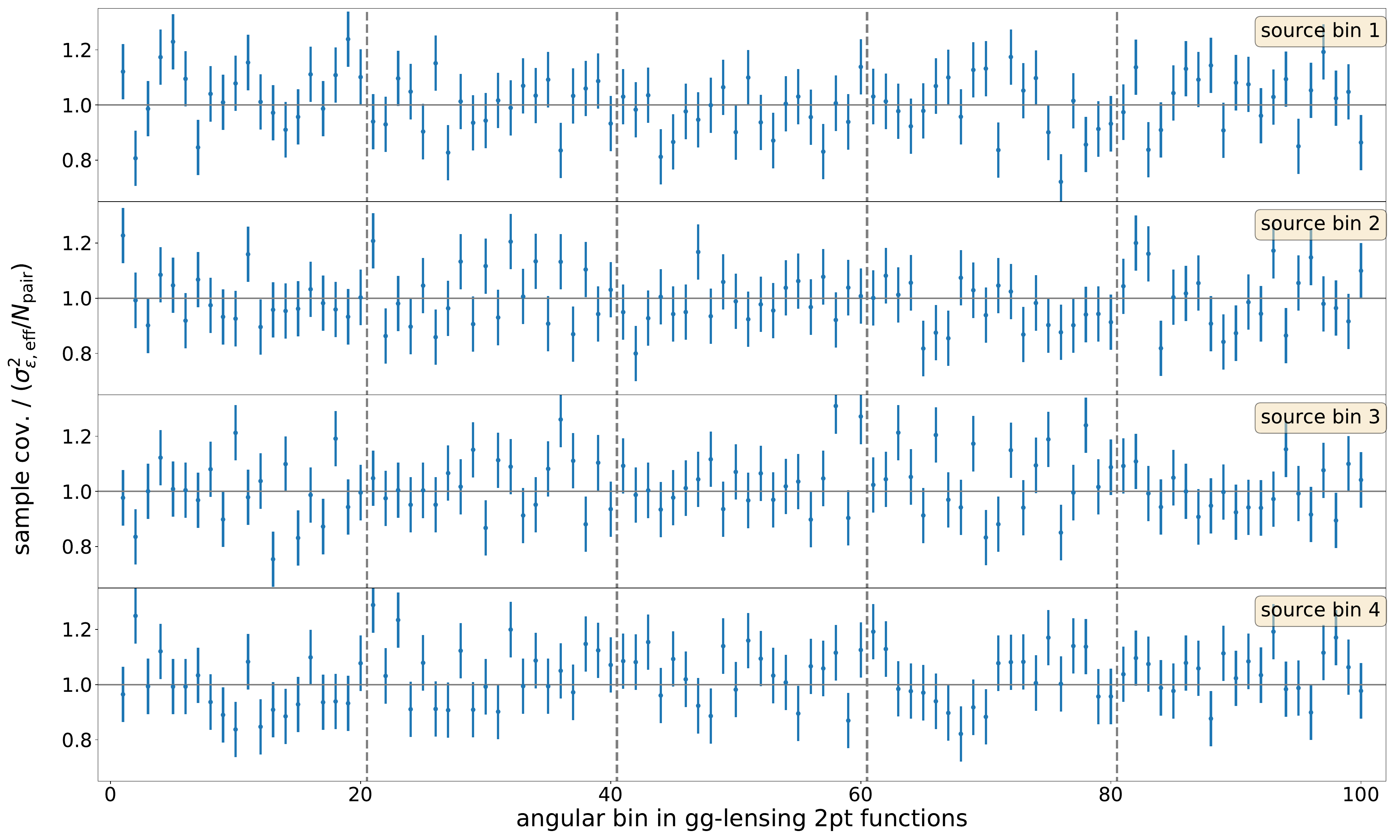}
   \caption{Ratio between the sample variance of $\hat\gamma_t$ measured in 200 randomly selected sub-samples of the DESY3 lens and sources catalogs and \eqnref{gammaT_noise_variance_with_sigmaEff} for the shape-noise contribution to the covariance (again using \eqnref{effective_shape-noise} to calculate the effective shape-noise dispersion $\sigma_{\epsilon,\mathrm{eff}}$). Each row displays the variances measured for a different source redshift bin and vertical dashed lines separate points belonging to different lens redshift bins (1-5 from left to right). Assuming that the covariances estimates have a Wishart distribution we calculate the covariance matrix of these ratios \citep[\cf ][]{Taylor2013} and find that they are consistent with 1 (both for the cosmic shear and galaxy-galaxy lensing variances).}
  \label{fig:gg_lensing_jackknife}
\end{figure*}

To test the validity of our expression for effective shape-noise in \eqnref{effective_shape-noise} we run a sub-sample covariance estimator on our data \citep[see \eg][]{Friedrich2016}. In particular, we divide all of our source and lens galaxy samples into 200 randomly chosen sub-samples and measure the galaxy-galaxy lensing correlation function of each source-lens bin combination. As a result we obtain 200 measurements of $\hat\gamma_t$ in each source-lens bin combination. Since we employ completely random sub-sampling, \ie\ without any regard for \eg\ a division of our footprint into sub-regions, the sample covariance of these 200 measurements will almost exclusively be dominated by shape-noise and shot-noise. This is even more so, because the lens and source densities of the sub-samples are very low.

In \figref{gg_lensing_jackknife} we show the ratio of the variances of the 200 galaxy-galaxy lensing measurements $\hat\gamma_t$ in the different lens-source bin combinations to \eqnref{gammaT_noise_variance_with_sigmaEff}. Assuming that the sub-sample covariances follow a Wishart distribution we find that these ratios are perfectly consistent with $1$. This indicates that \eqnref{effective_shape-noise} indeed yields an accurate effective shape-noise dispersion, and that one should indeed use the plain density of lens galaxies (as opposed to any notion of effective density) when evaluating covariance expressions.

\section{A simple $\chi^ 2$ test}
\label{sec:simplechi2}

\begin{figure}
  \includegraphics[width=0.5\textwidth]{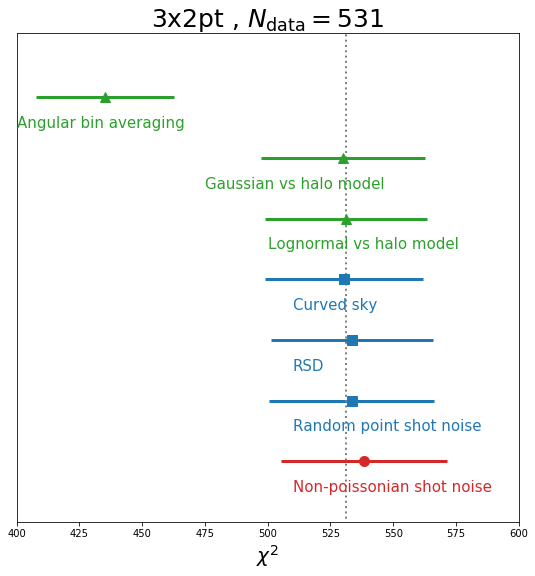}
   \caption{$\chi^ 2$ tests taking into account with different effects. Colors follow the scheme of Figure \ref{fig:chiSq_offsets}.}
  \label{fig:chi2simple}
\end{figure}

In this short section we present a simple $\chi^ 2$ test that does not rely on the linearized framework. However, it has the disadvantage of not addressing the impact on the estimation of parameters.
Here we generate a large number of ``contaminated"  data vectors (we use $1,000$) by a Gaussian sampling of a given covariance matrix that includes different effects and to compute a $\chi^ 2$ distribution from these data vectors using a fiducial covariance matrix. The resulting shifts in the mean value of $\chi^2$ and their standard deviations give another benchmark for the importance of the different effects considered here.  
We show the results of this test in \figref{chi2simple}. Note that the relative increases in $\chi^2$ follow closely what we obtained within the linearized likelihood framework in \figref{chiSq_offsets}. This indicates that the dominant way in which covariance errors cause $\chi^2$ offsets is not through the altered scatter of maximum posterior parameter locations but simply through using an erroneous inverse covariance when computing $\chi^2$. That also justifies our usage of the linearized likelihood framework since any impact of non-linear parameter dependencies on parameter fitting can be expected to be even less relevant then linear fitting in the first place. 

\section{Discussions and Conclusions}
\label{sec:summary}

In this paper we have presented the fiducial covariance model of the DES-Y3 joint analysis of cosmic shear, galaxy-galaxy lensing and galaxy clustering correlation functions (the 3x2pt analysis). We then investigated how the assumptions and approximations of that model (including the assumption of Gaussian statistical uncertainties) impact the distribution of maximum posterior $\chi^2$ and maximum posterior estimates of cosmological parameters.

The fiducial covariance matrix of the DES-Y3 3x2pt analysis uses the formalism of \citet{Krause:2016jvl} to model super-sample covariance as well as the trispectrum contribution to the covariance. The model for the Gaussian covariance part (\ie\ the contributions from the disconnected 4-point function) correctly takes into account sky curvature and includes analytical averaging over the finite angular bins in which the 2-point functions are measured. Furthermore, the galaxy clustering power spectra that enter our calculation of the Gaussian covariance part are computed using the non-Limber formalism of \citet{Fang20nonlimber} and also include modelling of redshift space distortions. The finite survey area of DES-Y3 is incorporated in the covariance model via the $f_{\mathrm{sky}}$ approximation
(except in the pure shape-noise and shot-noise terms where we follow \citealt{Troxel2018b}).

In order to perform our validation tests for the DES-Y3 covariance matrix we developed a plethora of new modelling ansatzes and testing strategies which are applicable in general. 
These new techniques are:
\begin{itemize}
    \item We have motivated and devised a way of drawing realisations of the 3x2pt data vector from a non-Gaussian distribution in order to test the accuracy of our Gaussian likelihood assumption.
    \item We have derived analytic expressions for angular bin averaging of all four types of 2-point correlation functions included in the 3x2pt vector ($\xi_+, \xi_-, \gamma_t, w$). These expressions correctly account for sky curvature. To the best of our knowledge, an analytic treatment of bin averaging for cosmic shear 2-point function has not been presented before (though we have shared our results with \citealt{Fang:2020vhc} who have used them for the fiducial covariance computations).
    \item We have extended the lognormal analytical model for the covariance of cosmic shear 2-point function of \citet{Hilbert2011} to the other 2-point functions present in the 3x2pt data vector.
    \item Within a linearized likelihood formalism we have analytically derived how covariance model inaccuracies influence the distribution of maximum-posterior $\chi^2$ and of maximum-posterior parameter estimates. The results we presented also allow for the possibility of including the Gaussian priors on certain model parameters and can be used to analytically estimate the impact of covariance errors on cosmological likelihood analyses.
    \item By fitting an effective number density to the high-$\ell$ plateau of galaxy clustering $C_\ell$ measurements we have estimated how much the assumption of Poissonian shot-noise influences our likelihood analysis. This is similar in spirit to the {\tt RASCALC} technique presented by \citet{Philcox2020}, and we agree with those authors that non-Poissonian shot noise can be viewed as an effective description of how short-scale non-linearities in galaxy clustering influence the covariance.
    \item We calculated covariance matrices for $100$ different sets of cosmological and nuisance parameters randomly drawn from a simulated likelihood chain. This allowed us to investigate whether calculating our covariance model at a reasonable, but wrong point in parameter space significantly impacts our analysis. This was done both within our linearised likelihood framework and by using importance sampling to quickly evaluate the $100$ non-linear likelihoods.
    \item We have derived how the 2-point correlation function of weight maps influence the covariance of galaxy clustering 2-point function measurements. In that context we have also found that traditional ways of deriving an effective number density for a given set of galaxy weights are erroneous when those weights are aimed at undoing a suspected depletion of galaxy density (\eg\ due to variations in observing conditions).
    \item We have derived an expression for the effective dispersion of intrinsic source shapes for the situation when source galaxy weights are correlated with galaxy ellipticity. We have also shown how metacalibation responses \citep{Sheldon2017} enter that expression for effective shape-noise.
    \item We have described a clean sub-sample covariance estimation scheme that directly measures the sampling noise contributions to the covariance from a given data set. We then used the resulting covariance estimates to test the validity of our assumed effective shape noise values.
    \item We have employed the hybrid covariance estimation technique PME \citep{Friedrich_Eifler} to efficiently evaluate the importance of individual contributions to the covariance from only a limited set of simulated data (in our case: 200 realisations of the 3x2pt data vector including shape-noise and 100 realisations without shape noise).
    \item We have devised a treatment of survey geometry in covariance modelling that improves upon existing approximations \citep[of \eg\ ][]{Efstathiou2004} and we have demonstrated how to carry those approximations from harmonic space to real space.
    
\end{itemize}

Using these results we perform several tests for the fiducial DES-Y3 3x2pt covariance matrix and likelihood model, with the following conclusions:
\begin{itemize}
    \item The assumption of Gaussian statistical uncertainties is sufficiently accurate (\cf\ \secrefnospace{Gauss}). Hence, knowledge of the covariance of the 3x2pt data vector is sufficient to model our statistical uncertainties. The main assumption made to arrive at this conclusion is that non-Gaussian error bars are primarily a large-scale problem and that at small scales the number of modes present within the DES-Y3 survey volume converges to a Gaussian distribution via the central limit theorem.
    \item The non-Gaussian part of the covariance has a negligible impact on both maximum posterior $\chi^2$ and parameter constraints (\cf\ \secref{impact_of_4point}). This statement is not a general one but only holds for the specific DES-Y3 3x2pt analysis setup. The main assumption made to arrive at that conclusion is that the CosmoLike model \citep{Krause2016b} or the log-normal model \citep{Hilbert2011} for the non-Gaussian covariance do not vastly underestimate the true covariance. Given the results of \citet{Sato2009, Hilbert2011} we find this a safe assumption.
    \item Of all covariance modelling assumptions investigated in this paper the $f_{\mathrm{sky}}$ approximation (made in the mixed term and cosmic variance term of our covariance model) has the largest effect on maximum posterior $\chi^2$. On average it increases $\chi^2$ between measurement and maximum posterior model by about $3.7\%$ ($\Delta\chi^2 \approx 18.9$) for the 3x2pt data vectors and by about $5.7\%$ ($\Delta\chi^2 \approx 16.0$) for the 2x2pt data vector (\cf\ \tabrefnospace{summary}).
    \item However, neither $f_{\mathrm{sky}}$ approximation nor any other covariance modelling detail tested in this paper (\cf\ \tabref{summary}; with the exception of finite bin width, see next point) has any significant impact on the location and width of constraints on the parameters $\Omega_m, \sigma_8, w$.
    \item The only exception to this statement is finite angular bin width which - if not taken into account in the mixed term and cosmic variance term of the covariance model - significantly increases the scatter of maximum posterior parameters (without increasing the inferred constraints accordingly). However, finite bin width has been taken into account in the past in an approximate manner - see e.g. \citet{Krause2017}.
    \item The fact that we do not know the true cosmological parameters of the Universe forces us to evaluate our covariance model at a wrong set of parameters. Even when iteratively adjusting those parameters to the maximum posterior parameters of the analysis, the parameters of the covariance model will scatter around the 'true' cosmological / nuisance parameters. We consider this an irreducible covariance error and find that it increases the maximum posterior scatter of $\Omega_m$ and $\sigma_8$ by about $3\%$ and that of the dark energy equation of state parameter $w$ by about $5\%$ (\cf\ \secref{impact_of_cov_cosmology} and \tabrefnospace{summary}). At the same time, we find this effect to have a negligible impact on maximum posterior $\chi^2$.
\end{itemize}

In summary, we have shown that our fiducial covariance and likelihood model underestimates the scatter of maximum posterior parameters by about $3$-$5\%$, which is mostly caused by uncertainty in the set of cosmological and nuisance parameters at which we evaluate that model. On average, the $\chi^2$ between maximum posterior model and measurement of the 3x2pt data vector will be $\sim 4\%$ higher than expected with perfect knowledge of the covariance matrix. This is mainly caused by our use of the $f_{\mathrm{sky}}$ approximation. We have devised an improved treatment of the full survey geometry (\cf\ Section \ref{sec:masking_tests}) but for the reason mentioned above this was only used to test the impact of the $f_{\mathrm{sky}}$ approximation on parameter constraints. 

Given the small impact that we estimated from the unaccounted effects in the covariance modelling, we conclude that the fiducial covariance model is adequate to be used in the 3x2pt DES-Y3 analysis.
While our validation of this covariance model has been carried out with a preliminary set of scale cuts and redshift distributions, we don't expect qualitative changes for the final DES-Y3 analysis setup. 

While the DESY3 specific outcomes of our study can not straightforwardly be transferred to other surveys and analyses, our methodological innovations will be useful tools in the covariance and likelihood validation of future experiments.

\section*{Acknowledgements}

OF gratefully acknowledges support by the Kavli Foundation and the International Newton Trust through a Newton-Kavli-Junior Fellowship and by Churchill College Cambridge through a postdoctoral By-Fellowship. The authors thank Henrique Xavier for very helpful discussions about mock simulations and the FLASK code. This research was partially supported by the Laborat\'orio Interinstitucional de e-Astronomia (LIneA), the Brazilian Funding agency CNPq, 
the INCT of the e-Universe and the Sao Paulo State Research Agency (FAPESP).
The authors acknowledge the use of computational resources from LIneA, the Center for Scientific Computing (NCC/GridUNESP) of the Sao Paulo State University (UNESP), and from the National Laboratory for Scientific Computing (LNCC/MCTI, Brazil), where the SDumont supercomputer ({\tt sdumont.lncc.br}) was used.

This paper has gone through internal review by the DES collaboration. Funding for the DES Projects has been provided by the U.S. Department of Energy, the U.S. National Science Foundation, the Ministry of Science and Education of Spain, 
the Science and Technology Facilities Council of the United Kingdom, the Higher Education Funding Council for England, the National Center for Supercomputing 
Applications at the University of Illinois at Urbana-Champaign, the Kavli Institute of Cosmological Physics at the University of Chicago, 
the Center for Cosmology and Astro-Particle Physics at the Ohio State University,
the Mitchell Institute for Fundamental Physics and Astronomy at Texas A\&M University, Financiadora de Estudos e Projetos, 
Funda{\c c}{\~a}o Carlos Chagas Filho de Amparo {\`a} Pesquisa do Estado do Rio de Janeiro, Conselho Nacional de Desenvolvimento Cient{\'i}fico e Tecnol{\'o}gico and 
the Minist{\'e}rio da Ci{\^e}ncia, Tecnologia e Inova{\c c}{\~a}o, the Deutsche Forschungsgemeinschaft and the Collaborating Institutions in the Dark Energy Survey. 

The Collaborating Institutions are Argonne National Laboratory, the University of California at Santa Cruz, the University of Cambridge, Centro de Investigaciones Energ{\'e}ticas, 
Medioambientales y Tecnol{\'o}gicas-Madrid, the University of Chicago, University College London, the DES-Brazil Consortium, the University of Edinburgh, 
the Eidgen{\"o}ssische Technische Hochschule (ETH) Z{\"u}rich, 
Fermi National Accelerator Laboratory, the University of Illinois at Urbana-Champaign, the Institut de Ci{\`e}ncies de l'Espai (IEEC/CSIC), 
the Institut de F{\'i}sica d'Altes Energies, Lawrence Berkeley National Laboratory, the Ludwig-Maximilians Universit{\"a}t M{\"u}nchen and the associated Excellence Cluster Universe, 
the University of Michigan, the National Optical Astronomy Observatory, the University of Nottingham, The Ohio State University, the University of Pennsylvania, the University of Portsmouth, 
SLAC National Accelerator Laboratory, Stanford University, the University of Sussex, Texas A\&M University, and the OzDES Membership Consortium.

Based in part on observations at Cerro Tololo Inter-American Observatory at NSF's NOIRLab (NOIRLab Prop. ID 2012B-0001; PI: J. Frieman), which is managed by the Association of Universities for Research in Astronomy (AURA) under a cooperative agreement with the National Science Foundation.

The DES data management system is supported by the National Science Foundation under Grant Numbers AST-1138766 and AST-1536171.
The DES participants from Spanish institutions are partially supported by MINECO under grants AYA2015-71825, ESP2015-66861, FPA2015-68048, SEV-2016-0588, SEV-2016-0597, and MDM-2015-0509, 
some of which include ERDF funds from the European Union. IFAE is partially funded by the CERCA program of the Generalitat de Catalunya.
Research leading to these results has received funding from the European Research
Council under the European Union's Seventh Framework Program (FP7/2007-2013) including ERC grant agreements 240672, 291329, and 306478.
We  acknowledge support from the Brazilian Instituto Nacional de Ci\^encia
e Tecnologia (INCT) e-Universe (CNPq grant 465376/2014-2).

This manuscript has been authored by Fermi Research Alliance, LLC under Contract No. DE-AC02-07CH11359 with the U.S. Department of Energy, Office of Science, Office of High Energy Physics.

This work made use of the software packages {\tt GetDist} \citep{getdist}, {\tt ChainConsumer} \citep{chainconsumer}, {\tt matplotlib} \citep{matplotlib}, and {\tt numpy} \citep{numpy}.

We would like to thank the anonymous journal referee for their helpful comments.

\section*{Data availability} 

The DES-Y3 3x2pt covariance matrix and likelihoods will be made public upon publication of our final data analysis. \verb|C++| and \verb|python| tools to configure \verb|FLASK| as described in \secref{FLASK} are available at \url{https://github.com/OliverFHD/CosMomentum} . Tools to compute Gaussian and halomodel covariance is available at \url{https://github.com/CosmoLike/CosmoCov} .

\appendix

\section{Curved sky formalism}
\label{app:curvedsky}
The angular clustering correlation function of galaxies $w (\theta) $ is given in terms of the galaxy clustering power spectrum $C_\ell^{gg} $ as
\begin{equation}
\label{eq:wtheta}
w (\theta ) = \sum_\ell \frac{2\ell + 1}{4\pi} P_\ell\left( \cos \theta \right) \left( C_\ell^{gg} + \frac{1}{n} \right) \ ,
\end{equation}
where $n$ is the galaxy density per steradian. The term proportional to $\smash{\frac{1}{n}}$ is usually ommited (cf. \citet{Ross2011}) since it sums up to\footnote{from $\smash{\sum_\ell \frac{2\ell + 1}{2} P_\ell(x) P_\ell(y) = \delta_D(x -y)}$ - see \citet{Bronstein1979} for this and other properties of Legendre polynomials.}
\begin{eqnarray}
\frac{1}{2\pi n} \sum_\ell \frac{2\ell + 1}{2} P_\ell\left( \cos \theta \right) &=& \frac{1}{2\pi n}\delta_D(\cos\theta -1 ) \nonumber \\
&=& \frac{\delta_D(\theta)}{2\pi n\sin \theta} \ ,
\end{eqnarray}
which has to be interpreted as a 2-dimensional Dirac delta function on the sphere.

According to \citet{dePutter2010} (see also \citealt{Stebbins1996}) the galaxy-galaxy lensing correlation function $\gamma_t(\theta)$ is given in terms of the galaxy-convergence cross-power spectrum $C_\ell^{g\kappa} $ as
\begin{equation}
\gamma_t (\theta ) = \sum_\ell \frac{2\ell + 1}{4\pi} \frac{P_\ell^2\left( \cos \theta \right)}{\ell(\ell + 1)} C_\ell^{g\kappa}\ ,
\end{equation}
where $P_\ell^m$ are the associated Legendre Polynomials.


Finally, he cosmic shear correlation functions $\xi_\pm(\theta)$ are given by
\begin{eqnarray}
\label{eq:xi_pm}
\xi_\pm(\theta) &=& \sum_{\ell \geq 2} \frac{2\ell + 1}{4\pi}\ \frac{2(G_{\ell, 2}^+(x) \pm G_{\ell, 2}^-(x))}{\ell^2(\ell + 1)^2}\ C_\ell^{E} \nonumber \ ,\\
\end{eqnarray}
where $x = \cos \theta$, $C_\ell^{E}$ is the E-mode power spectrum of shear and we assume B-modes to vanish. The functions $G_{\ell, 2}^\pm(x)$ are defined in eq. 4.18\footnote{Note that a factor of $1/i\sin(\theta)$ is missing in the second line of this equation.} of \citet{Stebbins1996}. Eq. \ref{eq:xi_pm} can be expressed in terms of associated Legendre polynomials by using eq. 4.19 of \citet{Stebbins1996}, which gives
\begin{eqnarray}
\label{eq:Gpm}
G_{\ell, 2}^+(x) \pm G_{\ell, 2}^-(x) &=& P_\ell^2(x) \left\lbrace \frac{4 - \ell \pm 2x(\ell-1)}{1-x^2} - \frac{\ell(\ell - 1)}{2} \right\rbrace \nonumber \\
&&\ + P_{\ell-1}^2(x) \frac{(\ell+2)(x\mp 2)}{1 - x^2}\ .
\end{eqnarray}
In appendix \ref{app:bin_averaging} we show how to obtain \ref{eq:xi_pm} from the notation in \citet{Stebbins1996}.

It can be seen from above that each of the considered 2-point correlation functions can be written in terms of the corresponding power spectra as
\begin{equation}
    \xi^A(\theta) = \sum_{\ell} \frac{2\ell + 1}{4\pi}\ F_\ell^A(\cos \theta)\ C_\ell^A\ .
\end{equation}

\section{Averaging correlation functions over finite bins}
\label{app:bin_averaging}

The area average can be performed for $\gamma_t$by replacing $P_\ell^2\left( \cos \theta \right)$ with
\begin{eqnarray}
\frac{\int_{\theta_{\min}}^{\theta_{\max}} \mathrm d \theta\ \sin \theta\ P_\ell^2\left( \cos \theta \right)}{\cos\theta_{\min } - \cos\theta_{\max }} & = & \frac{\int_{\cos\theta_{\max}}^{\cos\theta_{\min}} \mathrm d x\ P_\ell^2\left( x \right)}{\cos\theta_{\min } - \cos\theta_{\max }} \nonumber \\
\nonumber \\
& = & \frac{\int_{\cos\theta_{\max}}^{\cos\theta_{\min}} \mathrm d x\ (1-x^2) \frac{\mathrm{d}^2}{\mathrm{d} x^2}P_\ell\left( x \right)}{\cos\theta_{\min } - \cos\theta_{\max }} \ .\nonumber \\
\end{eqnarray}
Using integration by parts and various recursion relations of Legendre polynomials \citep[cf.][]{Bronstein1979}, this becomes:
\begin{eqnarray}
&& \frac{\int_{\theta_{\min}}^{\theta_{\max}} \mathrm d \theta\ \sin \theta\ P_\ell^2\left( \cos \theta \right)}{\cos\theta_{\min } - \cos\theta_{\max }} \nonumber \\
& = & \frac{1}{\cos\theta_{\min } - \cos\theta_{\max }}\left\lbrace\left( \ell + \frac{2}{2\ell +1} \right)\ \left[ P_{\ell-1}(x)\right]_{\cos\theta_{\max}}^{\cos\theta_{\min}} \right.\nonumber \\
& & +\ (2-\ell)\ \left[ x P_{\ell}(x)\right]_{\cos\theta_{\max}}^{\cos\theta_{\min}} \nonumber \\
& & \left. -\ \frac{2}{2\ell +1}\ \left[ P_{\ell+1}(x)\right]_{\cos\theta_{\max}}^{\cos\theta_{\min}}  \right\rbrace\ .
\end{eqnarray}

In his equation 4.26 \citet{Stebbins1996} defines the shear correlation function $C_\gamma(\theta, \phi_1, \phi_2)$ which, by inspection of equation 4.27 and figure 2 of this work, translates into the shear correlation functions $\xi_\pm(\theta)$ as
\begin{equation}
\xi_\pm(\theta) = C_\gamma(\theta, 0, 0) \pm C_\gamma(\theta, \pi/4, \pi/4)\ .
\end{equation}
This way one directly arives at our expression for the shear correlation functions, equation \ref{eq:xi_pm}.

To account for finite bin width in equation \ref{eq:xi_pm} and in the covariance of $\hat \xi_\pm$ one has to perform the area-weighted bin average of the functions $\left(G_{\ell, 2}^+(\cos \theta) \pm G_{\ell, 2}^-(\cos \theta)\right)$. To do so one can insert the relations
\begin{eqnarray}
\int_{x_1}^{x_2}\mathrm d x\ \frac{x\ P_\ell^2(x)}{1-x^2} &= & \left[x\frac{\mathrm d P_\ell(x)}{\mathrm d x} \right]_{x_1}^{x_2} -\ \  \left[P_\ell(x)\right]_{x_1}^{x_2}\nonumber \\
\nonumber\\
\int_{x_1}^{x_2}\mathrm d x\ \frac{P_\ell^2(x)}{1-x^2}&= & \left[\frac{\mathrm d P_\ell(x)}{\mathrm d x} \right]_{x_1}^{x_2}
\end{eqnarray}
into equation \ref{eq:Gpm}. In summary this means one has to exchange the functions $\left(G_{\ell, 2}^+(cos \theta) \pm G_{\ell, 2}^-(cos \theta)\right)$ as follows:
\begin{eqnarray}
&& \frac{\int_{\cos\theta_{\max}}^{\cos\theta_{\min}}\mathrm d x\ \left(G_{\ell, 2}^+(x) \pm G_{\ell, 2}^-(x)\right)}{\cos\theta_{\min} - \cos\theta_{\max}} \nonumber \\
\nonumber\\
&=& \left\lbrace-\ \left(\frac{\ell(\ell-1)}{2}\right)\left(\ell + \frac{2}{2\ell + 1}\right)\left[ P_{\ell-1}(x)\right]_{\cos\theta_{\max}}^{\cos\theta_{\min}} \right. \nonumber \\
\nonumber\\
&& -\ \frac{\ell(\ell-1)(2-\ell)}{2}\left[ xP_{\ell}(x)\right]_{\cos\theta_{\max}}^{\cos\theta_{\min}} \nonumber \\
\nonumber\\
&& +\ \frac{\ell(\ell-1)}{2\ell+1}\left[ P_{\ell+1}(x)\right]_{\cos\theta_{\max}}^{\cos\theta_{\min}} \nonumber \\
\nonumber\\
&& +\ (4-\ell)\left[\frac{\mathrm d P_\ell(x)}{\mathrm d x} \right]_{\cos\theta_{\max}}^{\cos\theta_{\min}} \nonumber \\
\nonumber\\
&& +\ (\ell+2)\left\lbrace \left[x\frac{\mathrm d P_{\ell-1}(x)}{\mathrm d x} \right]_{\cos\theta_{\max}}^{\cos\theta_{\min}} -\ \  \left[P_{\ell-1}(x)\right]_{\cos\theta_{\max}}^{\cos\theta_{\min}} \right\rbrace \nonumber \\
\nonumber\\
&& \pm\ 2(\ell-1)\left\lbrace \left[x\frac{\mathrm d P_\ell(x)}{\mathrm d x} \right]_{\cos\theta_{\max}}^{\cos\theta_{\min}} -\ \  \left[P_\ell(x)\right]_{\cos\theta_{\max}}^{\cos\theta_{\min}} \right\rbrace \nonumber \\
\nonumber\\
&& \left.\mp\ 2(\ell+2)\left[\frac{\mathrm d P_{\ell-1}(x)}{\mathrm d x} \right]_{\cos\theta_{\max}}^{\cos\theta_{\min}}\right\rbrace \frac{1}{\cos\theta_{\min} - \cos\theta_{\max}} \ .\nonumber\\
\end{eqnarray}
These expressions can be very efficiently calculated and pre-tabulated e.g. with the help of the \verb|gnu scientific library| \citep{Galassi2009}.


\section{Masking in real space covariances}
\label{app:masking}

DES observations don't cover the entire sky, but are located within a survey mask, described by a function $W(\mathbf{\hat n})$ which is $=1$ if we have observed the sky at location $\mathbf{\hat n}$ and zero otherwise\footnote{A map of the mask will come with a finally resolution, in which case the fractional values $0 < W < 1$ of the mask will describe the completeness of observations within the map resolution.}. In this appendix we derive how this masking changes the covariance of any measured 2-point statistics.

We start by computing, how many galaxy pairs we expect to find within our mask (assuming that galaxies do not cluster). If $n_1$, $n_2$ are the number densities of 2 different tracer samples, then the expected number of pairs $\dd N_{\mathrm{pair}}(\theta)$ with angular separation within $[\theta, \theta + \dd\theta]$ is given by \citep[\cf][]{Troxel2018b}
\begin{align}
    &\ \frac{\dd N_{\mathrm{pair}}(\theta)}{n_1 n_2} \nonumber \\
    =&\ \dd \theta \int \dd \Omega_1 \dd \Omega_2\ W(\mathbf{\hat n_1})\ W(\mathbf{\hat n_2})\ \delta_D(\arccos[\mathbf{\hat n}_1\cdot\mathbf{\hat n}_2] - \theta)\nonumber \\
    =&\ \dd \theta \int \dd \Omega_1 \dd \Omega_2\ W(\mathbf{\hat n_1})\ W(\mathbf{\hat n_2})\ \sqrt{1- x_{12}^2}\delta_D(x_{12} - \cos\theta)\nonumber \\
    =&\ \dd \theta \sin \theta\ \int \dd \Omega_1 \dd \Omega_2\ W(\mathbf{\hat n_1})\ W(\mathbf{\hat n_2})\ \delta_D(x_{12} - \cos\theta)\nonumber \\
    =&\ \dd \theta \sin \theta\ \sum_\ell \frac{2\ell+1}{2} P_\ell(\cos \theta) \int \dd \Omega_1 \dd \Omega_2\ W(\mathbf{\hat n_1})\ W(\mathbf{\hat n_2})P_\ell(x_{12})\ .
\end{align}
Using the fact that
\begin{equation}
\label{eq:Pell_in_terms_of_Yellm}
    P_\ell(\mathbf{\hat n}\cdot\mathbf{\hat m}) = \frac{4\pi}{2\ell+1} \sum_{m=-\ell}^\ell Y_{\ell m}(\mathbf{\hat n})Y_{\ell m}^*(\mathbf{\hat m})
\end{equation}
\begin{equation}
    \left(\Rightarrow \xi(\mathbf{\hat n}\cdot\mathbf{\hat m}) = \sum_{\ell} \frac{2\ell+1}{4\pi} C_\ell\ P_\ell(\mathbf{\hat n}\cdot\mathbf{\hat m}) = \sum_{\ell m} C_\ell Y_{\ell m}(\mathbf{\hat n})Y_{\ell m}^*(\mathbf{\hat m})\right)
\end{equation}
this becomes
\begin{align}
    &\ \frac{\dd N_{\mathrm{pair}}(\theta)}{n_1 n_2} \nonumber \\
    =&\ 2\pi\sin \theta\ \dd\theta\ \sum_{\ell m} P_\ell(\cos \theta)\ *\nonumber \\
    &\ *\ \left(\int \dd \Omega_1\ W(\mathbf{\hat n}_1)\ Y_{\ell m}^*(\mathbf{\hat n}_1)\right)\left(\int \dd \Omega_2\ W(\mathbf{\hat n}_2)\ Y_{\ell m}(\mathbf{\hat n}_2)\right)\nonumber \\
    =&\ 2\pi\sin \theta\ \dd\theta\ \sum_{\ell m} P_\ell(\cos \theta)\ |W_{\ell m}|^2\nonumber \\
    =&\ 2\pi\sin \theta\ \dd\theta\ \sum_{\ell} (2\ell+1) P_\ell(\cos \theta)\ C_\ell^W\nonumber \\
    =&\ 8\pi^2\sin \theta\ \dd\theta\ \xi^W(\theta)\ ,
\end{align}
where in the last steps we have defined the angular power spectrum of the mask through $(2\ell+1) C_\ell^W = \sum_m |W_{\ell m}|^2$ as well as the angular 2-point function of the mask, $\xi^W(\theta)$. The total number of galaxy pairs in a finite angular bin $[\theta_{\min},\theta_{\max}]$ is then
\begin{align}
    N_{\mathrm{pair}}[\theta_{\min},\theta_{\max}] =&\ (8\pi^2 n_1 n_2) \int_{\theta_{\min}}^{\theta_{\max}}\dd\theta\ \sin \theta\ \xi^W(\theta) \nonumber \\
    =&\ (8\pi^2 n_1 n_2) \sum_{\ell} \frac{\left[P_{\ell+1}(x) - P_{\ell-1}(x) \right]_{\theta_{\max}}^{\theta_{\min}}}{4\pi}\ C_\ell^W \ .
\end{align}
The 2-point correlation function of 2 scalar random fields, $\xi^{ab}(\theta) = \langle \delta_a(\mathbf{\hat n}_a) \delta_b(\mathbf{\hat n}_b) | \mathbf{\hat n}_a\cdot \mathbf{\hat n}_b = \cos \theta \rangle$, is in practice estimated within a finite angular bin as 
\begin{align}
\label{eq:xi_hat_with_masking}
    &\ \hat\xi^{ab}[\theta_{\min},\theta_{\max}]\cdot  \frac{N_{\mathrm{pair}}[\theta_{\min},\theta_{\max}]}{n_a n_b} \nonumber \\
    =&\ \int_{\theta_{\min}}^{\theta_{\max}}\dd \theta \sin \theta \int \dd \Omega_a \dd \Omega_b\ W(\mathbf{\hat n_a}) W(\mathbf{\hat n_b}) *\nonumber \\
    &\ *\delta_D(x_{ab} - \cos\theta)\ \delta_a(\mathbf{\hat n}_a) \delta_b(\mathbf{\hat n}_b)\nonumber \\
    =&\ 2\pi\sum_{\ell m}\int_{\theta_{\min}}^{\theta_{\max}}\dd \theta \sin \theta\ P_\ell(\theta)\ *\nonumber \\
    &\ *\ \int \dd \Omega_a \dd \Omega_b\ W(\mathbf{\hat n_a})\ W(\mathbf{\hat n_b})\ Y_{\ell m}(\mathbf{\hat n}_a)\ Y_{\ell m}^*(\mathbf{\hat n}_b)\ \delta_a(\mathbf{\hat n}_a) \delta_b(\mathbf{\hat n}_b)\nonumber \\
    =&\ 2\pi\sum_{\ell }\left[P_{\ell+1}(x) - P_{\ell-1}(x) \right]_{\theta_{\max}}^{\theta_{\min}}\ *\nonumber \\
    &\ *\ \frac{1}{2\ell+1} \sum_m\ \int \dd \Omega_a\ W(\mathbf{\hat n_a}) \delta_a(\mathbf{\hat n}_a) Y_{\ell m}(\mathbf{\hat n}_a)  \ *\nonumber \\
    &\ *\ \int \dd \Omega_b\ W(\mathbf{\hat n_b})\ Y_{\ell m}^*(\mathbf{\hat n}_b) \delta_b(\mathbf{\hat n}_b)\ .
\end{align}
From the last line it can be seen that $\hat\xi^{ab}$ - apart from the normalisation by $N_{\mathrm{pair}}[\theta_{\min},\theta_{\max}]/n_a n_b$ - is exactly the angular space counter part of the Pseudo-Cells estimator in the corresponding harmonic space \citep[\cf][]{Efstathiou2004}. Why this is the case can be understood most easily in the limit of an infinitesimal angular bin. In this limit
\begin{align}
    N_{\mathrm{pair}}(\theta) \propto&\ \sin \theta\ \xi^W(\theta)\nonumber \\
    N_{\mathrm{pair}}(\theta) \cdot \langle \hat\xi^{ab}(\theta)\rangle \propto&\ \sin \theta\ \xi^W(\theta)\ \xi^{ab}(\theta)\ ,
\end{align}
where the second line shows that the convolution of mask and signal power spectrum in harmonic space \citep{Efstathiou2004} becomes a simple multiplication in angular space. Especially, normalisation by $N_{\mathrm{pair}}(\theta)$ is the angular analog of multiplication with the inverse mode-coupling matrix that appears in harmonic space.

Let $\hat C_\ell^{ab}$ be the pseudo-$C_\ell$ estimator we identified in \eqnref{xi_hat_with_masking}, i.e. we write
\begin{align}
    &\ \hat\xi^{ab}[\theta_{\min},\theta_{\max}]\cdot  \frac{N_{\mathrm{pair}}[\theta_{\min},\theta_{\max}]}{n_a n_b}\nonumber \\
    =&\ 2\pi\sum_\ell \left[P_{\ell+1}(x) - P_{\ell-1}(x) \right]_{\theta_{\max}}^{\theta_{\min}}\ \hat C_\ell^{ab}
\end{align}
with
\begin{equation}
    \hat C_\ell^{ab} \equiv \frac{1}{2\ell+1}\sum_m \left(W\delta_a\right)_{\ell m}\left(W\delta_b\right)_{\ell m}^*\ .
\end{equation}
Then the covariance of the 2-point function measurements between the fields $\delta_a\&\delta_b$ and $\delta_c\&\delta_d$ within angular bins $[\theta_{-}^{ab},\theta_{+}^{ab}]$ and $[\theta_{-}^{cd},\theta_{+}^{cd}]$ is given by
\begin{align}
    &\ \mathrm{Cov}\left\lbrace\hat\xi^{ab}[\theta_{-}^{ab},\theta_{+}^{ab}],\hat\xi^{cd}[\theta_{-}^{cd},\theta_{+}^{cd}]\right\rbrace\ \frac{N_{\mathrm{pair}}^{ab}[\theta_{-}^{ab},\theta_{+}^{ab}] \ N_{\mathrm{pair}}^{cd}[\theta_{-}^{cd},\theta_{+}^{cd}]}{n_a n_b n_c n_d}\nonumber \\
    =&\ (2\pi)^2\sum_{\ell_1\ \ell_2} \left[P_{\ell_1+1}(x) - P_{\ell_1-1}(x) \right]_{\theta_{+}^{ab}}^{\theta_{-}^{ab}}\ \left[P_{\ell_2+1}(x) - P_{\ell_2-1}(x) \right]_{\theta_{+}^{cd}}^{\theta_{-}^{cd}}\cdot\nonumber\\
    &\ \mathrm{Cov}\left\lbrace\hat C_{\ell_1}^{ab}, \hat C_{\ell_2}^{cd} \right\rbrace\ .
\end{align}
Assuming that $\delta_a,\delta_b,\delta_c,\delta_d$ are Gaussian random fields and defining the symbols
\begin{equation}
    W_{\ell_1 \ell_2 m_1 m_2}\ \equiv \int \dd \Omega\ W(\mathbf{\hat n})\ Y_{\ell_1 m_1}(\mathbf{\hat n})\ Y_{\ell_2 m_2}^*(\mathbf{\hat n})
\end{equation}
it is straight forward to show that $\mathrm{Cov}\left\lbrace\hat C_{\ell_1}^{ab}, \hat C_{\ell_2}^{cd} \right\rbrace$ is given by \citep[\cf][]{Efstathiou2004}
\begin{align}
    &\ \mathrm{Cov}\left\lbrace\hat C_{\ell_1}^{ab}, \hat C_{\ell_2}^{cd} \right\rbrace \nonumber \\
    =&\ \frac{1}{(2\ell_1+1)(2\ell_2+1)} \sum_{m_1\ m_2} \int \dd\Omega_a\dd\Omega_b\dd\Omega_c\dd\Omega_d\ *\nonumber \\
    &\ *\ W(\mathbf{\hat n}_a)W(\mathbf{\hat n}_b)W(\mathbf{\hat n}_c)W(\mathbf{\hat n}_d)\ *\nonumber \\
    &\ *\ Y_{\ell_1 m_1}(\mathbf{\hat n}_a)Y_{\ell_1 m_1}^*(\mathbf{\hat n}_b)Y_{\ell_2 m_2}(\mathbf{\hat n}_c)Y_{\ell_2 m_2}^*(\mathbf{\hat n}_d)\ *\nonumber \\
    &\ *\ \left\lbrace\xi^{ac}(\theta^{ac})\xi^{bd}(\theta^{bd}) + \xi^{ad}(\theta^{ad})\xi^{bc}(\theta^{bc})\right\rbrace\nonumber \\
    =&\ \frac{1}{(2\ell_1+1)(2\ell_2+1)} \sum_{m_1\ m_2} \sum_{\ell_3\ m_3} \sum_{\ell_4\ m_4} \left( C_{\ell_3}^{ac} C_{\ell_4}^{bd} + C_{\ell_3}^{ad} C_{\ell_4}^{bc}\right)\ *\nonumber \\
    &\ *\ W_{\ell_1 \ell_3 m_1 m_3}W_{\ell_3 \ell_2 m_3 m_2}W_{\ell_2 \ell_4 m_2 m_4}W_{\ell_4 \ell_1 m_4 m_1}\nonumber \\
\end{align}
At this point \eg\ \citet{Efstathiou2004, Varshalovich1988} follow with the approximation
\begin{align}
    \approx&\ \ \frac{1}{2}\frac{C_{\ell_1}^{ac} C_{\ell_2}^{bd} + C_{\ell_2}^{ac} C_{\ell_1}^{bd} + C_{\ell_1}^{ad} C_{\ell_2}^{bc} + C_{\ell_2}^{ad} C_{\ell_1}^{bc}}{(2\ell_1+1)(2\ell_2+1)}\ *\nonumber \\
    &\ * \sum_{m_1 m_2} \sum_{\ell_3 m_3} \sum_{\ell_4 m_4} W_{\ell_1 \ell_3 m_1 m_3}W_{\ell_3 \ell_2 m_3 m_2}W_{\ell_2 \ell_4 m_2 m_4}W_{\ell_4 \ell_1 m_4 m_1}\ .
\end{align}
We however find that for fixed values of $\ell_1$ and $\ell_2$ the expression
\begin{equation}
    \sum_{m_1\ m_2\ m_3\ m_4} W_{\ell_1 \ell_3 m_1 m_3}W_{\ell_3 \ell_2 m_3 m_2}W_{\ell_2 \ell_4 m_2 m_4}W_{\ell_4 \ell_1 m_4 m_1}
\end{equation}
has four maxima at $[\ell_3 = \ell_1 , \ell_4 = \ell_2]$ , $[\ell_3 = \ell_2 , \ell_4 = \ell_1]$ , $[\ell_3 = \ell_1 , \ell_4 = \ell_1]$ and at $[\ell_3 = \ell_2 , \ell_4 = \ell_2]$ and that the area around each of these maxima contributes a similar amount to the sums over $\ell_3$ and $\ell_4$. Hence, we opt instead for the approximation 
\begin{align}
    \label{eq:Efstathiou_2.0}
    &\ \mathrm{Cov}\left\lbrace\hat C_{\ell_1}^{ab}, \hat C_{\ell_2}^{cd} \right\rbrace \approx \nonumber \\
    &\ \ \frac{1}{4}\left(\frac{C_{\ell_1}^{ac} C_{\ell_2}^{bd} + C_{\ell_2}^{ac} C_{\ell_1}^{bd} + C_{\ell_1}^{ac} C_{\ell_1}^{bd} + C_{\ell_2}^{ac} C_{\ell_2}^{bd}}{(2\ell_1+1)(2\ell_2+1)}\right.\ +\nonumber \\
    &\ \ +\left.\frac{C_{\ell_1}^{ad} C_{\ell_2}^{bc} + C_{\ell_2}^{ad} C_{\ell_1}^{bc} + C_{\ell_1}^{ad} C_{\ell_1}^{bc} + C_{\ell_2}^{ad} C_{\ell_2}^{bc}}{(2\ell_1+1)(2\ell_2+1)}\right)\ *\nonumber \\
    &\ *\ \sum_{m_1, m_2, m_3, m_4}W_{\ell_1 \ell_3 m_1 m_3}W_{\ell_3 \ell_2 m_3 m_2}W_{\ell_2 \ell_4 m_2 m_4}W_{\ell_4 \ell_1 m_4 m_1}\ .
\end{align}
In practice, both approximations yield very similar results and they are valid on scales $\ell_1, \ell_2$ which are much smaller than the typical scales of the mask $W$ \citep{Efstathiou2004, Varshalovich1988}. Unfortunately, the DES-Y3 analysis mask has features and holes over a large range of scales. Hence, the angular scales of interest in the 3x2pt analysis are never strictly smaller than the scales of our mask. Hence, \eqnref{Efstathiou_2.0} is not sufficiently accurate in our case and infact significantly overestimates our covariance matrix. In \figref{masking_and_fsky} we explain a simple scheme that can be used to correct for this: Calculating the covariance of $\hat\xi^{ab}[\theta_{-}^{ab}$ and $\theta_{+}^{ab}],\hat\xi^{cd}[\theta_{-}^{cd},\theta_{+}^{cd}]$ within the Gaussian covariance model (see also \secrefnospace{model}) requires integration over all pairs of locations within our survey mask that fall into the angular bins $[\theta_{-}^{ab},\theta_{+}^{ab}]$ and $[\theta_{-}^{cd},\theta_{+}^{cd}]$. Schematically, the covariance then depends on expressions of the form
\begin{align}
    &\mathrm{Cov} \propto \nonumber \\
    &\underset{(ab)\in\mathrm{mask,bin}}{\int} \dd \Omega^{a}\dd \Omega^{b} \underset{(cd)\in\mathrm{mask,bin}}{\int} \dd \Omega^{a}\dd \Omega^{b} \xi^{ac}(\theta^{ac})\xi^{bd}(\theta^{bd}) + \dots\ .
\end{align}
\Figref{masking_and_fsky} visualizes this for the mixed terms in the covariance, where one of the correlation functions $\xi^{ac}$ or $\xi^{bd}$ is due to sampling noise such as shape-noise of shot-noise and hence is proportional to a Dirac delta function. In that case, the integration is over pairs that share one end point. Now the approximation made \eg\ in \citet{Efstathiou2004} or by our \eqnref{Efstathiou_2.0} assumes that also the correlation function between the other two end points effectively acts as a delta function - at least with respect to the smallest scale features in the survey mask. We find that this is not the case for the DES-Y3 mask and that it contains features on all scales relevant to our analysis. But \figref{masking_and_fsky} also indicates a simple way to fix this: approximating the integrand over the distance of the two remaining endpoints by a delta function roughly overestimates the integral by a factor equal to one over the fraction of small scale hole in the survey footprint compared to the average scale at which the 2-point function between the 2 end points decays. And multiplying the the mixed terms in the covariance by this fraction can serve as a next-to-leading order correction to our \eqnrefnospace{Efstathiou_2.0}. By similar arguments one can deduce that the cosmic variance terms (terms where none of the end points must be joined) can be corrected by performing this multiplication twice.

To implement this correction we draw circles within the DES-Y3 survey footprint with radii ranging from $5$arcmin to $20$arcmin and measure the masking fraction in these circles. We find that this fraction is $\approx 90\%$ across the considered scales. Multiplying the mixed terms in the covariance by that fraction and the cosmic variance terms by the square of that fraction (and using \eqnrefnospace{Efstathiou_2.0}) we indeed find significant improvement of the maximum posterior $\chi^2$ obtained for the FLASK simulations (\cf\ lower panel of \figref{masking_and_fsky} as well as \figref{chiSq_offsets}).

We end this appendix by further simplifying \eqnrefnospace{Efstathiou_2.0}. Using the completeness of the $Y_{\ell m}$ as well as the fact that $W(\mathbf{\hat n})^2 = W(\mathbf{\hat n})$ one can show that \citep{Efstathiou2004}
\begin{align}
    &\ \sum_{m_1, m_2, m_3, m_4}W_{\ell_1 \ell_3 m_1 m_3}W_{\ell_3 \ell_2 m_3 m_2}W_{\ell_2 \ell_4 m_2 m_4}W_{\ell_4 \ell_1 m_4 m_1}\nonumber \\
    =&\ |W_{\ell_1 \ell_2 m_1 m_2}|^2\ .
\end{align}
Then re-writing $W_{\ell_1 \ell_2 m_1 m_2}$ as
\begin{align}
    W_{\ell_1 \ell_2 m_1 m_2} =&\ \int \dd \Omega\ W(\mathbf{\hat n})\ Y_{\ell_1 m_1}(\mathbf{\hat n})\ Y_{\ell_2 m_2}^*(\mathbf{\hat n})\nonumber \\
    =&\ \sum_{\ell m} W_{\ell m}\int \dd \Omega\ Y_{\ell m}(\mathbf{\hat n})\ Y_{\ell_1 m_1}(\mathbf{\hat n})\ Y_{\ell_2 m_2}^*(\mathbf{\hat n})\nonumber \\
    =&\ (-1)^{m_2} \sum_{\ell m} W_{\ell m} \sqrt{\frac{(2\ell + 1)(2\ell_1 + 1)(2\ell_2 + 1)}{4\pi}}\ *\nonumber \\
    &\ *\ \begin{pmatrix} \ell & \ell_1 & \ell_2 \\ 0 & 0 & 0 \end{pmatrix} \begin{pmatrix} \ell & \ell_1 & \ell_2 \\ m & m_1 & -m_2 \end{pmatrix}\ .
\end{align}
and using orthogonality properties of Wigner 3j symbols one can see that
\begin{align}
    \frac{\sum_{m_1 m_2} |W_{\ell_1 \ell_2 m_1 m_2}|^2}{(2\ell_1+1)(2\ell_2+1)} =&\ \sum_{\ell m} \frac{|W_{\ell m}|^2}{4\pi} \begin{pmatrix} \ell & \ell_1 & \ell_2 \\ 0 & 0 & 0 \end{pmatrix}^2\nonumber \\
    =&\ \sum_{\ell} \frac{2\ell + 1}{4\pi} C_\ell^W \begin{pmatrix} \ell & \ell_1 & \ell_2 \\ 0 & 0 & 0 \end{pmatrix}^2\ .
\end{align}

\begin{align}
    &\ \Rightarrow \mathrm{Cov}\left\lbrace\hat C_{\ell_1}^{ab}, \hat C_{\ell_2}^{cd} \right\rbrace \approx \nonumber \\
    &\ \ \frac{1}{4}\left(C_{\ell_1}^{ac} C_{\ell_2}^{bd} + C_{\ell_2}^{ac} C_{\ell_1}^{bd} + C_{\ell_1}^{ac} C_{\ell_1}^{bd} + C_{\ell_2}^{ac} C_{\ell_2}^{bd}\right.\ +\nonumber \\
    &\ \ +\left. C_{\ell_1}^{ad} C_{\ell_2}^{bc} + C_{\ell_2}^{ad} C_{\ell_1}^{bc} + C_{\ell_1}^{ad} C_{\ell_1}^{bc} + C_{\ell_2}^{ad} C_{\ell_2}^{bc}\right)\ *\nonumber \\
    &\ *\ \sum_{\ell} \frac{2\ell + 1}{4\pi} C_\ell^W \begin{pmatrix} \ell & \ell_1 & \ell_2 \\ 0 & 0 & 0 \end{pmatrix}^2\nonumber
\end{align}
\begin{align}
    \equiv&\ \ \frac{1}{4}\left(C_{\ell_1}^{ac} C_{\ell_2}^{bd} + C_{\ell_2}^{ac} C_{\ell_1}^{bd} + C_{\ell_1}^{ac} C_{\ell_1}^{bd} + C_{\ell_2}^{ac} C_{\ell_2}^{bd}\right.\ +\nonumber \\
    &\ \ +\left. C_{\ell_1}^{ad} C_{\ell_2}^{bc} + C_{\ell_2}^{ad} C_{\ell_1}^{bc} + C_{\ell_1}^{ad} C_{\ell_1}^{bc} + C_{\ell_2}^{ad} C_{\ell_2}^{bc}\right)\ \mathcal{M}_{\ell_1 \ell_2}\ .
\end{align}
For an efficient numerical evaluation of the above sum we point out the following useful relation of Wigner-$3j$ symbols (following from \url{functions.wolfram.com/HypergeometricFunctions/ThreeJSymbol/}):
\begin{align}
    &\ \begin{pmatrix} \ell & \ell_1 & \ell_2 \\ 0 & 0 & 0 \end{pmatrix}^2\ \left/\ \begin{pmatrix} \ell-2 & \ell_1 & \ell_2 \\ 0 & 0 & 0 \end{pmatrix}^2 \right.\nonumber \\
    \nonumber \\
    =&\ \frac{(\ell_2 - \ell_1 +\ell -1)(\ell_1 - \ell_2 +\ell -1)(\ell_1 + \ell_2 -\ell + 2)(\ell_1 + \ell_2 + \ell)}{(\ell_2 - \ell_1 +\ell)(\ell_1 - \ell_2 +\ell)(\ell_1 + \ell_2 -\ell + 1)(\ell_1 + \ell_2 + \ell+1)}\ .
\end{align}

\section{Motivation for our re-scaling ansatz for masking effects}
\label{app:motivation_for_rescaling}

In this appendix we make some of the arguments presented in \secref{masking_tests} more precise. We have stated there that calculating the covariance of $\hat\xi^{ab}[\theta_{-}^{ab},\theta_{+}^{ab}]$ and $\hat\xi^{cd}[\theta_{-}^{cd},\theta_{+}^{cd}]$ amounts to integration over all pairs of locations within our survey mask that fall into the angular bins $[\theta_{-}^{ab},\theta_{+}^{ab}]$ and $[\theta_{-}^{cd},\theta_{+}^{cd}]$. Schematically, this leads to terms of the form
\begin{align}
\label{eq:covariance_real_space_integrals_app}
    &\mathrm{Cov} = \nonumber \\
    &\frac{1}{\mathcal{N}}\underset{(ab)\in\mathrm{mask,bin}}{\int} \dd \Omega^{a}\dd \Omega^{b} \underset{(cd)\in\mathrm{mask,bin}}{\int} \dd \Omega^{c}\dd \Omega^{d} \xi^{ac}(\theta^{ac})\xi^{bd}(\theta^{bd})\nonumber \\
    &+ \dots\ ,
\end{align}
where compared to \eqnref{covariance_real_space_integrals} we have now explicitly included a normalisation factor which is proportional to the product of the number of pairs of locations within our mask that fall into the angular bins $[\theta_{-}^{ab},\theta_{+}^{ab}]$ and $[\theta_{-}^{cd},\theta_{+}^{cd}]$,
\begin{eqnarray}
\mathcal{N} \propto N_{\mathrm{pair}}^{ab}[\theta_{-}^{ab},\theta_{+}^{ab}] \ N_{\mathrm{pair}}^{cd}[\theta_{-}^{cd},\theta_{+}^{cd}]\ .
\end{eqnarray}
We have argued in \secref{masking_tests} that the approximation of \citet{Efstathiou2004} for evaluating the impact of masking on the 2-point function covariance can roughly be understood as making the replacements
\begin{align}
\label{eq:replacement_1}
    &\ \int \dd \Omega^{a}\ \dots\ W(\Omega^{a})\xi^{ac}(\theta^{ac})\nonumber \\
    \approx&\ \bar{\xi}^{ac}\int \dd \Omega^{a}\ \dots\ W(\Omega^{a})\delta_{\mathrm{Dirac}}^2(\Omega^{a}-\Omega^{c})
\end{align}
and
\begin{align}
\label{eq:replacement_2}
    &\ \int \dd \Omega^{b}\ \dots\ W(\Omega^{b})\xi^{bd}(\theta^{bd})\nonumber \\
    \approx&\ \bar{\xi}^{bd}\int \dd \Omega^{b}\ \dots\ W(\Omega^{b})\delta_{\mathrm{Dirac}}^2(\Omega^{b}-\Omega^{d})\ ,
\end{align}
where $\bar{\xi}^{ac}$ is the integral of $\xi^{ac}(\theta^{ac})$ over the 2-dimensional angular distance vector $\boldsymbol\theta^{ac}$ and $\bar{\xi}^{bd}$ is the integral of $\xi^{bd}(\theta^{bd})$ over $\boldsymbol\theta^{bd}$.

Within these approximations the right side of \eqnref{covariance_real_space_integrals_app} can only be non-zero if the angular bins $[\theta_{-}^{ab},\theta_{+}^{ab}]$ and $[\theta_{-}^{cd},\theta_{+}^{cd}]$ are identical. If that is the case, then the covariance becomes
\begin{align}
\label{eq:covariance_real_space_integrals_app_2}
    &\mathrm{Cov} \approx \frac{\bar{\xi}^{ac}\bar{\xi}^{bd}}{\mathcal{N}}\underset{(ab)\in\mathrm{mask,bin}}{\int} \dd \Omega^{a}\dd \Omega^{b} + \dots\ .
\end{align}
But the integral on the right side of this equation is nothing but $N_{\mathrm{pair}}^{ab}[\theta_{-}^{ab},\theta_{+}^{ab}]$, so we have
\begin{equation}
    \mathrm{Cov}\left\lbrace\hat{\xi}^{ac}[\theta_{-},\theta_{+}],\hat{\xi}^{bd}[\theta_{-},\theta_{+}]\right\rbrace \propto \frac{1}{N_{\mathrm{pair}}^{ab}[\theta_{-},\theta_{+}]}\ ,
\end{equation}
with proportionality coefficients that do not depend on the survey mask. So if our above understanding of the approximation proposed by \citet{Efstathiou2004} is (at least approximately) correct, then its ratio with respect to the $f_{\mathrm{sky}}$ approximation (\ie\ the approximation where one computes the covariance of a full-sky survey and then re-scales it with the sky fraction $f_{\mathrm{sky}}$ of the survey footprint) should be given by the inverse ratio of the exact value of $N_{\mathrm{pair}}^{ab}[\theta_{-},\theta_{+}]$ to the $f_{\mathrm{sky}}$ calculation
\begin{equation}
\label{eq:naive_pairs}
N_{\mathrm{pair,}f_{\mathrm{sky}}}^{ab}[\theta_{-},\theta_{+}] = 4\pi^2(\theta_{+}^2 - \theta_{-}^2) f_{\mathrm{sky}}n_a n_b\ .
\end{equation}

\begin{figure}
  \includegraphics[width=0.5\textwidth]{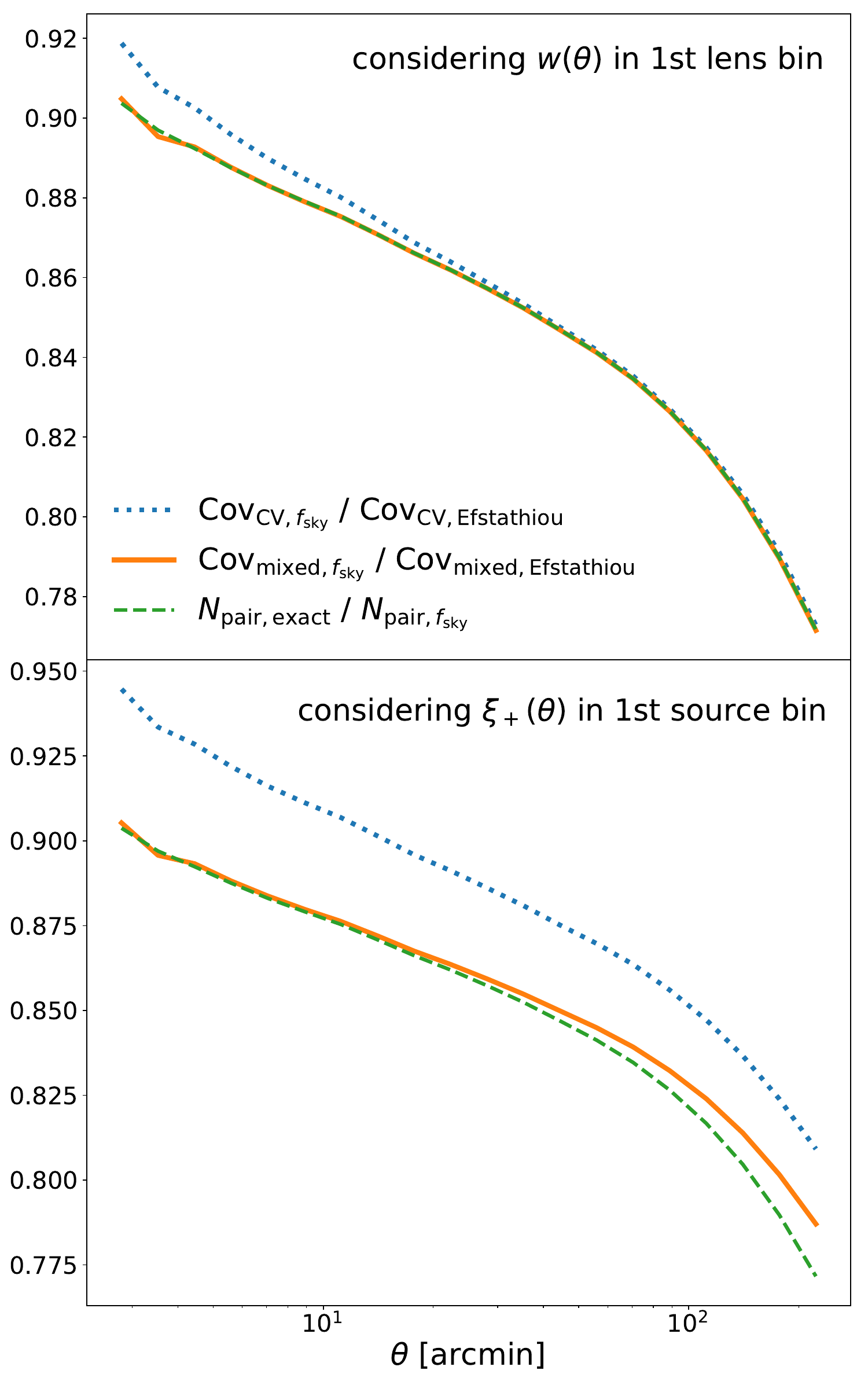}
   \caption{Ratio of exact galaxy pair counts $N_{\mathrm{pair}}$ to pair counts computed using \eqnref{naive_pairs} (green dashed lines) compared to the ratios of different covariance terms when computed with either the $f_{\mathrm{sky}}$ approximation or the approximation of \citet{Efstathiou2004} (cosmic variance term: blue dotted lines; mixed term: solid orange lines).}
  \label{fig:masking_ratios}
\end{figure}

In \figref{masking_ratios} we show the ratio of $N_{\mathrm{pair}}$ to $N_{\mathrm{pair,}f_{\mathrm{sky}}}$ (green dashed lines) and compare it to the ratios of different covariance terms when computed with either the $f_{\mathrm{sky}}$ approximation or the approximation of \citet{Efstathiou2004} (cosmic variance term: blue dotted lines; mixed term: solid orange lines). The upper panel of the figure computes these ratios for the galaxy clustering 2-point function $w(\theta)$ in the first lens bin of our fiducial configuration while the lower panel considers the cosmic shear correlation function $\xi_+(\theta)$ for our first source bins (other redshift bins and 2-point functions behave similarly). For $w(\theta)$ these different ratios indeed closely agree with each other. The agreement between the ratio of pair counts and the ratio of the different approximations for the mixed terms is especially striking. It is most likely caused by the fact that for the mixed covariance terms one of the two replacements in Equations~\ref{eq:replacement_1} and \ref{eq:replacement_2} is actually exact. Even for $\xi_+(\theta)$ the ratio of the different approximations for the mixed term is well described by the ratio of pair counts on most scales. For the cosmic variance of $\xi_+(\theta)$ one can on the other hand observe a strong deviation. This does not necessarily signify a breakdown of our arguments but may be caused by the fact that cosmic shear is a spin-2 field. The calculations of \citet{Efstathiou2004} do in fact only hold for scalar fields and our extension of their formulae to shear correlation functions is only approximate \citep[see \eg][for more general calculations]{Challinor2005}. Since we have identified the mixed terms to carry the strongest impact of masking on the total covariance, we do not address this any further. Instead, we consider the agreement between pair count ratios and mixed term ratios observed in \figref{masking_ratios} as sufficient justification for the re-scaling ansatz of the different covariance terms presented in \secrefnospace{masking_tests}.

\section{Precision matrix expansion to investigate the impact of masking on individual covarince terms}
\label{app:PME}

In this appendix we briefly summarize the PME method that went into \figrefnospace{masking_chiSq_model_vs_PME}. The covariance of the 2x2pt (\ie\ non-cosmic-shear) part of our data vector has contributions from shape-noise because of the presence of the mixed term described in \secref{covariance_modelling}. To pinpoint further which parts of our analytic covariance contribute to the elevation in $\chi^2$ (and to further motivate our heuristic modelling ansatz for masking effect in the covariance presented in \secrefnospace{masking_tests}), we re-run 100 of the \verb|FLASK| simulations with shape-noise turned off. We then use the covariances estimated from the different \verb|FLASK| runs to derive corrections to our covariance model. This can be done - even with only a limited number of simulations - with the method of \emph{precision matrix expansion} (PME) that was described by \citet{Friedrich_Eifler}. At the 1st order their expansion estimates the precision matrix $\boldsymbol{\Psi}$ (\ie\ the inverse covariance matrix) as
\begin{equation}
\label{eq:1st_order_PME}
    \boldsymbol{\hat\Psi} = \mathbf{C}_{\mathrm{model}}^{-1} - \mathbf{C}_{\mathrm{model}}^{-1} (\mathbf{\hat B}-\mathbf{B}_{\mathrm{model}}) \mathbf{C}_{\mathrm{model}}^{-1}\ .
\end{equation}
Here, the matrix $\mathbf{B}_{\mathrm{model}}$ can be either the full covariance model, in which case $\mathbf{\hat B}$ is the full covariance estimated from \verb|FLASK| or it could be the shape-noise free part of the covariance, in which case $\mathbf{\hat B}$ will be the covariance estimated from the shape-noise free \verb|FLASK| simulations. \citet{Friedrich_Eifler} have also derived a 2nd order correction to \eqnref{1st_order_PME}, but given the small magnitude of our observed $\chi^2$ elevation we restrict ourselves to the 1st order expansions which should also reduce the noise of the PME. Note that \eqnref{1st_order_PME} does not contain the inverse of any noisy matrix. This is why PME works well even in the presence of only few numerical simulations (a benefit that is even further boosted because the matrix $\mathbf{B}$ can be chosen to represent only sub-parts of the covariance).

For each \verb|FLASK| measurement of the 2x2pt data vector we estimate the 1st order PME from the remaining 196 \verb|FLASK| data vectors (respectively from the $\sim 100$ shape-noise free data vectors). The average resulting $\chi^2$ values between each data vector and the mean of all data vectors are displayed in \figref{masking_chiSq_model_vs_PME} and compared to the $\chi^2$ values obtained when applying the analytic masking corrections presented in \secref{masking_tests} to either the shape-noise free covariance terms or the full covariance. The average $\chi^2$ when using the analytic, best-guess covariance matrix is $\approx 318.8$ for a total of $302$ data points in the 2x2pt data vector. This corresponds to a bias in $\chi^2$ of about $5.5\%$. The PME estimate of the inverse covariance manages to push this down to $\approx 307.9$ ($\approx 304.8$ with our analytic ansatz) hence decreasing the bias in $\chi^2$ to about $1.9\%$ ($< 1\%$ for the analytic anasatz). If the PME correction term is computed with the shape-noise free \verb|FLASK| covariance, then the bias is only slightly reduces to $\langle \chi^2 \rangle \approx 314.3$ ($\approx 316.6$ with our analytic ansatz). Hence, the shape-noise dependent mixed terms in the covariance indeed seem to be the main cause of our remaining $\chi^2$ offset. This was also found by \citet{Joachimi2020} for the latest analysis of the Kilo Degree Survey. These mixed terms do not depent on the connected 4-point function of the density field (\cf\ \secrefnospace{model}) and the only approximation we make in their calculation is the treatment of our survey footprint through the $f_{\mathrm{sky}}$ approximation. Hence, we follow \citet{Joachimi2020} in our conclusion that this approximation is the main driver of the residual errors in our covariance model.

\section{Impact of extreme cosmologies on parameter constraints}
\label{app:extreme_cosmo}

\begin{figure}
  \includegraphics[width=0.50\textwidth]{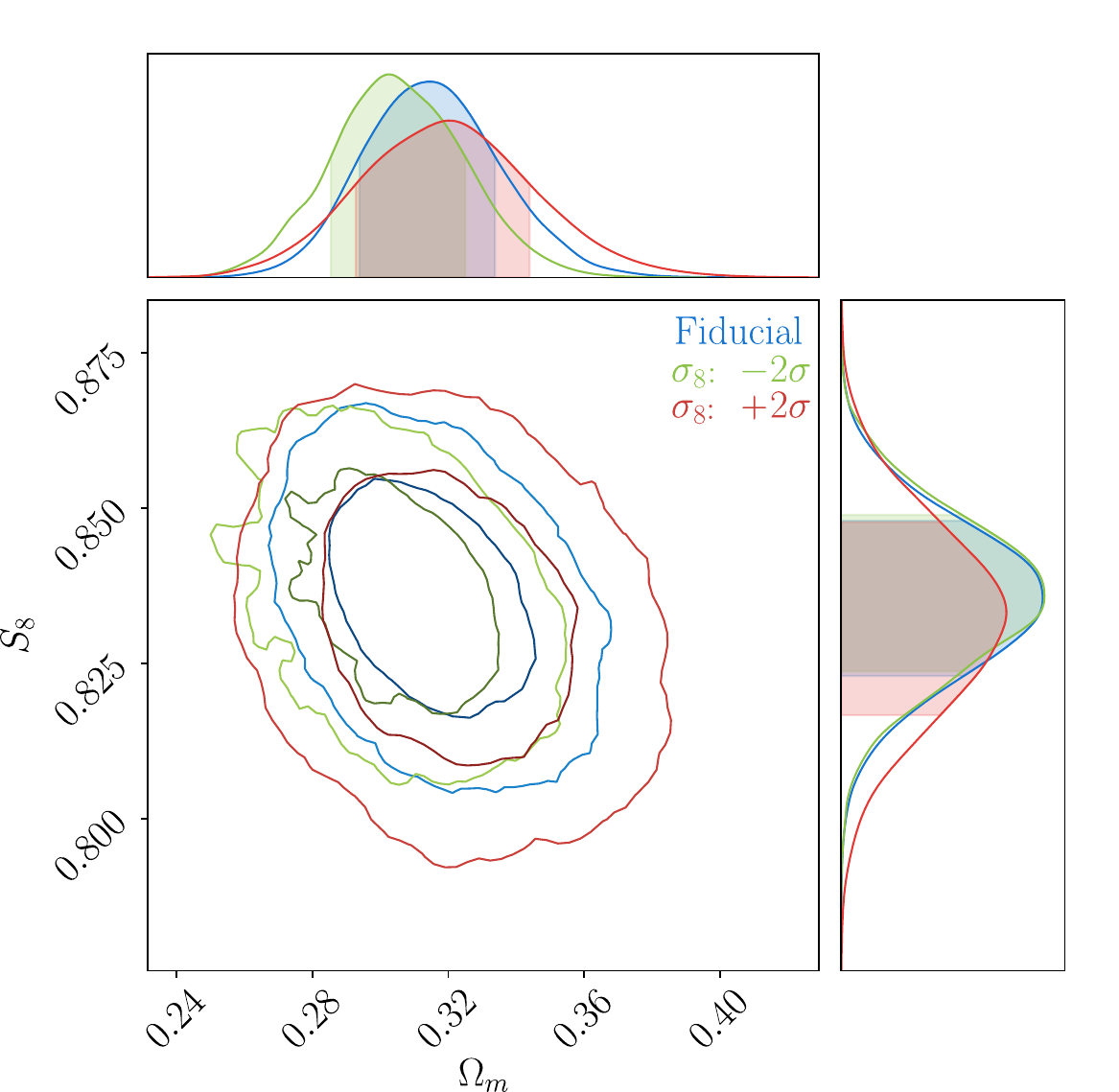}
   \caption{$(S_8, \Omega_m)$ constraints for a given noisy realization of the DES Y3 3x2pt data vector analyzed using: a fiducial covariance matrix (blue); a covariance matrix evaluated at $\sigma_8 = \sigma_8^\text{fiducial} - 2\sigma_{\sigma_8}$ (green); a covariance matrix evaluated at $\sigma_8 = \sigma_8^\text{fiducial} + 2\sigma_{\sigma_8}$ (red). The green and red posteriors were obtained by importance sampling the fiducial samples.}
  \label{fig:extre_sigma8}
\end{figure}

To demonstrate that the importance sampling technique employed in \secref{impact_of_cov_cosmology} indeed manages to capture even strong changes in the likelihood, we repeat the tests presented there with covariance matrices that drastically differ from our fiducial covariance model. In particular we shift the value of $\sigma_8$ for which the covariance model is evaluated by $\pm 2\sigma$ of the marginalised $\sigma_8$ constraints expected from DES-Y3. Note that is a radical change because it ignores parameter degeneracies, \ie\ such a shift of $\sigma_8$ without changes in other parameters would be detected at high significance. \Figref{extre_sigma8} shows the likelihood contours in the $S_8$-$\Omega_m$ plane obtained from both our fiducial covariance and from importance sampling with the altered covariance matrices. One can now clearly see a change in contour width. But as we have show in \secref{impact_of_cov_cosmology}, this effect is far less significant for realistic parameter uncertainties in the covariance model.

\section{Effective shape noise when using metacalibration}
\label{app:metacal}

In \secref{effective_quantities} we have considered how the sampling noise contribution to covariance of the galaxy-galaxy lensing 2-point function can be expressed in terms of an effective shape-noise when each source galaxy is weighted by a certain weight (\eg\ weight $w_j$ for the $j$th galaxy, with the average weight $\langle w_j \rangle_j = 1$).
The expressions derived there have to change when taking into account responses $R_j$ of a shape catalog generated with metacalibation \citep{Sheldon2017}. In that case a measurement of $\hat \gamma_t[\theta_1, \theta_2]$ becomes
\begin{equation}
    \hat \gamma_t[\theta_1, \theta_2] = \frac{\sum_{\mathrm{pxl}\ i,\ \mathrm{source}\ j} \Delta_{[\theta_1, \theta_2]}^{ij}\ \delta_{l,i}\ \epsilon_{t,j\rightarrow i}\ w_j^s}{\sum_{\mathrm{pxl}\ i,\ \mathrm{source}\ j} \Delta_{[\theta_1, \theta_2]}^{ij}\ w_j^s R_j}\ .
\end{equation}
One can re-write this to conform with the derivations of \secref{effective_quantities} by defining
\begin{align}
    \hat \gamma_t[\theta_1, \theta_2] =&\ \frac{\sum_{\mathrm{pxl}\ i,\ \mathrm{source}\ j} \Delta_{[\theta_1, \theta_2]}^{ij}\ \delta_{l,i}\ \frac{\epsilon_{t,j\rightarrow i}}{R_j}\ w_j^sR_j}{\sum_{\mathrm{pxl}\ i,\ \mathrm{source}\ j} \Delta_{[\theta_1, \theta_2]}^{ij}\ w_j^s R_j}\nonumber \\
    \equiv&\ \frac{\sum_{\mathrm{pxl}\ i,\ \mathrm{source}\ j} \Delta_{[\theta_1, \theta_2]}^{ij}\ \delta_{l,i}\ \tilde\epsilon_{t,j\rightarrow i}\ \tilde w_j^s}{\sum_{\mathrm{pxl}\ i,\ \mathrm{source}\ j} \Delta_{[\theta_1, \theta_2]}^{ij}\ \tilde w_j^s}\ .
\end{align}
Now the transformed weights $\tilde w_j^s$ should be normalised to $\langle \tilde w_j^s \rangle = 1$ and then be used together with the transformed ellipticities $\boldsymbol{\tilde\epsilon}_j$ to calculate $\sigma_{\epsilon,\mathrm{eff}}$ from \eqnrefnospace{effective_shape-noise}. This is what we have done for \figrefnospace{gg_lensing_jackknife}.

\def\aj{AJ}%
\def\araa{ARA\&A}%
\def\apj{ApJ}%
\def\apjl{ApJ}%
\def\apjs{ApJS}%
\def\ao{Appl.~Opt.}%
\def\apss{Ap\&SS}%
\def\aap{A\&A}%
\def\aapr{A\&A~Rev.}%
\def\aaps{A\&AS}%
\def\azh{AZh}%
\def\baas{BAAS}%
\def\jrasc{JRASC}%
\def\memras{MmRAS}%
\def\mnras{MNRAS}%
\def\pra{Phys.~Rev.~A}%
\def\prb{Phys.~Rev.~B}%
\def\prc{Phys.~Rev.~C}%
\def\prd{Phys.~Rev.~D}%
\def\pre{Phys.~Rev.~E}%
\def\prl{Phys.~Rev.~Lett.}%
\def\pasp{PASP}%
\def\pasj{PASJ}%
\def\qjras{QJRAS}%
\def\skytel{S\&T}%
\def\solphys{Sol.~Phys.}%
\def\sovast{Soviet~Ast.}%
\def\ssr{Space~Sci.~Rev.}%
\def\zap{ZAp}%
\def\nat{Nature}%
\def\iaucirc{IAU~Circ.}%
\def\aplett{Astrophys.~Lett.}%
\def\apspr{Astrophys.~Space~Phys.~Res.}%
\def\bain{Bull.~Astron.~Inst.~Netherlands}%
\def\fcp{Fund.~Cosmic~Phys.}%
\def\gca{Geochim.~Cosmochim.~Acta}%
\def\grl{Geophys.~Res.~Lett.}%
\def\jcap{JCAP}%
\def\jcp{J.~Chem.~Phys.}%
\def\jgr{J.~Geophys.~Res.}%
\def\jqsrt{J.~Quant.~Spec.~Radiat.~Transf.}%
\def\memsai{Mem.~Soc.~Astron.~Italiana}%
\def\nphysa{Nucl.~Phys.~A}%
\def\physrep{Phys.~Rep.}%
\def\physscr{Phys.~Scr}%
\def\planss{Planet.~Space~Sci.}%
\def\procspie{Proc.~SPIE}%

\bibliographystyle{mnras}
\bibliography{literature}

\section*{Author Affiliations}
\label{app:affiliations}

$^{1}$ Kavli Institute for Cosmology, University of Cambridge, Madingley Road, Cambridge CB3 0HA, UK\\
$^{2}$ Churchill College, University of Cambridge, CB3 0DS Cambridge, UK\\
$^{3}$ Instituto de F\'{i}sica Te\'orica, Universidade Estadual Paulista, S\~ao Paulo, Brazil\\
$^{4}$ Laborat\'orio Interinstitucional de e-Astronomia - LIneA, Rua Gal. Jos\'e Cristino 77, Rio de Janeiro, RJ - 20921-400, Brazil\\
$^{5}$ Department of Physics, University of Michigan, Ann Arbor, MI 48109, USA\\
$^{6}$ ICTP South American Institute for Fundamental Research\\ Instituto de F\'{\i}sica Te\'orica, Universidade Estadual Paulista, S\~ao Paulo, Brazil\\
$^{7}$ Fermi National Accelerator Laboratory, P. O. Box 500, Batavia, IL 60510, USA\\
$^{8}$ Department of Astronomy/Steward Observatory, University of Arizona, 933 North Cherry Avenue, Tucson, AZ 85721-0065, USA\\
$^{9}$ Jet Propulsion Laboratory, California Institute of Technology, 4800 Oak Grove Dr., Pasadena, CA 91109, USA\\
$^{10}$ Department of Astronomy and Astrophysics, University of Chicago, Chicago, IL 60637, USA\\
$^{11}$ Kavli Institute for Cosmological Physics, University of Chicago, Chicago, IL 60637, USA\\
$^{12}$ Kavli Institute for Particle Astrophysics \& Cosmology, P. O. Box 2450, Stanford University, Stanford, CA 94305, USA\\
$^{13}$ Department of Physics and Astronomy, Watanabe 416, 2505 Correa Road, Honolulu, HI 96822\\
$^{14}$ Center for Cosmology and Astro-Particle Physics, The Ohio State University, Columbus, OH 43210, USA\\
$^{15}$ Department of Physics, The Ohio State University, Columbus, OH 43210, USA\\
$^{16}$ Institut d'Estudis Espacials de Catalunya (IEEC), 08034 Barcelona, Spain\\
$^{17}$ Institute of Space Sciences (ICE, CSIC),  Campus UAB, Carrer de Can Magrans, s/n,  08193 Barcelona, Spain\\
$^{18}$ Argonne National Laboratory, 9700 South Cass Avenue, Lemont, IL 60439, USA\\
$^{19}$ Physics Department, 2320 Chamberlin Hall, University of Wisconsin-Madison, 1150 University Avenue Madison, WI  53706-1390\\
$^{20}$ Department of Physics and Astronomy, University of Pennsylvania, Philadelphia, PA 19104, USA\\
$^{21}$ SLAC National Accelerator Laboratory, Menlo Park, CA 94025, USA\\
$^{22}$ Department of Physics, Carnegie Mellon University, Pittsburgh, Pennsylvania 15312, USA\\
$^{23}$ Instituto de Astrofisica de Canarias, E-38205 La Laguna, Tenerife, Spain\\
$^{24}$ Universidad de La Laguna, Dpto. Astrofísica, E-38206 La Laguna, Tenerife, Spain\\
$^{25}$ Department of Astronomy, University of Illinois at Urbana-Champaign, 1002 W. Green Street, Urbana, IL 61801, USA\\
$^{26}$ National Center for Supercomputing Applications, 1205 West Clark St., Urbana, IL 61801, USA\\
$^{27}$ Jodrell Bank Center for Astrophysics, School of Physics and Astronomy, University of Manchester, Oxford Road, Manchester, M13 9PL, UK\\
$^{28}$ Department of Astronomy, University of California, Berkeley,  501 Campbell Hall, Berkeley, CA 94720, USA\\
$^{29}$ Santa Cruz Institute for Particle Physics, Santa Cruz, CA 95064, USA\\
$^{30}$ Department of Physics \& Astronomy, University College London, Gower Street, London, WC1E 6BT, UK\\
$^{31}$ Institut de F\'{\i}sica d'Altes Energies (IFAE), The Barcelona Institute of Science and Technology, Campus UAB, 08193 Bellaterra (Barcelona) Spain\\
$^{32}$ Department of Physics, Stanford University, 382 Via Pueblo Mall, Stanford, CA 94305, USA\\
$^{33}$ D\'{e}partement de Physique Th\'{e}orique and Center for Astroparticle Physics, Universit\'{e} de Gen\`{e}ve, 24 quai Ernest Ansermet, CH-1211 Geneva, Switzerland\\
$^{34}$ Department of Applied Mathematics and Theoretical Physics, University of Cambridge, Cambridge CB3 0WA, UK\\
$^{35}$ Centro de Investigaciones Energ\'eticas, Medioambientales y Tecnol\'ogicas (CIEMAT), Madrid, Spain\\
$^{36}$ Brookhaven National Laboratory, Bldg 510, Upton, NY 11973, USA\\
$^{37}$ Department of Physics, Duke University Durham, NC 27708, USA\\
$^{38}$ Departamento de F\'isica Matem\'atica, Instituto de F\'isica, Universidade de S\~ao Paulo, CP 66318, S\~ao Paulo, SP, 05314-970, Brazil\\
$^{39}$ Instituto de Fisica Teorica UAM/CSIC, Universidad Autonoma de Madrid, 28049 Madrid, Spain\\
$^{40}$ Institute of Cosmology and Gravitation, University of Portsmouth, Portsmouth, PO1 3FX, UK\\
$^{41}$ CNRS, UMR 7095, Institut d'Astrophysique de Paris, F-75014, Paris, France\\
$^{42}$ Sorbonne Universit\'es, UPMC Univ Paris 06, UMR 7095, Institut d'Astrophysique de Paris, F-75014, Paris, France\\
$^{43}$ Department of Physics and Astronomy, Pevensey Building, University of Sussex, Brighton, BN1 9QH, UK\\
$^{44}$ INAF-Osservatorio Astronomico di Trieste, via G. B. Tiepolo 11, I-34143 Trieste, Italy\\
$^{45}$ Institute for Fundamental Physics of the Universe, Via Beirut 2, 34014 Trieste, Italy\\
$^{46}$ Observat\'orio Nacional, Rua Gal. Jos\'e Cristino 77, Rio de Janeiro, RJ - 20921-400, Brazil\\
$^{47}$ Department of Physics, IIT Hyderabad, Kandi, Telangana 502285, India\\
$^{48}$ Department of Astronomy, University of Michigan, Ann Arbor, MI 48109, USA\\
$^{49}$ Institute of Theoretical Astrophysics, University of Oslo. P.O. Box 1029 Blindern, NO-0315 Oslo, Norway\\
$^{50}$ Institute of Astronomy, University of Cambridge, Madingley Road, Cambridge CB3 0HA, UK\\
$^{51}$ School of Mathematics and Physics, University of Queensland,  Brisbane, QLD 4072, Australia\\
$^{52}$ Center for Astrophysics $\vert$ Harvard \& Smithsonian, 60 Garden Street, Cambridge, MA 02138, USA\\
$^{53}$ Australian Astronomical Optics, Macquarie University, North Ryde, NSW 2113, Australia\\
$^{54}$ Lowell Observatory, 1400 Mars Hill Rd, Flagstaff, AZ 86001, USA\\
$^{55}$ Instituci\'o Catalana de Recerca i Estudis Avan\c{c}ats, E-08010 Barcelona, Spain\\
$^{56}$ Department of Astrophysical Sciences, Princeton University, Peyton Hall, Princeton, NJ 08544, USA\\
$^{57}$ School of Physics and Astronomy, University of Southampton,  Southampton, SO17 1BJ, UK\\
$^{58}$ Computer Science and Mathematics Division, Oak Ridge National Laboratory, Oak Ridge, TN 37831\\
$^{59}$ Max Planck Institute for Extraterrestrial Physics, Giessenbachstrasse, 85748 Garching, Germany\\
$^{60}$ Universit\"ats-Sternwarte, Fakult\"at f\"ur Physik, Ludwig-Maximilians Universit\"at M\"unchen, Scheinerstr. 1, 81679 M\"unchen, Germany\\

\end{document}